\documentclass[journal]{IEEEtran}

\usepackage{graphicx}
\usepackage{amsmath,amssymb} 
\usepackage{color}
\usepackage{overpic}
\usepackage[ruled]{algorithm2e}
\usepackage{soul}
\usepackage{xcolor,colortbl}
\usepackage{multirow}
\usepackage{subfigure}
\usepackage{booktabs}
\usepackage{threeparttable}
\usepackage{array}
\usepackage{cite}
\newcommand{\tabincell}[2]{\begin{tabular}{@{}#1@{}}#2\end{tabular}}
\makeatletter
\renewcommand{\@thesubfigure}{\hskip\subfiglabelskip}
\makeatother

\begin{document}
	\title{Towards Low Light Enhancement with \\ RAW Images}
	\author{
		Haofeng Huang,~\IEEEmembership{Student Member, IEEE},
		Wenhan Yang,~\IEEEmembership{Member, IEEE},
		Yueyu Hu,~\IEEEmembership{Student Member, IEEE},
		Jiaying Liu,~\IEEEmembership{Senior Member, IEEE}
		and Ling-Yu Duan,~\IEEEmembership{Member, IEEE}
	\thanks{			
	        This work was supported by the National Key Research and Development Program of China under Grant No. 2018AAA0102702, the National Natural Science Foundation of China under Grant 62088102 and 62172020, State Key Laboratory of Media Convergence Production Technology and Systems, and in part by the PKU-NTU Joint Research Institute (JRI) sponsored by a donation from the Ng Teng Fong Charitable Foundation.
			The associate editor coordinating the review of this manuscript and approving it for publication was Dr. Rafal Mantiuk.
			(Corresponding author: Ling-Yu Duan.)
		}
		\thanks{
		    The authors are with Peking University, Beijing 100871, China. (e-mail: \{hhf, yangwenhan, huyy, liujiaying, lingyu\}@pku.edu.cn)
		}
	}
    \markboth{IEEE TRANSACTIONS ON IMAGE PROCESSING}
	{Huang \MakeLowercase{\textit{et al.}}: Towards Low Light Enhancement with RAW Images}
	\maketitle

	\begin{abstract}
		In this paper, we make the first benchmark effort to elaborate on the superiority of using RAW images in the low light enhancement {and develop a novel alternative route to utilize RAW images in a more flexible and practical way}.
		{Inspired by a full consideration on the typical image processing pipeline, we are inspired} to develop a new evaluation framework, \textit{Factorized Enhancement Model} (\textit{FEM}), which decomposes the properties of RAW images into measurable factors {and provides a tool} for exploring how properties of RAW images affect the enhancement performance {empirically}.
		The {empirical benchmark} results show that the \textit{Linearity} of data and \textit{Exposure Time} recorded in meta-data play the most critical role, which brings {distinct performance gains} {in various measures} over the approaches taking the {sRGB} images as input.
		With the insights obtained from the benchmark results in mind, a RAW-guiding Exposure Enhancement Network (REENet) is developed, which makes trade-offs between {the advantages and inaccessibility of RAW images in real applications in a way of using RAW images only in the training phase.}
		REENet projects {sRGB} images into linear RAW domains to apply constraints with corresponding RAW images to reduce the difficulty of modeling training.
		After that, in the testing phase, our REENet does not rely on RAW images.
		Experimental results demonstrate not only the superiority of REENet to state-of-the-art {sRGB}-based methods and but also the effectiveness of the RAW guidance and all components.
	\end{abstract}
	

	\section{Introduction}
	\label{sec:Intro}
	\IEEEPARstart{L}{ow}-light environments cause a series of degradation in imaging, including intensive noise, low visibility,  color cast, \textit{etc}.
	More sophisticated shooting equipment and advanced specialized photographic systems pay a premium to alleviate the degradation to some extent.
	Modern digital cameras make efforts in tackling the problem by adjusting the shooting parameters but also incur accompanying issues.
	For instance, high ISO introduces amplified noise, and long-exposure time results in blurring.
	Hence, it is economical and desirable to enhance the low-light images by software.

	%
	In most applications, two kinds of images\footnote{https://en.wikipedia.org/wiki/Raw\_image\_format} are taken as the input of the enhanced approaches: \textit{RAW images}~\cite{SID,Jiang_2019_ICCV,CIEXYZNet,Xu_2020_CVPR};
	\textit{RGB images}~\cite{BPDHE,arici2009wahe,dong2011fast,Multi_scale_retinex,LLNet,Kind}, which are processed from raw images via several procedures, \textit{e.g.} demosaicing, white balance, tone mapping, \textit{etc}., in consideration of human vision preference and system requirement, \textit{e.g.} the storage limit.
	As reported in these prevailing works~\cite{SID,Chen_2019_ICCV,Jiang_2019_ICCV}, the low-light enhancement methods that take the RAW data as input usually achieve significantly superior performance to those taking {sRGB} data as their input.
	On one hand, compared with sRGB images, RAW data possesses two inherent advantages: 
	1) \textit{Primitive}: RAW data nearly is obtained directly from the sensor, and records the meta-data related to the hardware and shooting settings, whereas {sRGB} images have been processed for human vision preference and system requirement, which inevitably causes information loss.
	2) \textit{Linear}: As RAW data is directly captured by sensors, RAW data's relationship at different exposure levels keeps linear, while that dependency in the {sRGB} domain is nonlinear as processed by the processing system.

	On the other hand, in real applications, it might be more difficult to obtain RAW images from real applications.
	First, RAW images include abundant information that is stored costly, therefore many devices choose to only store {sRGB} images.
	Second, from the user side, a devastating display of RAW images relies on a series of professional processing operations and expert knowledge.
	Therefore, more casual users prefer a pocket device~\cite{Brown_ICCV_2019}, \textit{e.g.} mobile phone, instead of advanced devices for shooting, \textit{e.g.} digital single-lens reflex (DSLR).
	Therefore, more user-friendly {sRGB} image-based applications are becoming a trend.
	The advantages and disadvantages of using RAW data will be illustrated in detail in Sec.~\ref{sec:Method}-B.
	
	Based on the above discussion, two critical issues are revealed:
	\begin{itemize}
		\item
		What are the properties of RAW files that really contribute to the low-light image enhancement?
		\item
		Is there an alternative way to utilize RAW files for real applications instead of changing existing commonly used image processing systems? 
		For example, can we make full use of advantages of RAW files but get rid of them in testing?
	\end{itemize}
	
	To address these two issues, we start from a benchmark effort.
	Centering at the procedures of the image processing pipeline, we describe the low-light enhancement with a newly proposed \textit{Factorized Exposure Model} (\textit{FEM}). 
	FEM decomposes the ambiguity of low-light image enhancement into several measurable factors,
	\textit{e.g.} a simulation of exposure time adjustment in the image acquisition before processing.
	With the benefits of this framework, we compare several schemes of using RAW data with different combinations of inputs and guidance
	to reveal critical properties of RAW data that make real merits to the low-light image enhancement.
	The benchmark results demonstrate that, among all factors,  \textit{Linearity} of data and \textit{Exposure Time} recorded in meta-data play the most important role in quantitative measures.
	Inspired by this insight, a novel RAW-guiding Exposure Enhancement Network~(REENet) is proposed to show an alternative route that not only utilizes the RAW images but also is user-friendly to sRGB-based applications.
	{Different from previous RAW-based approaches, our REENet takes processed sRGB images as the input and only adopts RAW images as the guidance in the training process,
	while getting rid of them in the testing process.}
	Extensive experimental results demonstrate that our approach outperforms state-of-the-art sRGB-based approaches both quantitatively and qualitatively.
	
	The contributions of this work are summarized as follows,
	\begin{itemize}	
		\item
		To the best of our knowledge, our work is the first benchmark effort to elaborate on the superiority of using RAW images (different inputs/different supervision) quantitatively in the low light enhancement.
		With a detailed analysis, the benchmark results reveal meaningful insights, which inspire us to explore the new route to fill in the gap between sRGB-based and RAW-based approaches.
		\item
		We follow the image processing pipeline and introduce a newly proposed \textit{Factorized Exposure Model} (\textit{FEM}) to describe the low-light enhancement process with several measurable factors that lead to ambiguity, 
		\textit{e.g.} simulating exposure time adjustment in the image acquisition before processing, for benchmarking characteristics of RAW images and the way to utilize them.
		\item
		Inspired by the insights from the benchmark, we further propose a novel RAW-guiding Exposure Enhancement Network (REENet) for low-light enhancement that only needs RAW images as input during the training phase.
		Experimental results show that, the proposed method outperforms state-of-the-art sRGB-based methods when RAW input images are not available.
	\end{itemize}
	The rest of this paper is organized as follows.
	Section~\ref{sec:RW} briefly reviews the related sRGB-based and RAW-based work.
	Section~\ref{sec:Method} shows the benchmark results of various approach for the proposed evaluation framework called~\textit{Factorized Enhancement Model}.
	Section~\ref{sec:Net} introduces the proposed RAW-guiding exposure enhancement network and provide experimental results for comparison, and ablation study.
	Conclusions are summarized in Section~\ref{sec:conc}.
	
	\section{Literature Review}
	\label{sec:RW}	
	
	\renewcommand\arraystretch{0.9}
	\begin{table*}[!t]
		\begin{center}
			\caption{
				An overview of low-light image enhancement methods. Top: sRGB image based methods. Bottom: RAW image based methods. Gray: gray image; sRGB: sRGB image; RAW: RAW image; RAWV: RAW video; $\gamma$: exposure time ratio.
			}
			\begin{threeparttable}
				\begin{tabular}{p{2.4cm}<{\centering}ccp{9.3cm}<{\centering} }
					\toprule
					\specialrule{0em}{2pt}{2pt} 
					Category & Method & Input & Highlight \\
					\specialrule{0em}{2pt}{2pt} 
					\hline
					\specialrule{0em}{2pt}{2pt} 
					\multirow{10}{3cm}{
						\shortstack[l]{
							\textbf{sRGB}: \\
							\\
							- Nonlinearity; \\
							- No meta-data; \\
							- Coarse-grained \\ quantization level; \\
							-
							Easily stored.
						}
					} &
					\tabincell{c}{
						Histogram \\ Equalization
					} & \multicolumn{1}{m{2.5cm}}{ \centering
						Gray/RGB (Test)
					}
					& \multicolumn{1}{m{9cm}}{
						Adjust the illumination via expanding an image's dynamic range by manipulating its histogram
						globally or adaptively in local regions.
					}
					\\
					\specialrule{0em}{2pt}{2pt} 
					&\tabincell{c}{
						Invert \\ Dehazing
					}					
					& \multicolumn{1}{m{2.5cm}}{ \centering
						\tabincell{c}{
							RGB (Test)}
					}
					& \multicolumn{1}{m{9cm}}{
						Improve the visibility by dehazing approaches on the inverted versions of the low-light images.
					}
					\\
					\specialrule{0em}{2pt}{2pt} 
					&\tabincell{c}{
						Statistical \\ Model
					}
					& -- 
					&\multicolumn{1}{m{9cm}}{
						Optimize statistical structural constraints towards desirable properties of images, \textit{e.g.} gradient, and context priors, and environmental light.
					}
					\\
					\specialrule{0em}{2pt}{2pt} 
					&\tabincell{c}{
						Retinex \\ Model
					} & -- 
					& \multicolumn{1}{m{9cm}}{
						Improve the visual quality of low-light images via decomposing them into reflectance and illumination representations and use elaborately designed priors on them.
					} 
					\\ 
					\specialrule{0em}{1pt}{1pt} 
					&\tabincell{c}{
						Deep \\ Learning
					}& 
					\multicolumn{1}{m{2.5cm}}{ \centering
						\tabincell{c}{
							\ \,RGB (Train) \\
							RGB (Test) \\ 
						}
					}
					& \multicolumn{1}{m{9cm}}{
						\vspace{1.5mm}
						Restore normal-light images and pursuit better quality via injecting various kinds of priors and constraints into deep networks with diverse architectures.
					} 
					\\
					\specialrule{0em}{1pt}{1pt} 
					\hline
					\specialrule{0em}{1pt}{1pt} 
					\multirow{17}{3cm}{
						\shortstack[l]{
							\textbf{RAW}: \\
							\\
							- Linear; \\
							- With meta-data; \\
							- Fine-grained \\
							quantization level; \\
							- Costly stored.
						}
					}
					&SID~\cite{SID} & 
					\multicolumn{1}{m{2.5cm}}{ \centering
						\tabincell{c}{
							\ \,RAW, $\gamma$ (Train) \\
							RAW, $\gamma$ (Test) \\ 
						}
					}
					&
					\multicolumn{1}{m{9cm}}{
						\vspace{1mm}
						The first work that takes an end-to-end learnable structure to act as the image processing pipeline for generating normal-light sRGB images with a dataset with extremely dark RAW data and well-exposed sRGB.
					} 
					
					\\
					\specialrule{0em}{1pt}{1pt} 
					&DeepISP~\cite{DeepISP} &
					\multicolumn{1}{m{2.5cm}}{ \centering
						\tabincell{c}{
							\ \,RAW (Train) \\
							RAW (Test) \\ 
						}
					}
					& \multicolumn{1}{m{9cm}}{
						\vspace{1.5mm}
						Propose a new deep network with state-of-the-art results to perform denoising and demosaicing jointly with end-to-end image processing.
					} 
					\\
					\specialrule{0em}{1pt}{1pt} 
					&SMD~\cite{Chen_2019_ICCV} &
					\multicolumn{1}{m{2.5cm}}{  \centering
						\tabincell{c}{
							\ \,RAW, $\gamma$ (Train) \\
							RAW, $\gamma$ (Test) \\ 
						}
					}
					& \multicolumn{1}{m{9cm}}{
						\vspace{1.5mm}
						Construct a static video dataset with the ground truth and proposes a Siamese network to suppress noise while keeping inter-frame stability.
					} 
					\\
					\specialrule{0em}{2pt}{2pt} 
					&SMOID~\cite{Jiang_2019_ICCV} &
					\multicolumn{1}{m{2.5cm}}{ 
						\tabincell{c}{
							Paired RAWV (Train) \\
							RAWV (Test) \\ 
						}
					}
					& \multicolumn{1}{m{9cm}}{
						Develop a novel optical system that is capable of  capturing paired low/normal-light videos and 
						construct
						a fully convolutional network consisting of 3D and 2D miscellaneous operations for image enhancement.
					} 
					\\
					\specialrule{0em}{2pt}{2pt} 
					&ELD~\cite{Wei_2020_CVPR} &
					\multicolumn{1}{m{2.5cm}}{ \centering
						\tabincell{c}{ 
							\ \,RAW, $\gamma$ (Train) \\
							RAW, $\gamma$ (Test) \\ 
						}
					}
					& \multicolumn{1}{m{9cm}}{
						A novel noise formation model to synthesize more realistic extremely dark data for data augmentation that helps the trained models perform better.
					} 
					\\
					\specialrule{0em}{2pt}{2pt} 
					&EEMEFN~\cite{EEMEFN} &
					\multicolumn{1}{m{2.5cm}}{  \centering
						\tabincell{c}{
							\ \,RAW, $\gamma$ (Train) \\
							RAW, $\gamma$ (Test) \\ 
						}
					}
					& \multicolumn{1}{m{9cm}}{
						Take multi-exposed inputs to generate the well-exposed output, which is further enhanced by the edge enhancement module.
					} 
					
					\\
					\specialrule{0em}{2pt}{2pt} 
					&LRDE~\cite{Xu_2020_CVPR} &
					\multicolumn{1}{m{2.5cm}}{  \centering
						\tabincell{c}{
							\ \,RAW (Train) \\
							RAW (Test) \\ 
						}
					}
					& \multicolumn{1}{m{9cm}}{
						Recover objects in low-frequency layers first
						and enhance high-frequency details based on recovered objects later.
					} 
					\\
					%
					\specialrule{0em}{2pt}{2pt} 
					\bottomrule
				\end{tabular}
			\end{threeparttable}
			\label{table:relatedwork}
		\end{center}
	\end{table*}
	
	\renewcommand\arraystretch{1}
	\subsection{sRGB-based Methods}
	\label{sec:trad}
	
	The earlier methods mainly take sRGB images as input.
	The traditional \textit{histogram equalization} methods adjust the illumination via stretching the dynamic range of an image by manipulating its histogram, globally~\cite{HE1, HE2} or in a local adaptive way~\cite{BPDHE,arici2009wahe,HE3,DICM, nakai2013dheci}.
	These methods can effectively adjust the image contrast, but are incapable of changing visual structures of local regions, which inevitably leads to under/over-exposure and amplified noise.
	
	\textit{Inverted dehazing} methods~\cite{dong2011fast, dehaze2, dehaze1} invert low-light images to be haze ones, 
	improve the visibility via dehazing algorithms, 
	and then invert the processed result back as the output.
	Although achieving superior performance in some cases, these methods lack a convincing physical explanation.
	
	\textit{Statistical model} based methods optimize towards desirable properties of images,
	\textit{e.g.} perceptual quality measure~\cite{pdpf}, interpixel relationship~\cite{celik2011cvc}, physical lighting models~\cite{plm}, and
	imaging or visual perception guided models~\cite{BIMEF}.
	These methods show superior effectiveness in their focused aspects. Because of the absence of flexibility in injecting visual properties, these methods fail to handle extreme low-light environments where images are buried with intensive noise.
	
	\textit{Retinex model} based methods~\cite{Single_scale_retinex,Multi_scale_retinex,lee2013amsr,NPE,novel_retinex,FU201682} separate an image into two representations, \textit{i.e.} reflectance and illumination layers,
	and then the well-designed enhancement methods follow to enhance these two layers, respectively. 
	The works in~\cite{Li_2017_SRRM,Ren_2018_SD} enrich Retinex model-based methods with the robust constraint and
	an explicit noise term, which helps better capture and suppress noise.
	
	Since 2017, the low-light enhancement steps into the deep-learning era~\cite{LLNet}. 
	\textit{Deep learning} based methods bring in excellent enhancement performance and flexibility in injecting various kinds of priors and constraints via designing new architectures and training losses~\cite{MSRNet_2017,LLCNN,Lv2018MBLLEN,cai2018learning,Kind,deepUPE,DRD,deepUPE}.
	However, the performance of these methods is dependent on the distribution of the paired training images, which in fact limits the model's generality.
	Recently, learning-based enhancement methods with unpaired data, \textit{e.g.} EnlightenGAN~\cite{jiang2019enlightengan}, Zero-DCE~\cite{Zero-DCE} and DRBN~\cite{Yang_2020_CVPR},
	partially get rid of the issue with CycleGAN, self-learned curve adjustment, and quality guidance, respectively. {Besides, there are many works~\cite{Gharbi_2017_SIGGRAPH,Yang_2018_CVPR,Afifi_2021_CVPR,Wolf_2021_CVPR} dedicated to solving the composite tasks, \textit{e.g.} HDR and Blind Image Restoration, where the low light enhancement just acts as a single component in these pipelines.}

	However, as the image processing systems introduce nonlinearity and discard some fine-grained information when processing RAW images into sRGB ones, 
	the enhancement from sRGB images is highly ill-posed and hard to offer desirable results in the extremely dark condition.
	{
	Furthermore, most of these methods target to restore both illumination (estimating exposure level) and detailed signals (suppressing noise and revealing details). Comparatively, in our work, we target an image acquired with a longer exposure time, where the performance in the dimensions except for the exposure level is paid more attention to and the desired exposure level might be not unique and can be given by users at the testing time.
	}
	
	\subsection{RAW-based Methods}
	Some works make efforts in improving the image quality by building the learnable RAW image processing pipelines~\cite{SID,Chen_2019_ICCV,Jiang_2019_ICCV} or unprocessing the sRGB images back into the RAW domain for a more effective enhancement process~\cite{brooks2019unprocessing,CIEXYZNet,CycleISP,EEMEFN,Xu_2020_CVPR}.
	The signal values in RAW images are totally dependent on the photon number captured by the sensor and have a linear correlation with each other at different exposure levels.
	This property decreases the difficulty in manipulating the image pixel signal,
	and facilitates modeling and enhancing low-light images/videos.
	In \cite{SID}, Chen \textit{et al.} proposed a novel learnable processing strategy for the RAW data captured in extremely dark indoor/nighttime environments, and constructed the See-in-the-Dark (SID) dataset which is the first to provide short/long exposure RAW pairs.
	The successive RAW-based methods \cite{DeepISP,EEMEFN, Xu_2020_CVPR} develop more advanced architectures to further improve the low-light enhancement performance on SID.
	In~\cite{DeepISP}, a new deep network is designed for image enhancement to offer state-of-the-art results in turning the RAW image into a final high perceptual quality image.	
	In \cite{EEMEFN}, Zhu \textit{et al.} proposed a multi-exposure fusion module to combine the generated multi-exposure images with a set of exposure ratios and then adopted an edge enhancement module to produce high-quality results with sharp edges.
	The work in~\cite{Xu_2020_CVPR} also adopts a two-stage framework that introduces attention to context encoding blocks to deal with the restoration of low/high-frequency information at different stages.
	The work in \cite{Wei_2020_CVPR} focuses on noise suppression of RAW images captured in the low-light condition and improving the trained model's capacity via synthesizing more realistic data with the proposed noise model.
	The works in~\cite{Chen_2019_ICCV,Jiang_2019_ICCV} move one step forward to focus on low-light video enhancement.
	In~\cite{Chen_2019_ICCV}, a novel Dark Raw Video (DRV) dataset is created including paired low/normal-light RAW images in static scenes and unpaired low-light RAW images in dynamic scenes, and a new deep network fully considering generalization and temporal consistency is built jointly with VBM4D to effectively enhance the low-light videos while suppressing noise.
	In~\cite{Jiang_2019_ICCV}, Jiang and Zeng developed a novel optical system used to capture low/normal-light videos at the same scene, \textit{i.e.} See-Moving-Objects-in-the-Dark (SMOID) dataset, and built a learnable spatial-temporal transformation to turn the RAW videos into normal-light sRGB ones.

	Although adopting RAW images in learning-based methods leads to a large performance leap, it is still unclear what properties of RAW data contribute to those gains.
	Furthermore, the inaccessibility of RAW files limits their application scopes.
	{In our paper, we aim to benchmark the ways to utilize RAW images quantitatively and, different from RAW-based methods, we explore an alternative way to utilize RAW files for real applications without changing the existing ISP systems.}
	
	\begin{figure*}[t]
		\centering
		\subfigure{
			\includegraphics[width=180mm]{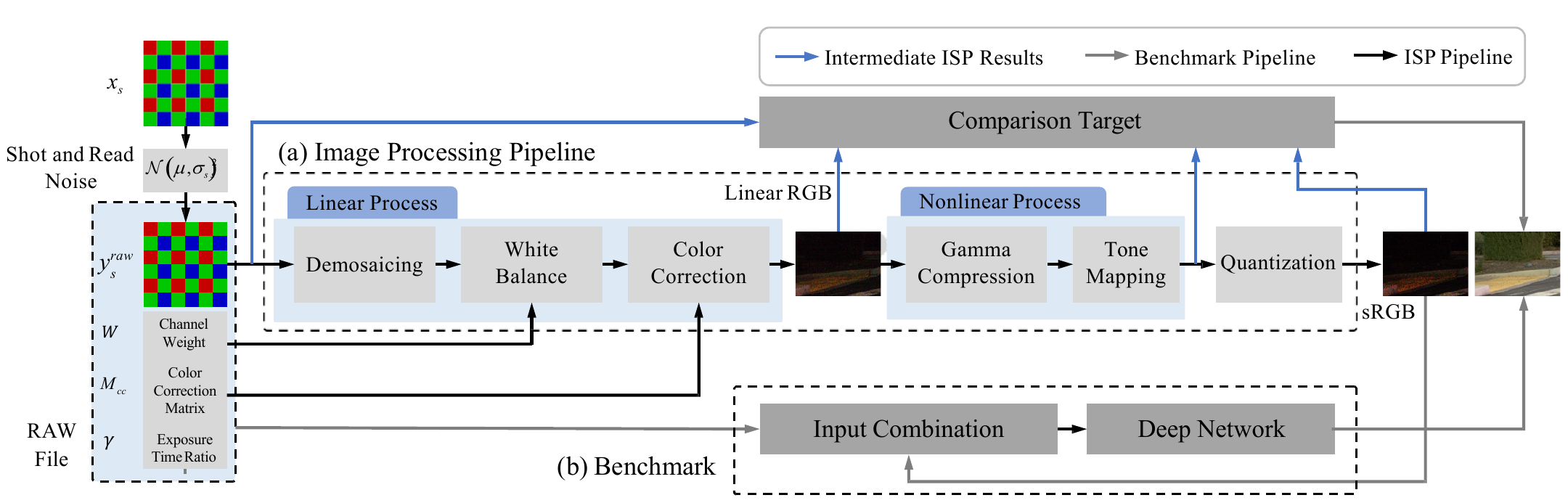}
		}
		\caption{
			An overview of the image processing pipeline and benchmark framework.
			(a) We use a simplified image processing conventional pipeline to model the process.
			The RAW file includes 
			the noisy RAW image $y_{s}^{raw}$,
			channel weight $\omega_c$ for white balance,
			color correction matrix $M_{cc}$ for color correction,
			and exposure time ratio $\gamma$.
			The workflow of the image processing pipeline is \textbf{denoted by black lines}.
			(b) In our benchmark, we aim to compare four output targets from the pipeline, \textit{i.e.} the RAW image $y_{s}^{raw}$, 
			linear RGB image $y^{lin}$, 
			the image before the quantization $y^{srgb}$, 
			and sRGB $f^{srgb}$ (\textbf{{\color{blue}denoted by blue lines}}).
			However, in the benchmark, we cannot access the image processing pipeline and therefore take the RAW image $y_{s}^{raw}$, 
			linear RGB image $y^{lin}$, 
			channel weight $\omega_c$ for white balance,
			color correction matrix $M_{cc}$ for color correction, 
			and exposure time ratio $\gamma$ as the input in our benchmark.
			These factors are combined for comparing the effects of different properties of RAW files on the low-light image enhancement.
			The related workflow is (\textbf{{\color{gray}denoted by gray lines}}).
		}
		\label{fig:pipe}
	\end{figure*}

	\section{Benchmarking RAW Data Utilization in Low-Light Image Enhancement}
	\label{sec:Method}
	\subsection{Motivation}
	Naturally, the low-light image enhancement problem taking the low-light sRGB image as the input image is highly ill-posed.
	Comparatively, restoring from RAW images is much less ambiguous especially when the exposure ratio in the meta-data of RAW files has provided much information about the illumination.
	To compare different methods from the perspective of RAW utilization,
	we formulate the image processing pipeline and propose a novel view to regard low-light image enhancement as the framework of \textit{Factorized Enhancement Model} (\textit{FEM}), 
	which decomposes that ambiguity into several measurable factors, and
	facilitates comparing the effects of various properties of RAW files on low-light image enhancement. 
	
	\subsection{Characteristics of RAW Files}
	\label{sec:property}
	Modern digital shooting systems with the image processing pipeline proceed the sensor data into a more visually pleasant image with less  noise, 
	which is stored as an RGB file~(\textit{e.g.}~sRGB image in JPEG or PNG format).
	Compared with the processed {sRGB} image, the RAW file has the following good properties:
	\vspace{1mm}
	
	\begin{itemize}
		\item{\textbf{Access to meta-data}}. 
		{During image acquisition, cameras record the shooting parameters as the meta-data $d^{meta}$ for original sensor data $d^{sens}$. 
		Influenced by the hardware, the sensor data is highly camera-specific, \textit{e.g.} adopting different black levels, saturation, and lens distortion and being modeled by a camera-specific real-world noise model~\cite{Wei_2020_CVPR}.
		A RAW file $f^{raw}$ consists of sensor data $d^{sens}$ and meta-data $d^{meta}$.
		}
		\vspace{1mm}
		
		\item{\textbf{Linearity of data}}. 
		In a linear image, the pixel values are directly related to real-world signal, \textit{i.e.} the number of photons received at that location on the sensor and therefore keep a linear correlation at different exposure levels.
		To restore \textit{linear RAW data} $y^{raw}$ from the sensor data $d^{sens}$, the hardware calibration operations $F_{cali}\left(\cdot\right)$ such as linearization and lens calibration are applied.
		A theoretically perfect calibration can decouple sensor data with its capture equipment, making the calibrated signal linearly depend on the real-world signal:
		\begin{equation}
			y^{raw} = F_{cali}(d^{meta}, d^{sens}, \alpha)\,,
		\end{equation}
		{Note that because sensor data $d^{sens}$ is stored discretely, the restored $y^{raw}$ is discrete as well. As the distributions of noise and bias induced by the hardware are quite complex and data-dependent~\cite{Wei_2020_CVPR}, a perfect calibration is hard to obtain.}
		\vspace{1mm}
		
		\item{\textbf{Fine-grained quantization level (\textit{i.e.} more abundant intensities and colors)}}.
		Most RAW files contain much abundant information, due to their high resolution and wide signal range capturing more fine-grained intensities and colors.
		However, the RAW images are stored costly and unfriendly to be displayed to the human vision (a nonlinear perception system), which limits the application scopes of RAW images.
		The final output of a processing system is usually an 8-bit sRGB image. 
	\end{itemize}
	\vspace{1mm}
	
	The aforementioned characteristics disappear when the RAW files are processed into final sRGB images.
	Most image processing systems serve human vision perceptual quality, therefore the successive adjustment stages in processing based on human vision are conducted,~\textit{e.g.} white balance, tone mapping and gamma correction.
	A standard sRGB system produces a nonlinear sRGB image $y^{srgb}$ as follows:
	\begin{equation}
		\label{eq:proc}
		f^{srgb} = F_{proc}(d^{meta}, y^{raw}, \beta)\,,
	\end{equation}
	where $F_{proc}(\cdot)$ denotes the processing stage, 
	and $\beta$ denotes the configuration.
	In Section~\ref{sec:pipe}, we will provide a more detailed analysis of $F_{proc}(\cdot)$.
	After the processing, a nonlinear 8-bit image $f^{srgb}$ is obtained.
	Note that the meta-data might be also available for {sRGB} files as well, \textit{e.g.} EXIF in JPEG format or a coupled metadata file directly obtained from the digital camera. In more common cases, \textit{e.g.} the images on the Internet and social networks, or the edited images by post-processing or editing, the perfect meta-data is hardly available. Therefore, in our paper, the final version of the proposed method does not rely on access to the meta-data in the final sRGB files. However, to make our paper more comprehensive, we also discuss situations where sRGB files are coupled with the perfect metadata recorded or not to see how it benefits the enhancement.
	
	To summarize, characteristics of RAW files include the access to meta-data, linearity of the data, as well as fine-grained quantization level (\textit{i.e.} more abundant intensities and colors).
	These properties disappear when the RAW files are projected into the final sRGB images via the image processing systems.
	\vspace{1mm}
	
	\subsection{Image Processing Pipeline}
	\label{sec:pipe}
	{
	In this section, we describe the image processing pipeline $F_{proc}\left( \cdot \right)$  in Eqn.~\eqref{eq:proc}. 
	As the specific pipelines and configurations of the processing systems in each kind of camera are kept as commercial secrets, in our discussion, we treat these details as black boxes. 
	Despite this, the conventional image processing system~\cite{brooks2019unprocessing} also helps establish a concise mathematical model as shown in Fig.~\ref{fig:pipe}~(a), which we can make use of as the framework to evaluate the properties of RAW that benefit low-light image enhancement as shown in Fig.~\ref{fig:pipe}~(b).
	Note that all sRGB images used in benchmark as the final targets are processed by Libraw, which is regarded as a black box in our discussion.
	Our defined simplified processing pipeline, including a simplified demosaicking module, only provides the intermediate supervision in the RAW domain and does NOT actually influence benchmark results.
	}
	\vspace{1mm}
	
	\noindent \textbf{Shot and Read Noise}.
	The real-world signal recorded in RAW files is mixed with physically caused noise. Compared with sRGB,
	the noise model in the RAW domain is seldom disturbed by the nonlinearity in the processing pipeline.
	Sensor noise in the RAW domain consists of two parts: shot noise and read noise~\cite{Hasinoff2014}.
	{By using the fixed aperture and ISO, the value of noise-free signal $x$ is linearly dependent on the exposure time.}
	Specifically, to simulate shooting in the low-light conditions, we utilize a short-exposure time, then the sensor data $y_{s}^{raw}$ in the RAW domain can be formulated:
	\begin{equation}
		y_{s}^{raw} = {x}_{s} + {n}_{shot}({x}_{s}) + {n}_{read}\,,
	\end{equation}
	where $x_{s}$ is the noise-free short-exposure RAW image and $y_{s}^{raw}$ is the noisy one, ${n}_{shot}$ and ${n}_{read}$ are shot noise and read noise. {Subscript $s$ denotes \textit{short-exposure} here.} As simplified in~\cite{Foi_TIP_2008},
	\begin{align}
		\label{equ:short}
		y_{s}^{raw} &\simeq {x}_{s} + {n}({x}_{s})\,, \nonumber\\
		{n}({x}_{s})[i] &\sim \mathcal N (0,\sigma_s^{2}[i])\,,	 \\
		\sigma_s^{2}[i] &= \lambda_{shot} {x}_{s}[i] + \lambda_{read}\,, \nonumber
	\end{align}
	where $\lambda_{shot}$ and $\lambda_{read}$ denote the noise levels for a camera, $i$ denotes location and $[\cdot]$ returns the value at location $i$. 
	\vspace{1mm}
	
	\begin{figure}[t]
		\centering
		\includegraphics[width=8cm]{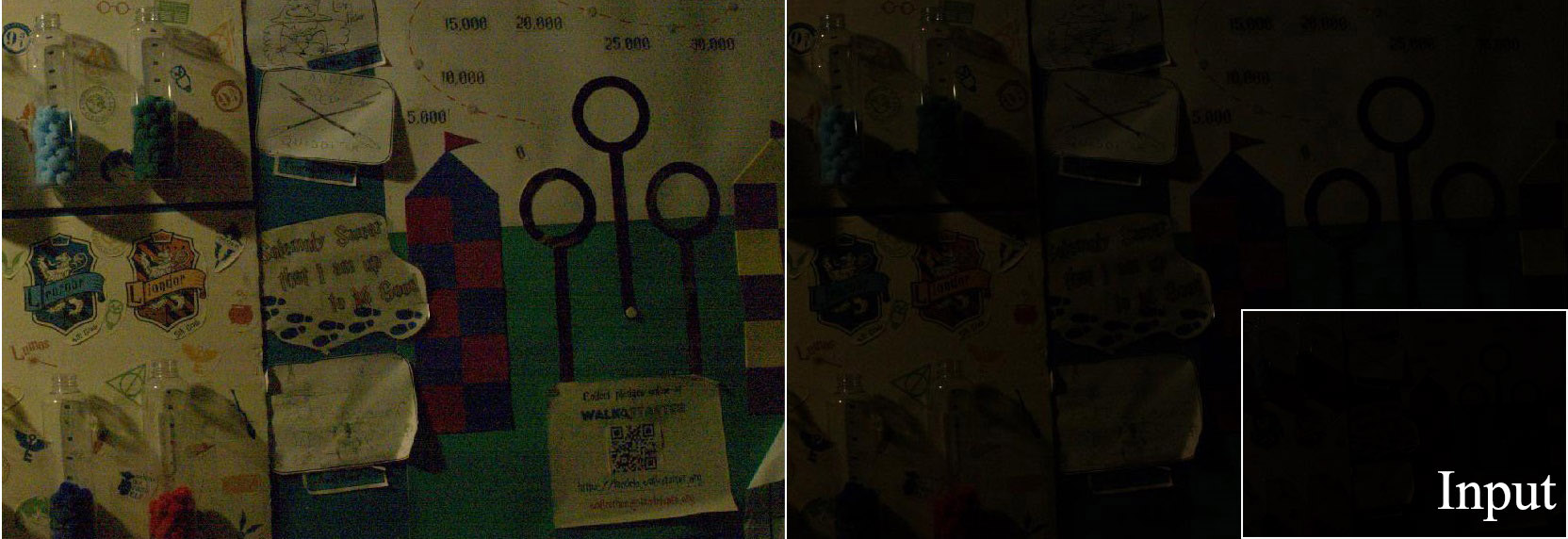}
		\caption{
			Visual comparison results of Gamma correction results with different settings.
			The sRGB-standard Gamma compression is utilized in the original image.
			\textbf{Left Panel}: The results brightened by inverting the sRGB-standard Gamma function.
			\textbf{Right Panel}: The results brightened by the inversion of Adobe RGB (1998) standard.
		}
		\label{fig:gamma}
		\vspace{-5mm}
	\end{figure}

	\noindent \textbf{Demosaicing}. 
	Since the sensor is only capable of capturing photons, not aware of the chromatic light, 
	to precept the chroma information, in the camera
	the pixels are covered by colored filters that are arranged with a certain pattern, \textit{e.g.} the R-G-G-B Bayer pattern.
	Demosaicing is one of the processing stages that helps reconstruct the full-size color image.
	In our implementation, the R-G-G-B pattern is converted into RGB channels via averaging green channels and adopting Bilinear interpolation to upsample the resolution to $m \times n \times 3$.
	
	\vspace{2mm}
	\noindent \textbf{White Balance and Color Correction}.
	Since the filtered sensor data is affected by the color temperature of the ambient light, 
	the camera applies the white balance to generate images under the normal illumination with the colors visually pleasing to human eyes.
	In this stage,
	three channels are multiplied with the weights $w_{c}\,(c=r,g,b)$, which are obtained from the RAW file.
	Note that, the light metering obtained from the low-light conditions might be inaccurate~\cite{Chen_2019_ICCV}, 
	those weights (denoted by $\widehat{w_{c}}$) are usually biased and need additional calibration.
	This module is formalized:
	\begin{align}
		\label{equ:wb}
		y_{s}^{wb} &= y_{s}^{raw} \circ \widehat{W}\,, \nonumber \\
		y_{l}^{wb} &= y_{l}^{raw} \circ W\,, \\
		\widehat{W} &= {\left[\left[\left[\widehat{w_{r}}, \widehat{w_{g}}, \widehat{w_{b}}\right]\right]\right]}_{1 \times 1 \times 3}\,, \nonumber \\
		W &= {\left[\left[\left[w_{r}, w_{g}, w_{b}\right]\right]\right]}_{1 \times 1 \times 3}\,, \nonumber
	\end{align}
	where $\circ$ means element-wise product {and subscripts $s$ and $l$ denote \textit{short-exposure} and \textit{long-exposure}, respectively.}
	A color correction follows to adopt a $3 \times 3$ color correction matrix (CCM) to transform the color space of the camera to the output one, namely sRGB. 
	We obtain the CCM $M_{cc}$ from the meta-data of RAW files.
	{To be specific, the matrix converting the camera color space into XYZ color space is usually recorded in RAW files or a configure file in the processing systems, \textit{e.g.} being stored in EXIF, and the matrix parameter converting the XYZ color space into sRGB color space is fixed.}
	This module is formalized:
	\begin{equation}
		y^{lin}=
		{y^{cc}}_{3 \times (m \times n)} = 
		\left( \begin{array}{c}
			y_{r}^{cc}\\ 
			\quad \\
			y_{g}^{cc}\\  
			\quad \\
			y_{b}^{cc}
		\end{array} \right) =
		\left( \begin{array}{c}
			y_{r}^{wb}\\ 
			\quad \\
			y_{g}^{wb}\\  
			\quad \\
			y_{b}^{wb}
		\end{array} \right)
		M_{cc}\,.
		\label{equ:cc}
	\end{equation}
	~\\
	For convenience, $y^{cc}$ is represented equivalently as $y^{lin}$.
	We call the procedure that converts $y^{raw}$ into $y^{lin}$ as \emph{linear process}.
	\vspace{1mm}
	
	\noindent \textbf{Gamma Compression and Tone Mapping}. 
	To make the images better perceived by humans, 
	nonlinear procedures are further conducted,
	including Gamma compression as well as tone mapping~\cite{brooks2019unprocessing}.
	For simplicity, more details about these two stages are skipped.
	We use a function $\sigma(\cdot)$ to denote the \emph{nonlinear process} consisting of these two stages as follows:
	\begin{equation}
		y^{srgb} = \sigma (y^{cc})\,.
		\label{equ:gamma}
	\end{equation}
	These nonlinear procedures introduce considerable  ambiguity for creating the inverse mapping of low-light image enhancement.
	For example, as shown in Fig.~\ref{fig:gamma}, if we cannot obtain the Gamma compression function accurately, 
	a huge gap between the brightened images\footnote{The brightened images are generated by sequential operations of inverted Gamma compression, multiplication with the ratio of exposure time, and Gamma compression.}
	by inverting two Gamma functions is incurred.
	It is demonstrated that, for different low-light images, the proper inverse Gamma functions should be adopted adaptively. 
	\vspace{1mm}
	
	\noindent \textbf{Quantization}.
	Finally, the quantization $Q(\cdot)$ comes to turn the data with more fine-grained quantization levels into 8bit to obtain a more compact representation for saving storage as follows:
	\begin{equation}
		f^{srgb} = Q(y^{srgb})\,.
		\label{equ:quantize}
	\end{equation}
	
	\subsection{Evaluation Framework: Factorized Enhancement Model}
	For the benchmark, we regard the low-light enhancement as a simulation of amplifying the exposure time during capturing, 
	which has a concise mathematical form and yields conveniences 
	for an accurate and controllable enhancement process of low-light images.
	{With exposure amplified $\gamma$ times, a corresponding long-exposure data $y_{l}^{raw}$,
	which is usually approximated as noise-free because of high Signal-Noise Ratio (SNR),
	can be represented as follows:} 
	\begin{equation}
		{y_{l}^{raw} \simeq x_{l}} = \gamma x_{s}\,,
		\label{equ:long}
	\end{equation}
	{where $x_{s}$, $x_{l}$ and $y_{l}^{raw} \in R^{m \times n \times 1}$ are the latent radiance value without any noise with a short-exposure shot, that with a long-exposure shot, and the measured noisy value captured with the long-exposure time, respectively.}
	Therefore, if the exposure ratio in the normal-light environment is given, 
	the low-light enhancement is intrinsically close to denoising on already properly brightened RAW images $  y_{b}^{raw}$:
	\begin{align}
		\label{equ:brightened}
		y_{b}^{raw} = \gamma y_{s}^{raw} &\simeq \gamma x_s + \gamma n(x_s)\,, \\
		\gamma n(x_s)[i] &\sim \mathcal{N}(0, \sigma_b^2[i])\,, \nonumber
	\end{align}
	{with $\sigma_b^{2}[i] = {\gamma\lambda_{shot}(\gamma x_{s}[i])+\gamma}^2\lambda_{read}\,$}, {where subscript $s$ signifies \emph{short-exposure}, $l$ represents \emph{long-exposure},
	and $b$ means \emph{brightened}.}
	Therefore, the enhancement model $F_{{enhance}}\left( \cdot \right)$ can be represented as follows,
	\begin{equation}
		\label{eq:enhance}
		\hat{f}^{srgb} = F_{{enhance}}\left( y_{b}^{raw} \right)\,.
	\end{equation}
	where $\hat{f}^{srgb}$ is the prediction of the enhancement model $F_{{enhance}}\left( \cdot \right)$. 
	Eqn.~\eqref{eq:enhance} provides a flexible way to benchmark RAW utilization as shown in Fig.~\ref{fig:pipe}~(b).
	That is, $y_{b}^{raw}$ can be replaced with any reasonable combination of images (RAW/sRGB images) and meta-data in the input combination module.
	After that, the input is feed-forwarded into a deep network for low-light image enhancement.
	From the performances of deep networks with different inputs, we can infer the importance of properties of RAW  files for the enhancement.  
	
	\begin{figure}[t]
		\centering
		\subfigure[Linearity]{
			\includegraphics[width=4cm]{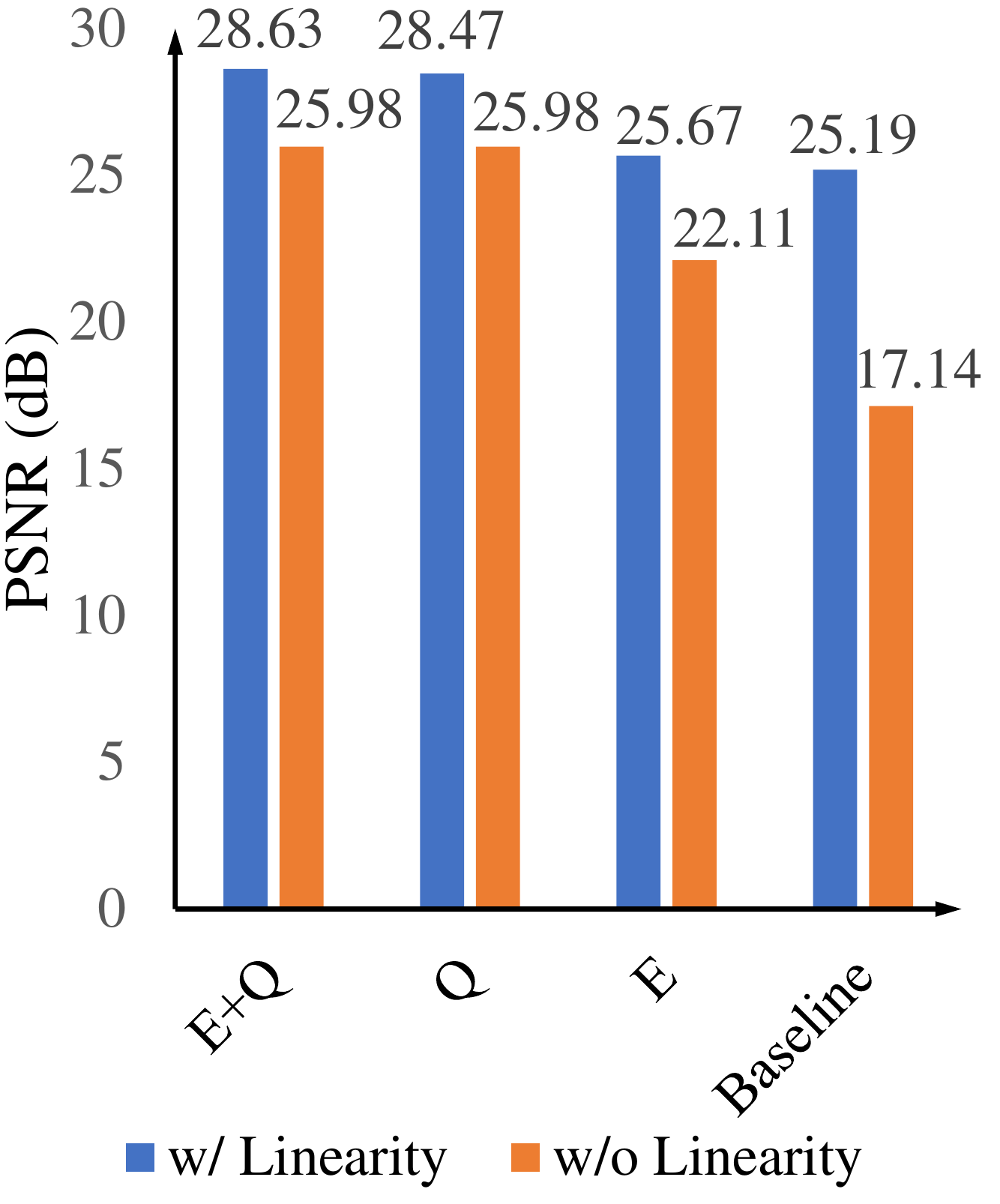}
		}
		\subfigure[Exposure Time]{
			\includegraphics[width=4cm]{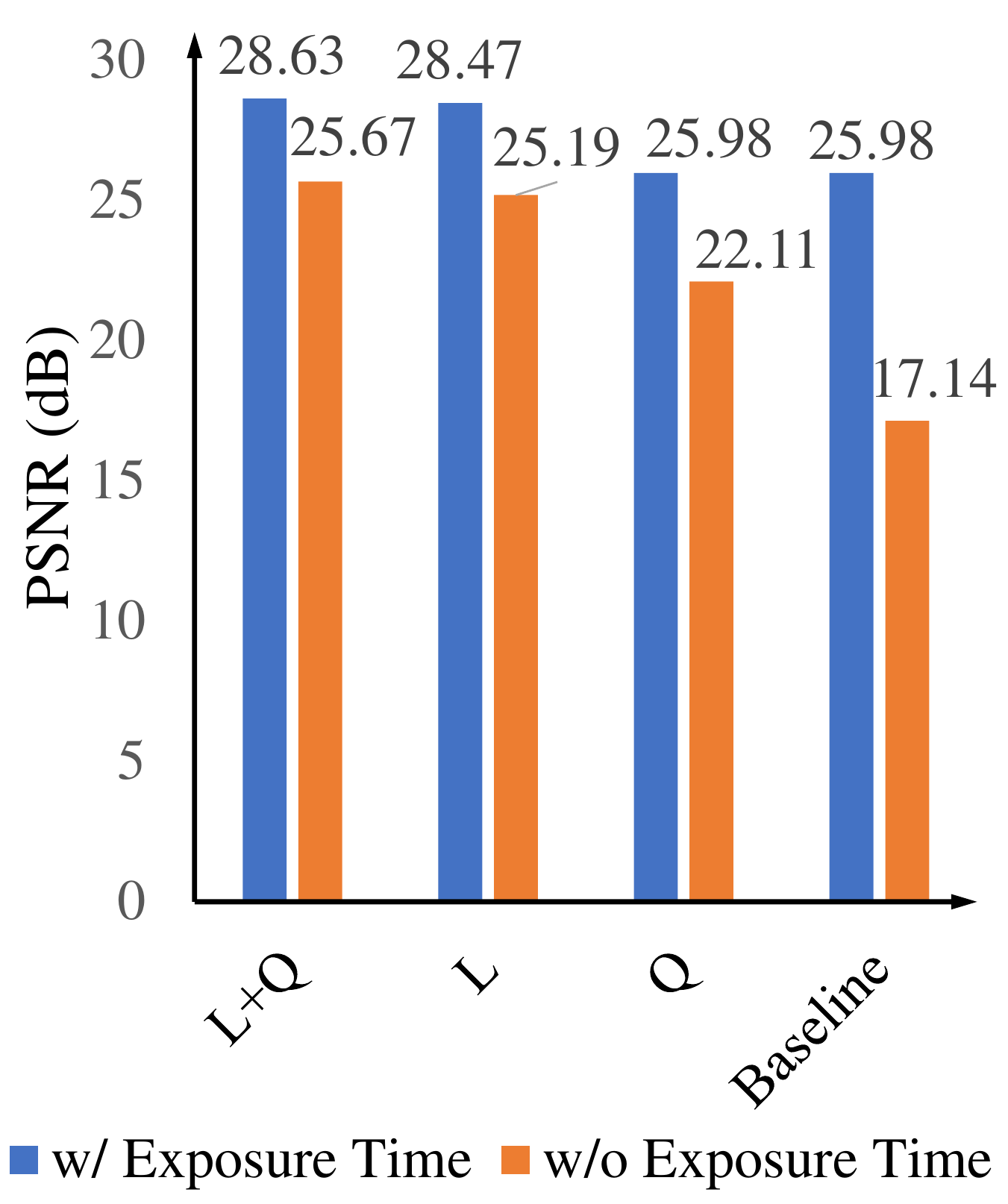}
		}
		\subfigure[Quantization Levels]{
			\includegraphics[width=4cm]{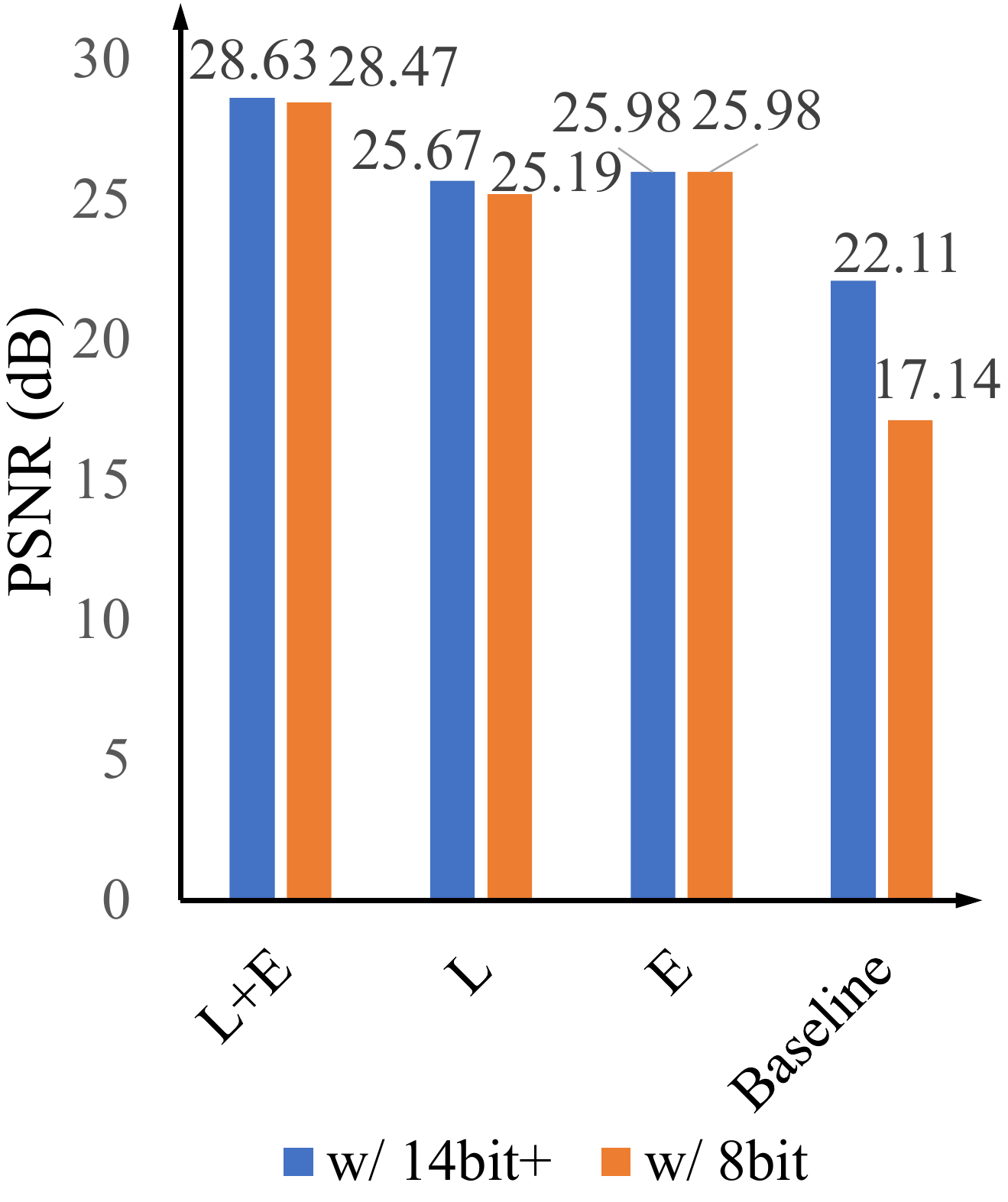}
		}
		\subfigure[White Balance]{
			\includegraphics[width=4cm]{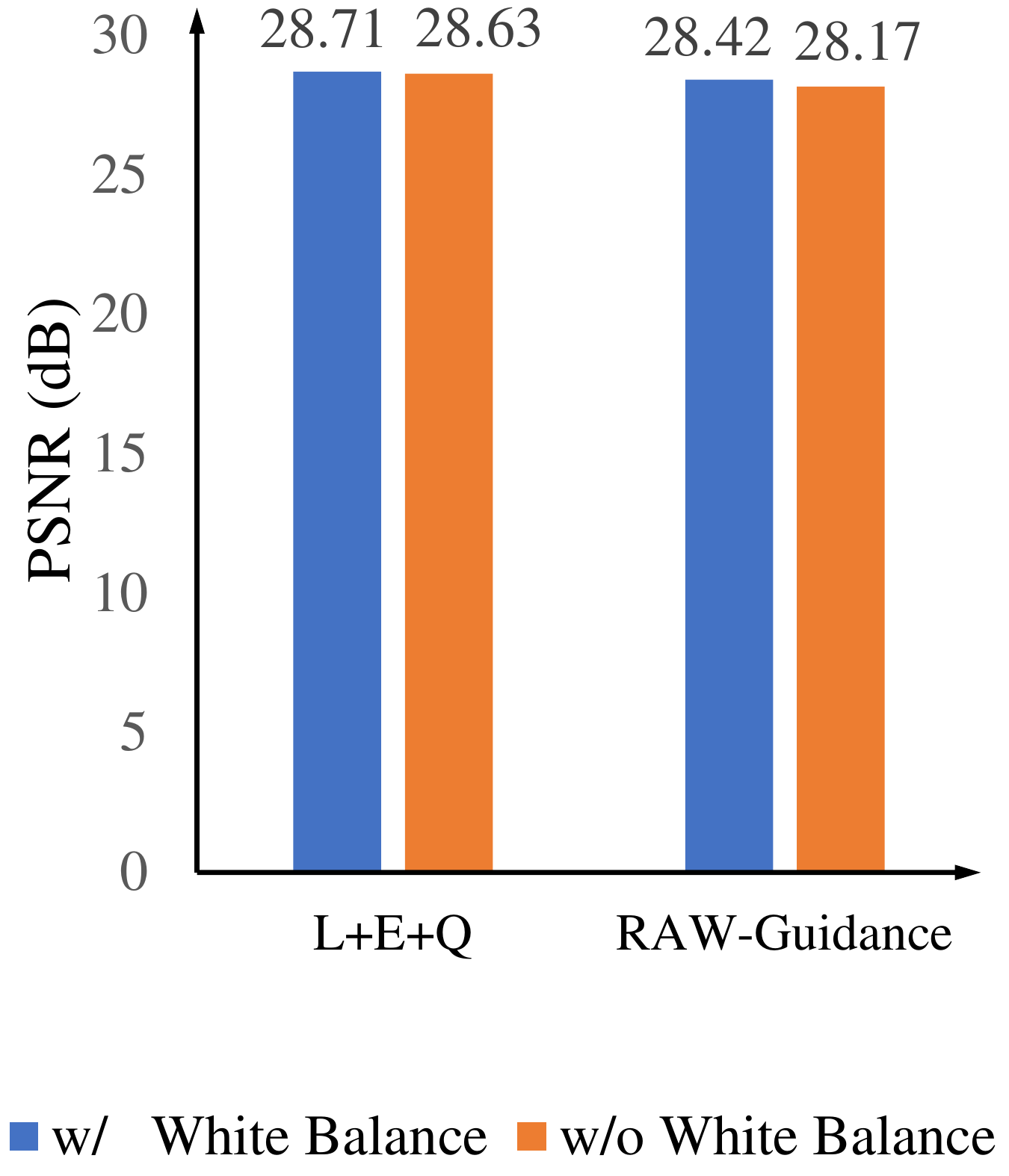}
		}	
		\caption{
			Evaluation of Effects of RAW properties on low-light image enhancement.
			E denotes \emph{exposure time}.
			{Q denotes \emph{dense quantizication levels}.}
			L denotes \textit{linearity}.
			Baseline denotes the method taking 8bit (sRGB) as the input.
			RAW-guidance denotes the proposed method in Sec.~\ref{sec:Net}.
		}
		\label{fig:L}
	\end{figure}
	
	\subsection{RAW Benchmarking}
	\label{sec:bench}
	In this benchmark, we compare several schemes of RAW data utilization with different inputs and guidance 
	to explore how many contributions the characteristics of RAW data can bring in to the low-light enhancement task.
	The effects of different characteristics including linearity, exposure time and white balance parameters recorded in metadata, and {quantization} levels, denoted by L, E, W and Q, are analyzed with experimental results.
	\vspace{1mm}

	\noindent \textbf{Experimental settings.}
	SID dataset~\cite{SID} is adopted for training and evaluation.
	We use Sony sub-dataset, constructed with a Sony $\alpha$7S II equipped with a Bayer sensor.
	The subset contains 409 paired low/normal-light RAW images.
	The training, testing, and validation sets include 280, 93, and 36 paired images.
	Based on characteristics of RAW files mentioned in Section~\ref{sec:property}, we employ different operations on input/target pairs and feed-forward them into the similar architecture~\textit{i.e.} U-Net~\cite{U-Net} for performance comparisons. 
	All approaches are trained from scratch on SID. For RAW based approaches, 
	the training settings follow the paradigm of~\cite{SID} \textit{i.e.}, unpacking the RAW data with Bayer pattern into 4 channels, linearizing the data, and normalizing it into [0, 1].
	Then, the data is fed into a U-Net~\cite{U-Net}.
	{
	For sRGB-based approaches, corresponding sRGB images are processed by Libraw, where the histogram stretching~\cite{SID} is not adopted because it will brighten images during processing, 
	which is far away from our both targets in benchmarking and developing a novel RAW utilization paradigm.
	}
	The network is trained with an $L_1$ loss with normal-light sRGB images as ground truths.
	{
	The benchmark results in PSNR ans SSIM are shown in Table~\ref{table:L}-\ref{table:meta}.
	The extended tables with more metrics are provided in Table I-V of the supplementary material due to the limited space.}
	\vspace{1mm}
	
		\begin{table}[t]
		\begin{center}
			\caption{
				Evaluation on effects of \emph{linearity}.
				Z denotes Zero-DCE, which aims to simulate the illumination adjustment guided by the exposure time ratio.
				$\hat{\gamma}$ denotes to the ratio calculated with the mean pixel value.
				$8bit(\cdot)$ denotes storing data with 255 quantification levels.
			}
			\footnotesize
			\resizebox{0.5\textwidth}{!}{
			\begin{tabular}{c|c|c|ccc|cc}
				\hline
				\multirow{2}{*}{Methods}& \multirow{2}{*}{Input}& \multirow{2}{*}{Model}& \multicolumn{3}{c|}{Characteristics}
				& \multirow{2}{*}{PSNR$\uparrow$}
				& \multirow{2}{*}{SSIM$\uparrow$}
				\\ \cline{4-6} 
				& & & L& E & Q& \\
				\hline
				L+E+Q&RAW$\times\gamma$
				& U-Net &$\checkmark$&$\checkmark$&$\checkmark$
				&$28.63$
				&$0.890$
				\\
				E+Q & $\left[ \text{sRGB}, \gamma \right]$
				& Z+U-Net & & $\checkmark$&$\checkmark$
				&$25.98$
				&$0.821$
				\\
				\hline			
				L+E& 8bit(RAW$\times\gamma$) & U-Net
				&$\checkmark$&$\checkmark$& 
				&$28.47$
				&$0.889$
				\\
				E&  $\left[ \text{sRGB}, \gamma \right]$& Z+$8bit$+U-Net
				& &$\checkmark$& 
				&$25.98$
				&$0.820$
				\\
				\hline
				L+Q & RAW$\times\hat{\gamma}$  
				& U-Net & $\checkmark$& &$\checkmark$
				&$25.67$
				&$0.855$
				\\
				Q& sRGB
				& U-Net &  & &$\checkmark$
				&$22.11$
				&$0.772$
				\\
				\hline
				L&8bit(RAW$\times\hat{\gamma}$)  & U-Net
				&$\checkmark$& & 
				&$25.19$
				&$0.846$
				\\
				Baseline&8bit(sRGB)  
				& U-Net & &  &
				&$17.14$
				&$0.374$
				\\
				\hline
			\end{tabular}
			}
			\label{table:L}
		\end{center}
	\end{table}
	
	\begin{table}[t]
		\begin{center}
			\caption{
				Evaluation on effects of \emph{exposure time}. 
				Z denotes Zero-DCE, which aims to simulate the illumination adjustment guided by the exposure time ratio.
				$\hat{\gamma}$ denotes to the ratio calculated with the mean pixel value.
				$8bit(\cdot)$ denotes storing data with 255 quantification levels.
			}
			\resizebox{0.5\textwidth}{!}{
			\begin{tabular}{c|c|c|ccc|cc}
				\hline
				
				\multirow{2}{*}{Methods}& \multirow{2}{*}{Input}& \multirow{2}{*}{Model}& \multicolumn{3}{c|}{Characteristics}
				& \multirow{2}{*}{PSNR$\uparrow$}
				& \multirow{2}{*}{SSIM$\uparrow$}
				\\ \cline{4-6} 
				& & & L& E & Q& \\
				\hline
				L+E+Q&RAW$\times\gamma$
				& U-Net
				&$\checkmark$&$\checkmark$&$\checkmark$
				&$28.63$
				&$0.890$
				\\
				L+Q&RAW$\times\hat{\gamma}$
				& U-Net
				&$\checkmark$& &$\checkmark$
				&$25.67$
				&$0.855$
				\\
				\hline
				L+E&8bit(RAW$\times\gamma$)
				& U-Net
				&$\checkmark$&$\checkmark$& 
				&$28.47$
				&$0.889$
				\\
				L&8bit(RAW$\times\hat{\gamma}$)
				&U-Net
				&$\checkmark$& &
				&$25.19$
				&$0.846$
				\\
				\hline
				E+Q & $\left[\text{sRGB},\gamma \right]$
				& Z+U-Net
				& &$\checkmark$&$\checkmark$
				&$25.98$
				&$0.821$
				\\
				Q&sRGB
				& U-Net
				&  & &$\checkmark$
				&$22.11$
				&$0.772$
				\\
				\hline
				E& $\left[\text{sRGB}, \gamma \right]$
				& Z+$8bit$+U-Net
				& &$\checkmark$& 
				&$25.98$
				&$0.820$
				\\
				Baseline & 8bit(sRGB) 
				& U-Net
				& & & 
				&$17.14$
				&$0.374$
				\\
				\hline
			\end{tabular}
			}
			\label{table:E}
		\end{center}
	\end{table}

	\begin{table}[t]
		\begin{center}
			\caption{
				Evaluation on effects of \emph{quantization}.
				Z denotes Zero-DCE, which aims to simulate the illumination adjustment guided by the exposure time ratio.
				$\hat{\gamma}$ denotes to the ratio calculated with the mean pixel value.
				$8bit(\cdot)$ denotes storing data with 255 quantification levels.
			}
			\resizebox{0.5\textwidth}{!}{
			\begin{tabular}{c|c|c|ccc|cc}
				\hline
				
				\multirow{2}{*}{Methods} & \multirow{2}{*}{Input} & \multirow{2}{*}{Model}
				& \multicolumn{3}{c|}{Characteristics}
				& \multirow{2}{*}{PSNR$\uparrow$}
				& \multirow{2}{*}{SSIM$\uparrow$}
				\\ \cline{4-6} 
				& & & L& E & Q& \\
				\hline
				L+E+Q&RAW$\times\gamma$
				& U-Net
				&$\checkmark$&$\checkmark$&$\checkmark$
				&$28.63$
				&$0.890$
			    \\
				L+E&8bit(RAW$\times\gamma$)
				& U-Net
				&$\checkmark$&$\checkmark$& 
				&$28.47$
				&$0.889$
				\\
				\hline
				
				
				L+Q&RAW$\times\hat{\gamma}$
				& U-Net
				&$\checkmark$& &$\checkmark$
				&$25.67$
				&$0.855$
				\\
				L&8bit(RAW$\times\hat{\gamma}$)
				&U-Net
				&$\checkmark$& &
				&$25.19$
				&$0.846$
				\\
				
				\hline
				E+Q & $\left[\text{sRGB},\gamma \right]$
				& Z+U-Net
				& &$\checkmark$&$\checkmark$
				&$ 25.98 $
				&$0.821$
				\\
				E& $\left[ \text{sRGB}, \gamma \right]$
				& Z+$8bit$+U-Net
				& &$\checkmark$& 
				&$ 25.98 $
				&$0.820$
				\\
				\hline
				Q&sRGB
				& U-Net
				&  & &$\checkmark$
				&$22.11$
				&$0.772$
				\\
				Baseline & 8bit(sRGB) 
				& U-Net
				& & & 
				&$17.14$
				&$0.374$
				\\
				\hline
			\end{tabular}
			}
			\label{table:Q}
		\end{center}
	\end{table}

	\begin{table}[t]
		\begin{center}
			\caption{
				Evaluation on the performance gap between \textit{Quantify then Brighten} and \textit{Brighten then Quantify}. xx/xx means PSNR$\uparrow$/SSIM$\uparrow$.
			}
			\begin{tabular}{c|c|c|c}
				\hline
				
				Low-Light Input
				&RAW& RAW & sRGB \\
				Brightening Method
				&$\times\gamma$ & $\times\hat{\gamma}$ &Zero-DCE\\
				\hline
				
				Brighten & $28.63/0.890$& $25.67/0.855$ & $25.98/0.821$\\
				Brighten then Quantify & $28.47/0.889$ & $25.19/0.846$ & $ 25.98/0.820$\\
				Quantify then Brighten & $12.36/0.137$ & --- & $16.89/0.285$\\
				\hline
				
			\end{tabular}
			\label{table:quan}
		\end{center}
	\end{table}

	\begin{table}[t]
		\begin{center}
			\caption{
				Effects of \emph{white balance}. White balance parameters (W) are recorded in meta-data of RAW files and linear process (P) is defined in Sec.~\ref{sec:pipe}. The REENet$_{raw}$ and REENet adopt proposed RAW guiding strategy and related details are illustrated in Sec.~\ref{sec:Exp}. Implementing white balance on RAW data helps bridge the gap between RAW and RGB.
			}
			\resizebox{0.5\textwidth}{!}{
			\begin{tabular}{c|c|c|cc}
				\hline
				
			    Methods & Input
				& Guidance
				& PSNR$\uparrow$
				& SSIM$\uparrow$ \\
				\hline
				L+E+Q+W&$\text{RAW} \times \gamma \times W$&--
				&28.71
				&0.890
				\\
				L+E+Q&$\text{RAW} \times \gamma$&--
				&28.63
				&0.890
				\\
				\hline
				REENet&sRGB&$\text{P}(\text{RAW} \times \gamma, W, M_{cc})$
				&28.42
				&0.883
				\\
				REENet$_{raw}$&sRGB&$\text{P}(\text{RAW} \times \gamma, 1, M_{cc})$
				&28.17
				&0.882
				\\
				
				\hline
			\end{tabular}
			}
			\label{table:W}
		\end{center}
	\end{table}
	
	\begin{table}[t]
		\begin{center}
			\caption{
				Evaluation on effects of using different white balance parameters for processing.
				L+E+Q adopts an end-to-end RAW-to-sRGB architecture.
				R2R$_l$ and R2R$_s$ learn to turn low-light RAW images into normal-light RAW images, via performing denoising on brightened RAW images, and then processing the RAW images.
				R2R$_s$ utilizes the white balance parameters of short-exposure RAW and R2R$_l$ uses those of long-exposure one.
				WB denotes white balance.
			}
			\resizebox{0.5\textwidth}{!}{
			\begin{tabular}{c|cc|c|cc}
				\hline		
				Methods&Input& Target & WB Parameters
				& PSNR$\uparrow$ & SSIM$\uparrow$  \\
				\hline
				L+E+Q&RAW$\times\gamma$ 
				&sRGB
				&---
				&$28.63$
				&$0.890$
				\\
				R2R$_s$&RAW$\times\gamma$ 
				&RAW
				&Short Exposure
				&$28.20$
				&$0.882$
				\\
				R2R$_l$&RAW$\times\gamma$ 
				&RAW
				&Long Exposure
				&$29.50$
				&$0.889$
				\\
				\hline
				
			\end{tabular}
			}
			\label{table:meta}
		\end{center}
	\end{table}

	\noindent \textbf{Linearity}.
	We compare several groups of versions that pre-process the signal in the linear and nonlinear domains, respectively, as shown in {Fig.~\ref{fig:L}~(a) and Table~\ref{table:L} (corresponding to Table I in the supplementary material).}
	For the methods working in the nonlinear sRGB domain, Zero-DCE~\cite{Zero-DCE} is adopted to adjust the illumination in the sRGB domain guided by the exposure time ratio, \textit{i.e.} taking the  the low-light input and the ratio as the input.
	{It is demonstrated that, the methods working in the linear domain, where the illumination can be directly adjusted via multiplication with the ratio, significantly outperform the ones working in the nonlinear domain with performance gains over 2.49 dB in PSNR, over 0.069 in SSIM, about 0.7 in NIQE and 0.03 in LPIPS.
	The results illustrate the critical role of linearity of RAW data in low-light image enhancement.}
	\vspace{1mm}
	
	\noindent \textbf{Exposure Time}.
	We also compare the methods that are assumed to obtain the exposure time or not, as shown in {Fig.~\ref{fig:L}~(b) and Table~\ref{table:E} (corresponding to Table II in the supplementary material)}.
	For the methods that do not have the ground truth exposure time ratio $\gamma$, {we use the estimated $\hat{\gamma}$ with the mean pixel values of short/long-exposure linear data.}
	{It is observed that, using the ground truth exposure time ratio $\gamma$ leads to significant performance improvement with performance gains over 2.96 dB in PSNR, over 0.035 in SSIM, about 0.2 in NIQE and over 0.03 in LPIPS, which demonstrates the exposure time recorded in meta-data as another dominating factor.}
	Apparently, the estimated ratio $\hat{\gamma}$ makes the adjusted low-light images biased, with over-exposed and dark regions, which increases ambiguity in the low-light enhancement.
	\vspace{1mm}
	
	\begin{table*}[t]
		\begin{center}
			\caption{
				An overview of Benchmark results on SID. L denotes Linearity, E denotes Exposure time, Q denotes dense 
				{Quantization} level and W denotes White balance. $\gamma$ is the ratio of exposure time and when meta-data is absent, it is replaced by $\hat{\gamma}$, the ratio of mean pixel value. The Zero-DCE (Z) implemented as pre-processing is trained from scratch with inputs concatenated with ratio and $L_1$ loss with the target. {R2R} means RAW-to-RAW approaches. KinD$\ast$ is trained from scratch on SID. The bold value denotes the best result in each category.
			}
			\begin{tabular}{c|c|c|c|ccccc}
				\hline
				
				Category & Methods & Input & Model
				& PSNR$\uparrow$ & SSIM$\uparrow$ & VIF$\uparrow$ & {NIQE}$\downarrow$ & {LPIPS}$\downarrow$\\ 
				\hline
				\multirow{9}{*}{RAW based}&L+E+Q+W
				& $\text{RAW} \times \gamma \times W$
				& U-Net
				&$\quad28.71\quad$&$\quad0.890\quad$&$\quad0.137\quad$&$\quad4.99\quad$
				&$\quad0.276\quad$
				\\
				&R2R$_l$
				& $\text{RAW} \times \gamma$
				& U-Net
				&$\textbf{\quad29.50\quad}$&$\quad0.889\quad$&$\quad0.140\quad$&$\quad5.10\quad$
				&$\quad0.265\quad$
				\\
				&R2R$_s$
				& $\text{RAW} \times \gamma$
				& U-Net
				&$\quad28.20\quad$&$\quad0.882\quad$&$\quad0.139\quad$&$\quad5.10\quad$
				&$\quad0.278\quad$
				\\
				
				&L+E+Q
				& $\text{RAW} \times \gamma$
				& U-Net
				&$\quad28.63\quad$&$\quad0.890\quad$&$\quad0.137\quad$&$\quad 5.04\quad$
				&$\quad0.272\quad$
				
				\\
				
				&L+E
				&$\text{8bit} (\text{RAW}\times\gamma)$ 
				& U-Net
				&$\quad28.47\quad$&$\quad0.889\quad$&$\quad0.137\quad$&$\quad5.03\quad$
				&$\quad0.273\quad$
				\\
				
				&L+Q
				&$\text{RAW}\times\hat{\gamma}$ 
				& U-Net
				&$\quad25.67\quad$&$\quad0.855\quad$&$\quad0.136\quad$&$\quad4.81\quad$
				&$\quad0.303\quad$
				\\
				
				&L
				&$\text{8bit}(\text{RAW}\times\hat{\gamma})$
				& U-Net
				&$\quad25.19\quad$&$\quad0.846\quad$&$\quad0.135\quad$&$\quad4.85\quad$
				&$\quad0.310\quad$
				\\
				
				&EEMEFN~\cite{EEMEFN}
				& $\text{RAW}\times1, \times\gamma/2$
				& EEMEFN
				&$\quad29.21\quad$&$\textbf{\quad0.910\quad}$&$\quad0.144\quad$&$\quad5.52\quad$
				&$\quad0.280\quad$
				\\
				
				&ELD~\cite{Wei_2020_CVPR}
				& $\text{RAW} \times \gamma$
				& U-Net
				&$\quad28.72\quad$&$\quad0.873\quad$&$\quad\textbf{0.149}\quad$&$\textbf{\quad4.48\quad}$
				&$\textbf{\quad0.255\quad}$
				\\
				
				\hline
				\multirow{21}{*}{sRGB-based}&REENet
				&$\left[\text{sRGB}, \gamma \right]$
				& REENet
				&$\textbf{\quad28.42\quad}$&$\textbf{\quad0.883\quad}$&$\quad0.137\quad$&$\textbf{\quad5.60\quad}$
				&$\quad0.322\quad$
				\\
				
				&REENet$_{raw}$
				& $\left[\text{sRGB}, \gamma \right]$
				& REENet
				&$\quad28.17\quad$&$\quad0.882\quad$&$\quad0.141\quad$&$\quad5.88\quad$
				&$\quad\textbf{0.286}\quad$
				\\

				&E+Q
				&$\left[\text{sRGB}, \gamma \right]$
				&Z+U-Net 
				&$\quad 25.98 \quad$&$\quad0.821\quad$&$\textbf{\quad0.143\quad}$&$\quad5.70\quad$
				&$\quad0.315\quad$
				\\
				
				&REENet$_{8bit}$
				& $\left[\text{8bit}(\text{sRGB}), \gamma \right]$
				& REENet
				&$\quad25.75\quad$&$\quad0.808\quad$&$\quad0.135\quad$&$\quad6.23\quad$
				&$\quad0.424\quad$
				\\
				
				&E
				&$\left[\text{sRGB}, \gamma \right]$ 
				&Z+$8bit$+U-Net
				&$\quad 25.98 \quad$&$\quad0.820\quad$&$\textbf{\quad0.143\quad}$&$\quad5.71\quad$
				&$\quad0.314\quad$
				\\
				
				&REENet$_{\hat{\gamma}}$
				& $\left[\text{sRGB}, \hat{\gamma} \right]$
				& REENet
				&$\quad24.18\quad$&$\quad0.811\quad$&$\quad0.138\quad$&$\quad6.82\quad$
				&$\quad0.410\quad$
				\\
				
				&Q
				&sRGB
				&U-Net
				&$\quad22.11\quad$&$\quad0.772\quad$&$\quad0.139\quad$&$\quad5.71\quad$
				&$\quad0.372\quad$
				\\
				
				&REENet$_{base}$
				& $\left[\text{8bit}(\text{sRGB}), \hat{\gamma} \right]$
				& REENet
				&$\quad22.42\quad$&$\quad0.802\quad$&$\quad0.135\quad$&$\quad6.21\quad$
				&$\quad0.422\quad$
				\\

				&baseline
				&$\text{8bit}(\text{sRGB})$
				&U-Net
				&$\quad17.14\quad$&$\quad0.374\quad$&$\quad0.078\quad$&$\quad6.87\quad$
				&$\quad0.533\quad$
				\\
				&HE~\cite{HE1}
				&sRGB
				& --
				&$\quad 5.90 \quad$&$\quad0.028\quad$&$\quad0.095\quad$&$\quad7.71\quad$
				&$\quad0.968\quad$
				\\
				&BPDHE~\cite{BPDHE}
				&sRGB
				& --
				&$\quad 10.67 \quad$&$\quad0.072\quad$&$\quad0.051\quad$&$\quad16.65\quad$
				&$\quad0.969\quad$
				\\
				&Dehazing~\cite{dong2011fast}
				&sRGB
				& --
				&$\quad 12.81 \quad$&$\quad0.103\quad$&$\quad0.077\quad$&$\quad8.09\quad$
				&$\quad0.784\quad$
				\\
				&MSR~\cite{Multi_scale_retinex}
				&sRGB
				& --
				&$\quad 10.04 \quad$&$\quad0.070\quad$&$\quad0.116\quad$&$\quad 6.33\quad$
				&$\quad1.031\quad$
				\\
				&MF~\cite{FU201682}
				&sRGB
				& --
				&$\quad 13.87 \quad$&$\quad0.111\quad$&$\quad0.108\quad$&$\quad6.34\quad$
				&$\quad0.950\quad$
				\\
				&LIME~\cite{Guo_2017_Lime}
				&sRGB
				& --
				&$\quad 12.59 \quad$&$\quad0.102\quad$&$\quad0.118\quad$&$\quad6.06\quad$
				&$\quad0.980\quad$
				\\
				&BIMEF~\cite{BIMEF}
				&sRGB
				& --
				&$\quad 13.06 \quad$&$\quad0.110\quad$&$\quad0.086\quad$&$\quad7.67\quad$
				&$\quad0.798\quad$
				\\
				&LLNet~\cite{LLNet}
				&sRGB
				&LLNet
				&$\quad 14.21 \quad$&$\quad0.221\quad$&$\quad0.047\quad$&$\quad7.65\quad$
				&$\quad0.693\quad$
				\\
				&SICE~\cite{cai2018learning}
				&sRGB
				&SICE
				&$\quad 14.26 \quad$&$\quad0.366\quad$&$\quad0.011\quad$&$\quad13.25\quad$
				&$\quad0.766\quad$
				\\
				&KinD~\cite{Kind}
				&sRGB
				&KinD
				&$\quad 13.50 \quad$&$\quad0.109\quad$&$\quad0.048\quad$&$\quad9.68\quad$
				&$\quad0.718\quad$
				\\
				&KinD$\ast$~\cite{Kind}
				&sRGB
				&KinD
				&$\quad 17.87 \quad$&$\quad0.611\quad$&$\quad0.081\quad$&$\quad9.71\quad$
				&$\quad0.629\quad$
				\\
				&DeepUPE~\cite{deepUPE}
				&sRGB
				&DeepUPE
				&$\quad 12.10 \quad$&$\quad0.070\quad$&$\quad0.028\quad$&$\quad11.28\quad$
				&$\quad0.772\quad$
				\\
				&Zero-DCE~\cite{Zero-DCE}
				&sRGB
				&Zero-DCE
				&$\quad 12.38 \quad$&$\quad0.137\quad$&$\quad0.050\quad$&$\quad10.82\quad$
				&$\quad0.768\quad$
				\\
				\hline
			\end{tabular}
			\label{table:benchmark}
		\end{center}
	\end{table*}
	
		\begin{figure*}[tbp]
		\centering
		\subfigure[Ground Truth]{
			\includegraphics[width=18mm]{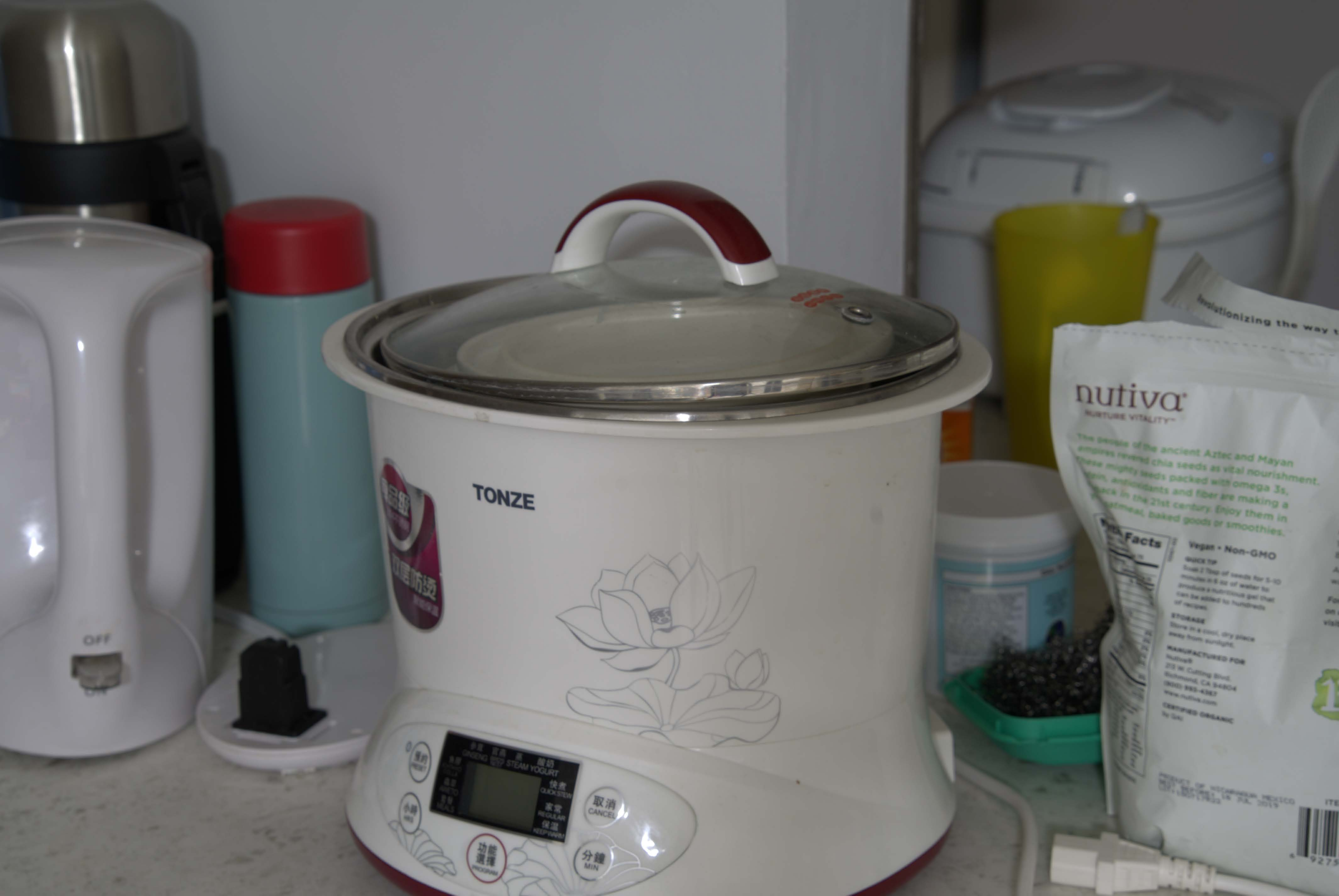}
		}
		\subfigure[Input]{
			\includegraphics[width=18mm]{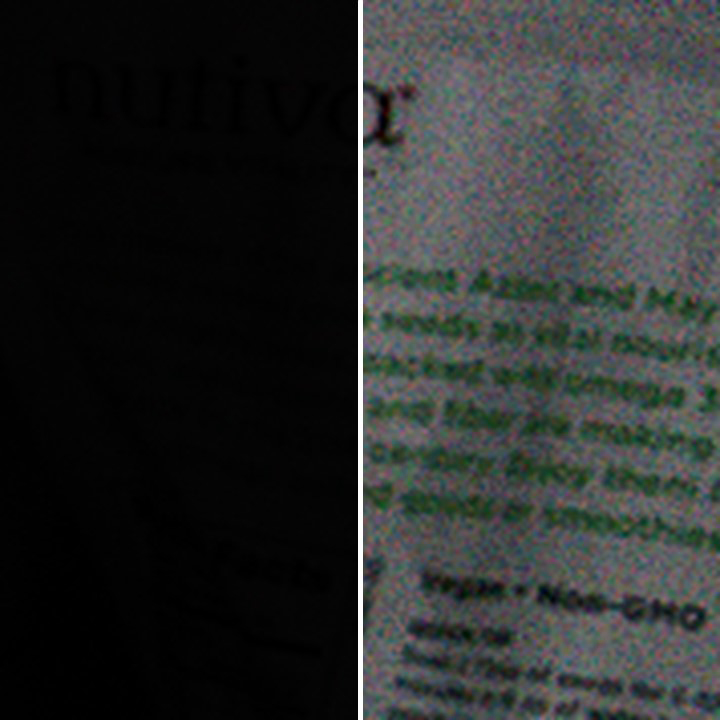}
		}
		\subfigure[L+E+Q+W]{
			\includegraphics[width=18mm]{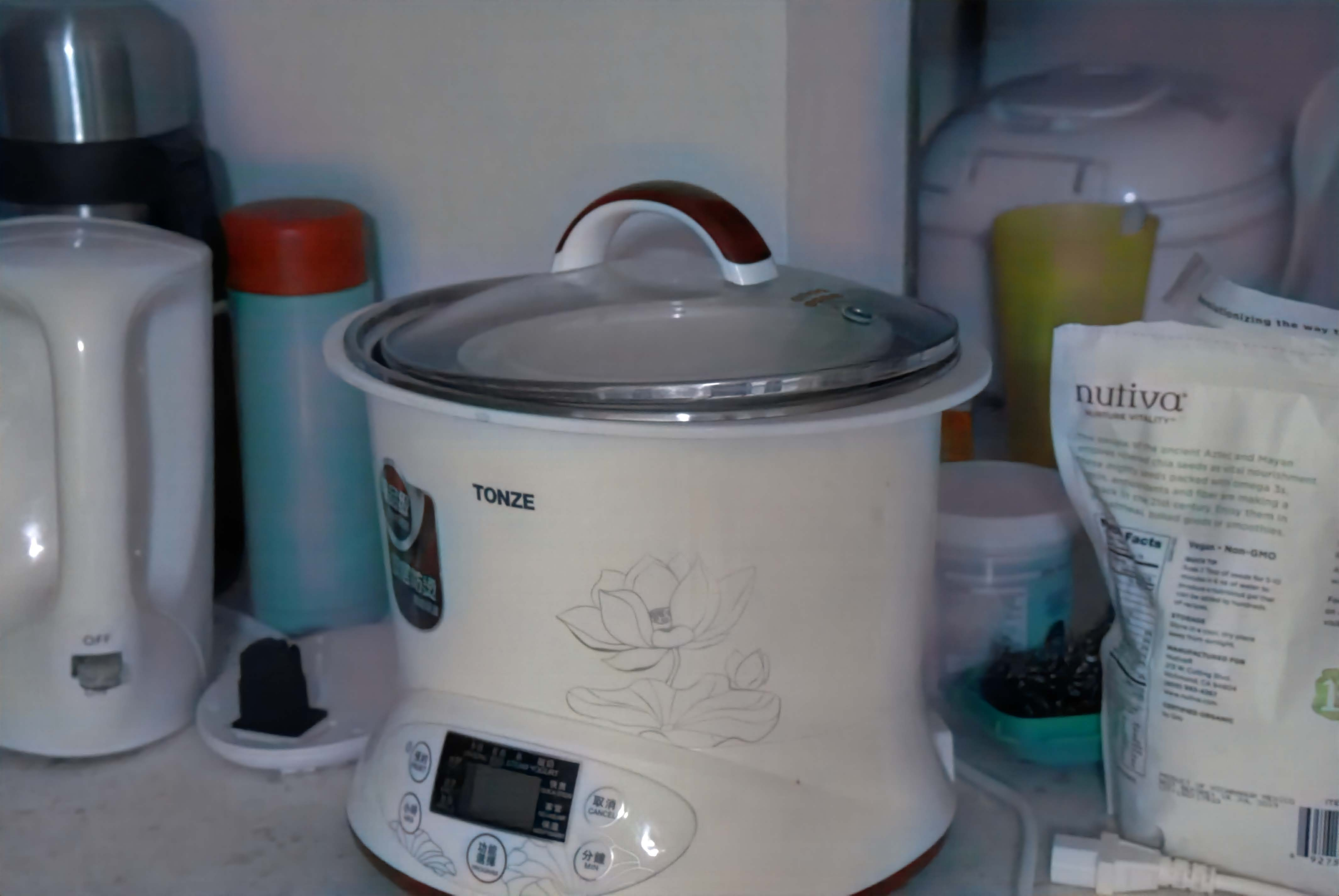}
		}
		\subfigure[R2R$_l$]{
			\includegraphics[width=18mm]{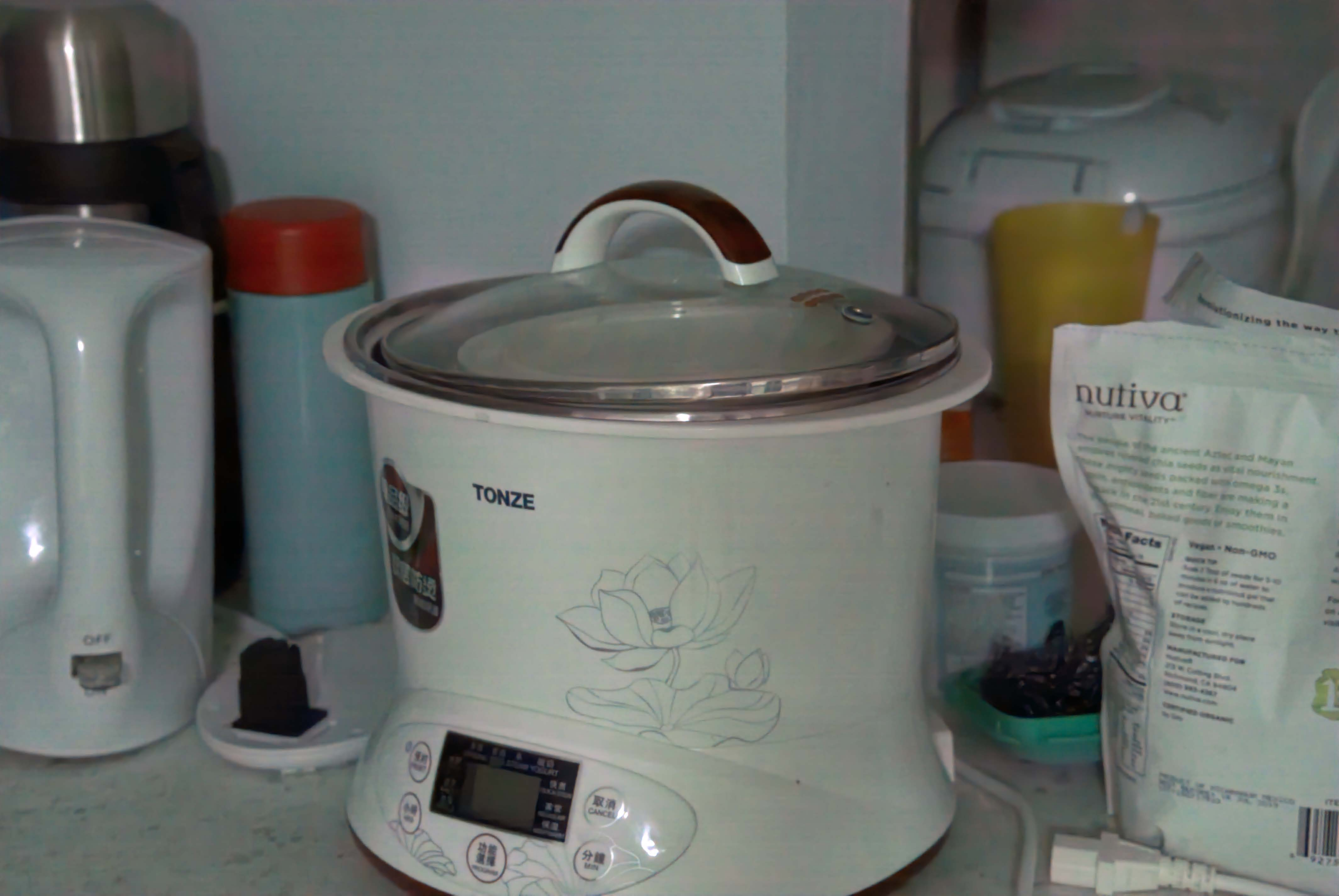}
		}
		\subfigure[R2R$_s$]{
			\includegraphics[width=18mm]{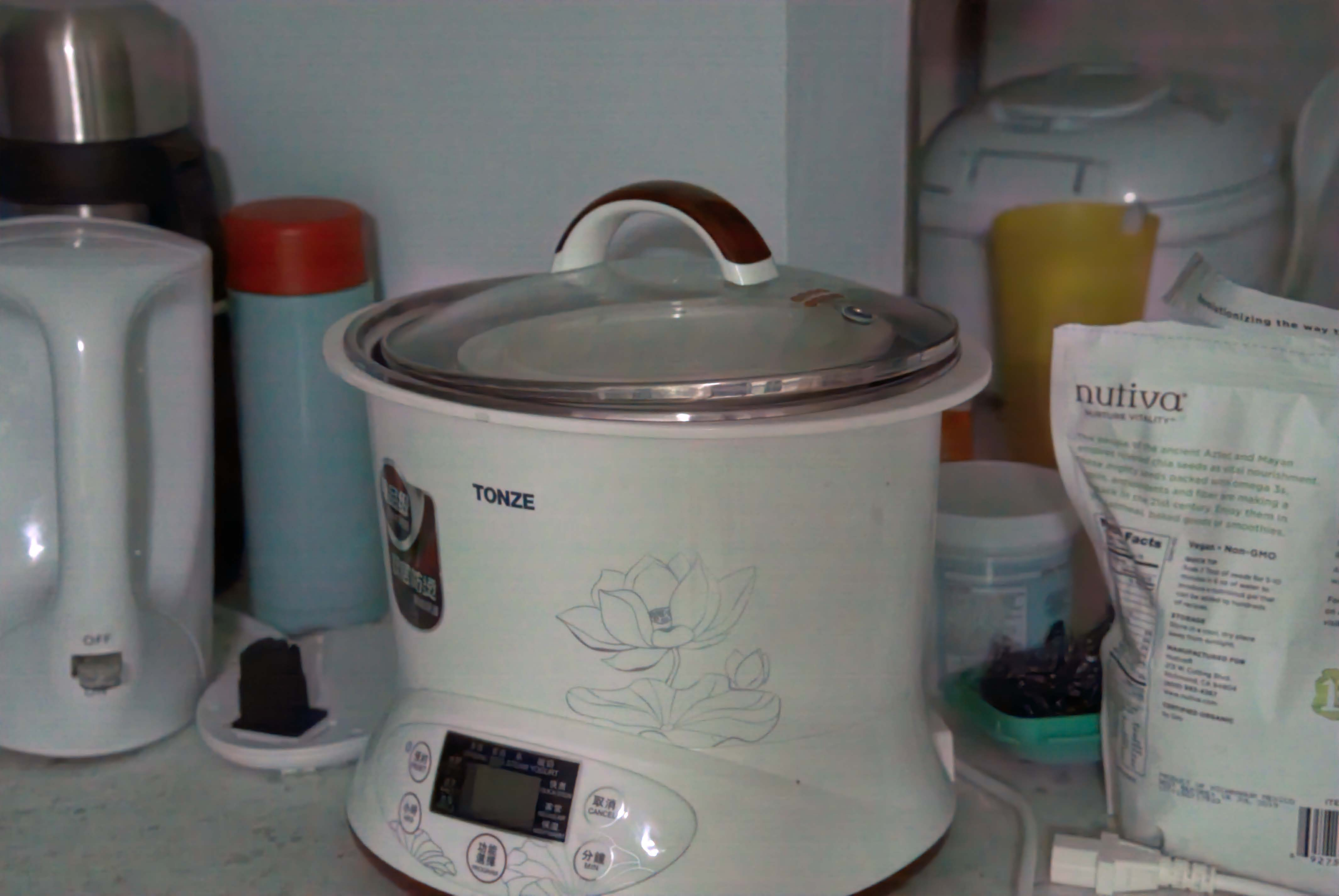}
		}
		\subfigure[L+E+Q]{
			\includegraphics[width=18mm]{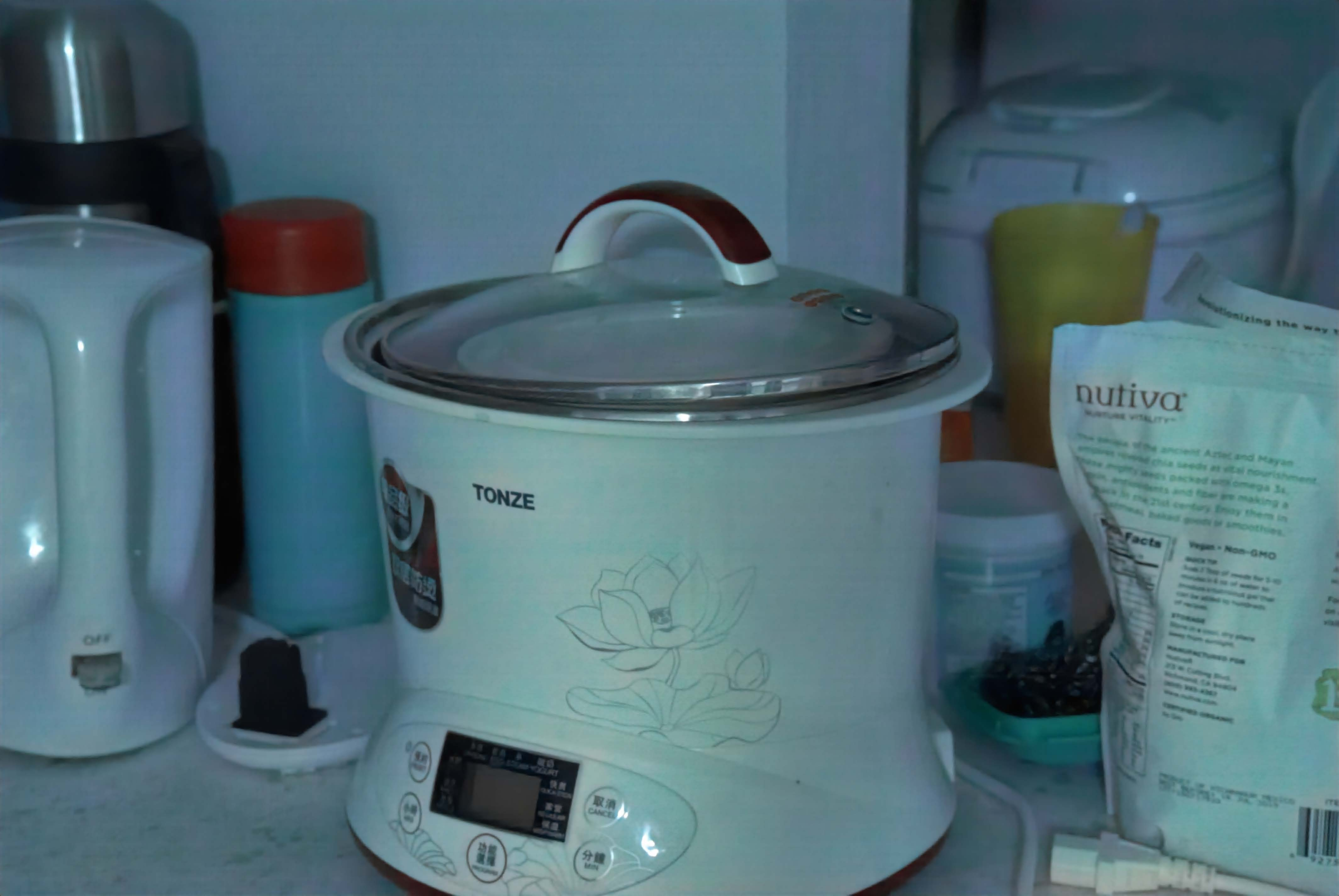}
		}
		\subfigure[L+E]{
			\includegraphics[width=18mm]{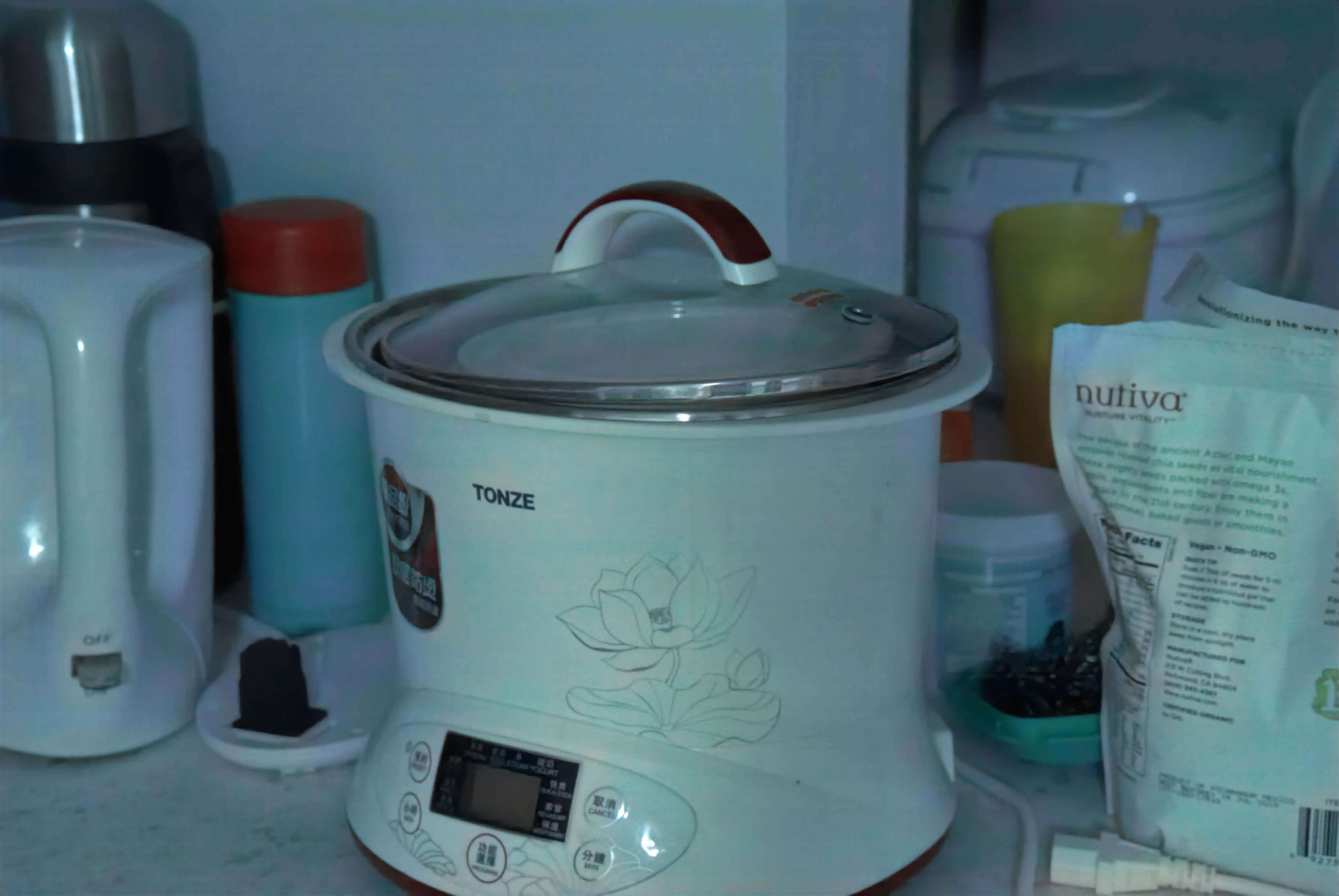}
		}
		\subfigure[L+Q]{
			\includegraphics[width=18mm]{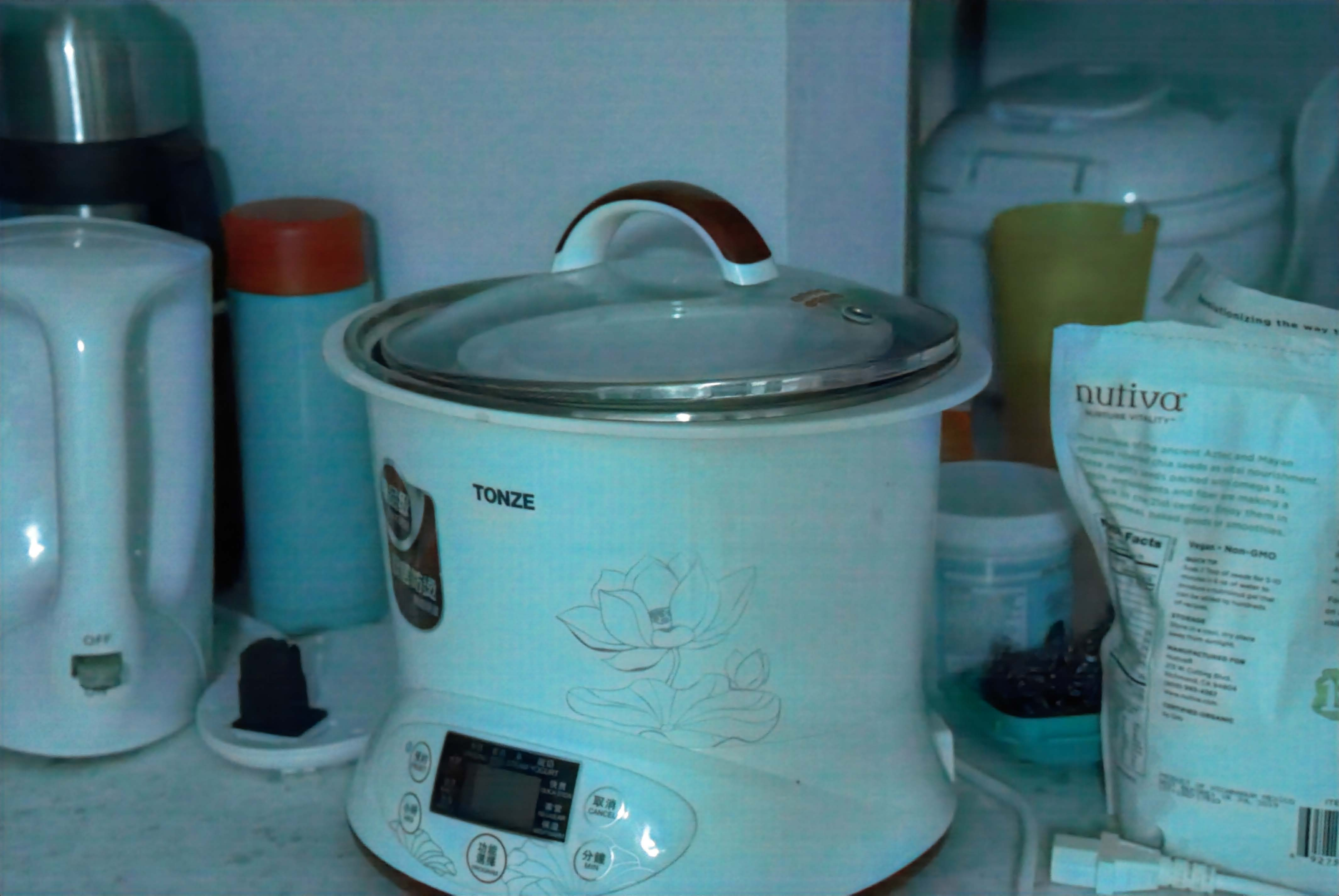}
		}
		\subfigure[E+Q]{
			\includegraphics[width=18mm]{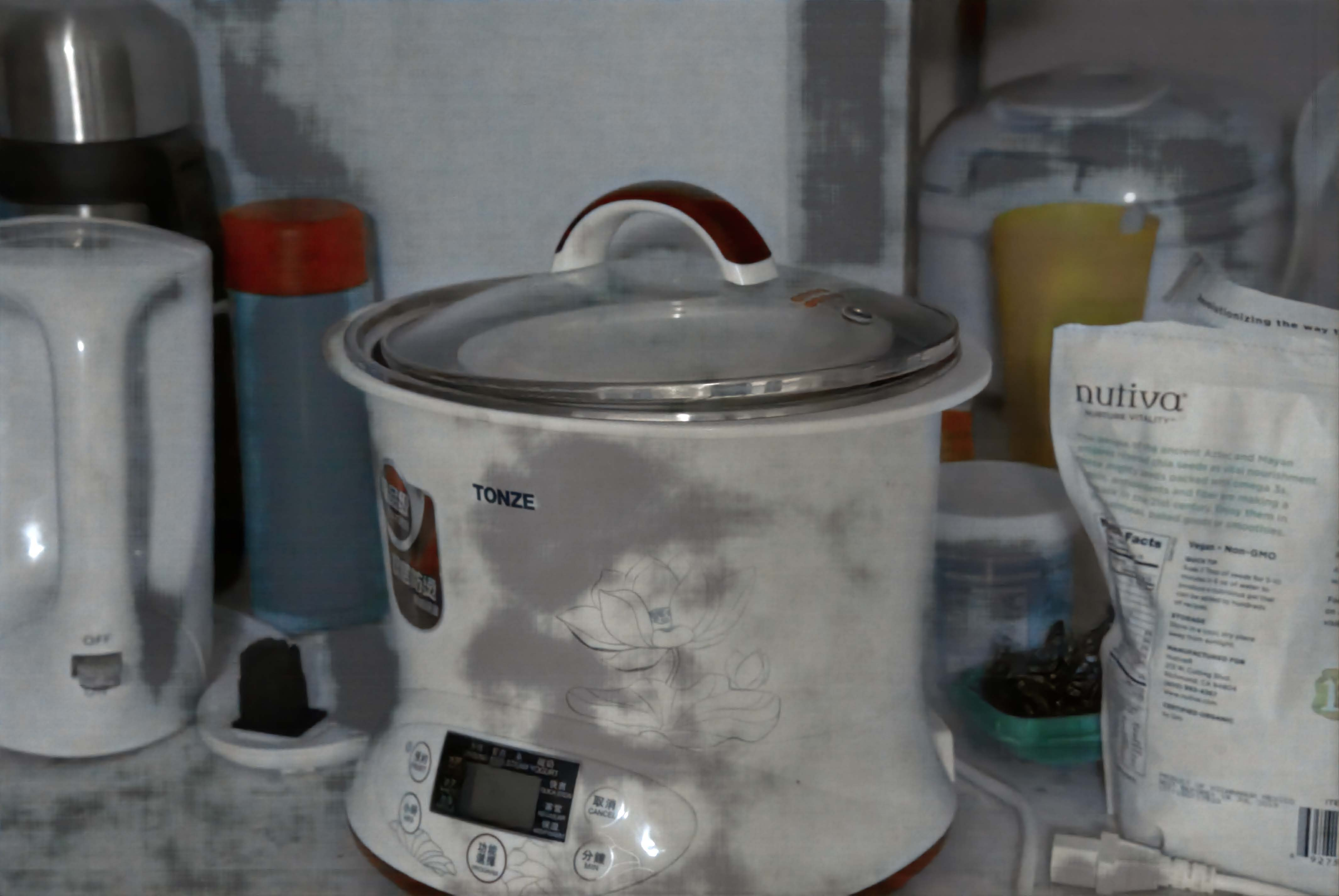}
		}
		\subfigure[L]{
			\includegraphics[width=18mm]{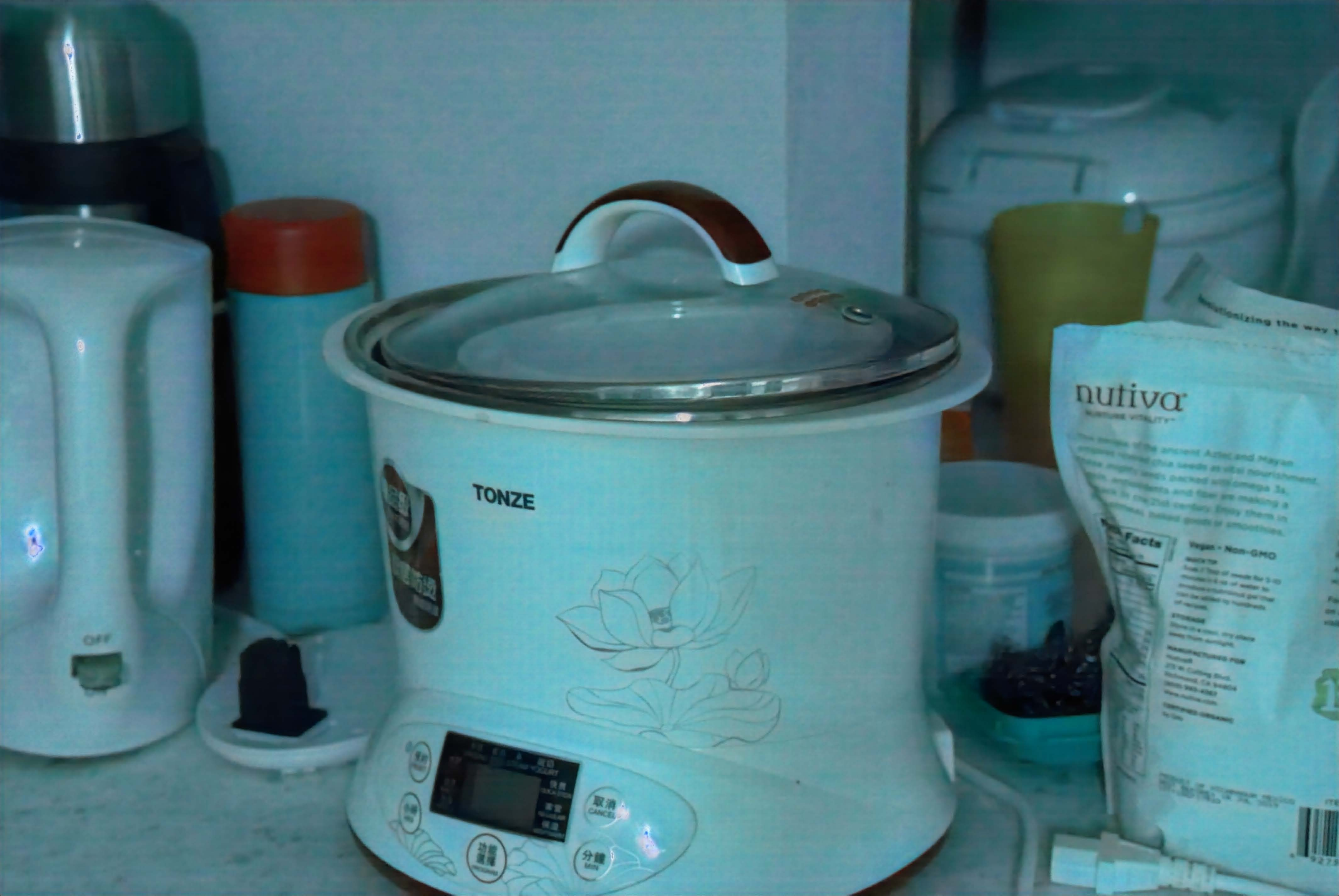}
		}
		\subfigure[E]{
			\includegraphics[width=18mm]{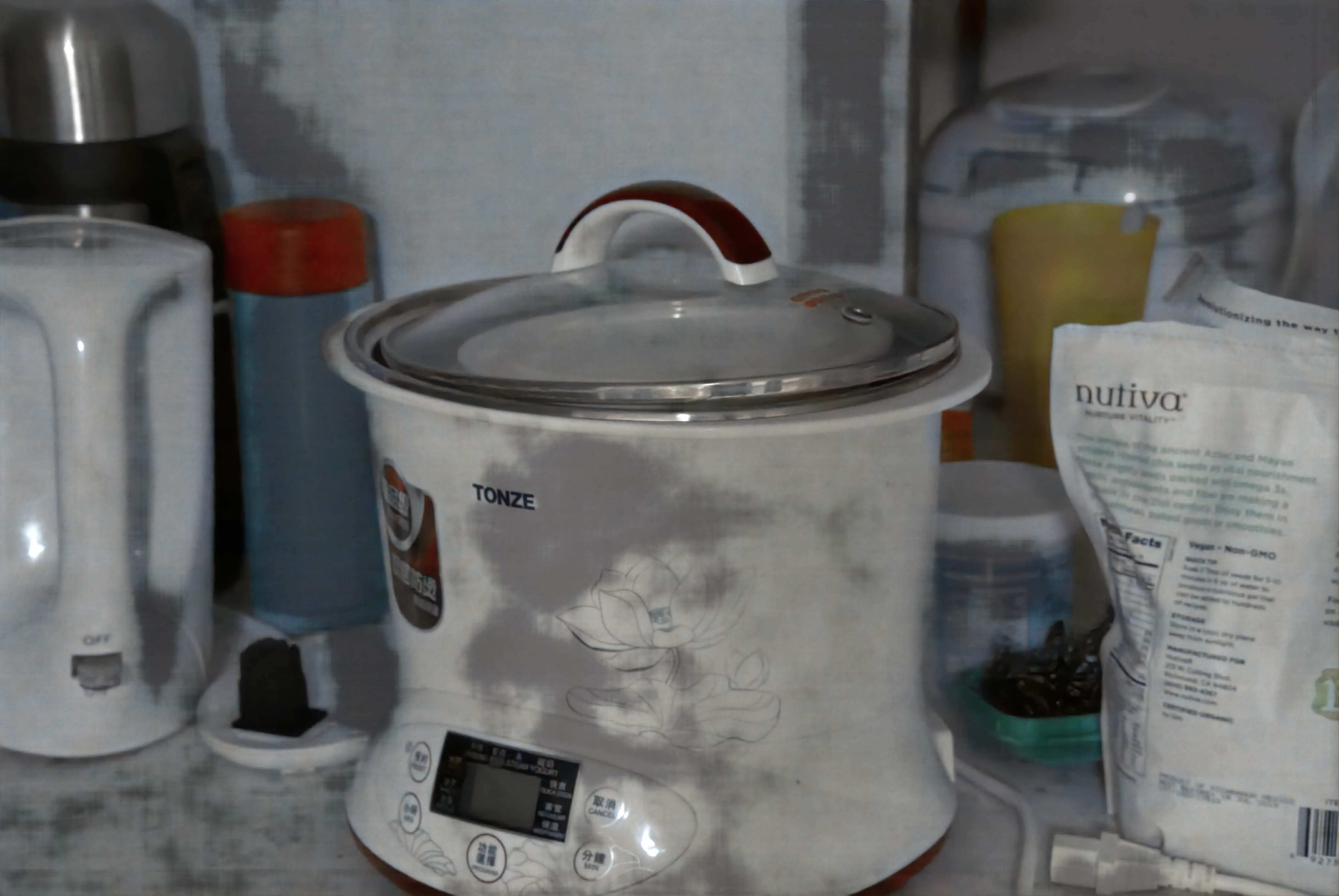}
		}
		\subfigure[Q]{
			\includegraphics[width=18mm]{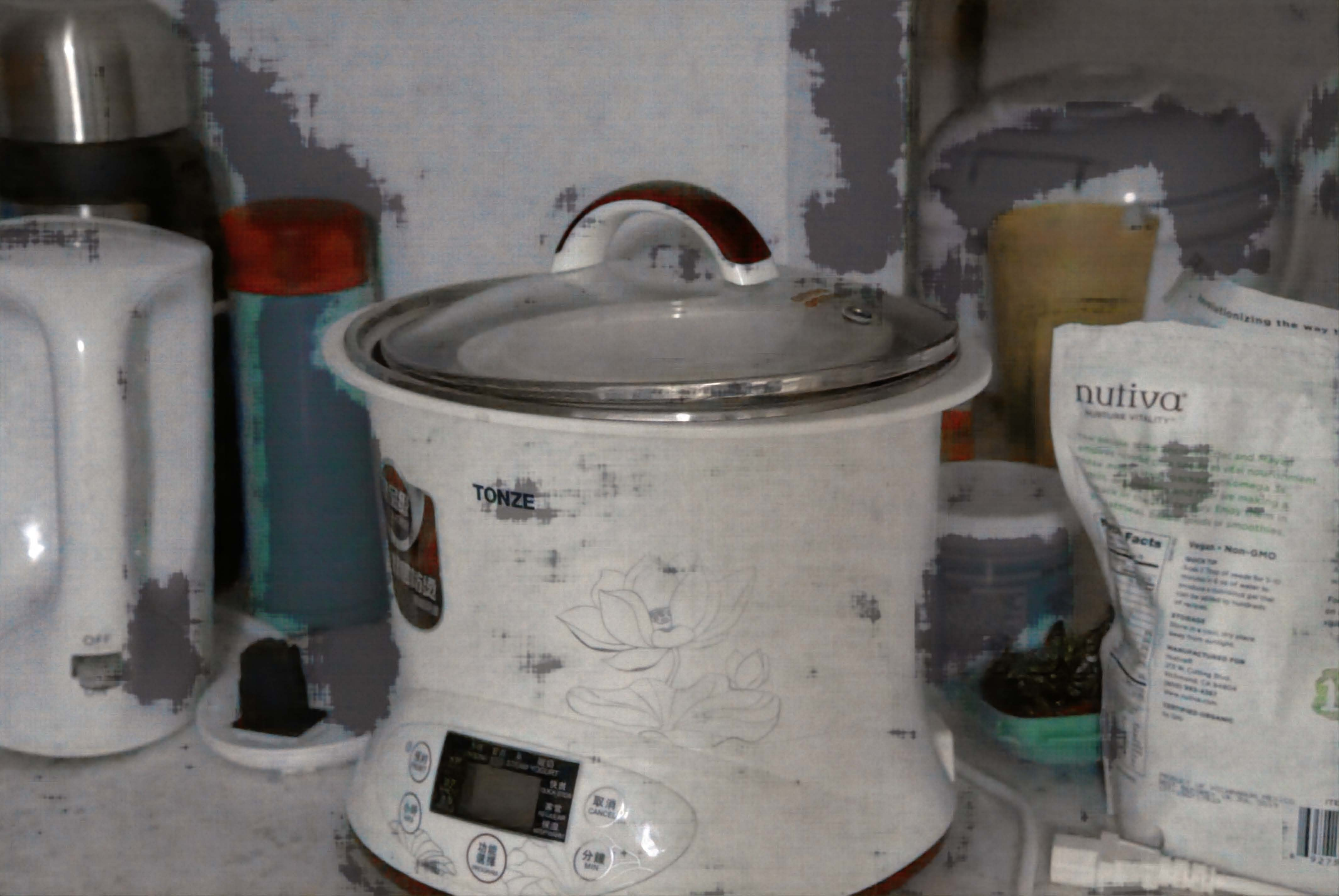}
		}
		\subfigure[Baseline]{
			\includegraphics[width=18mm]{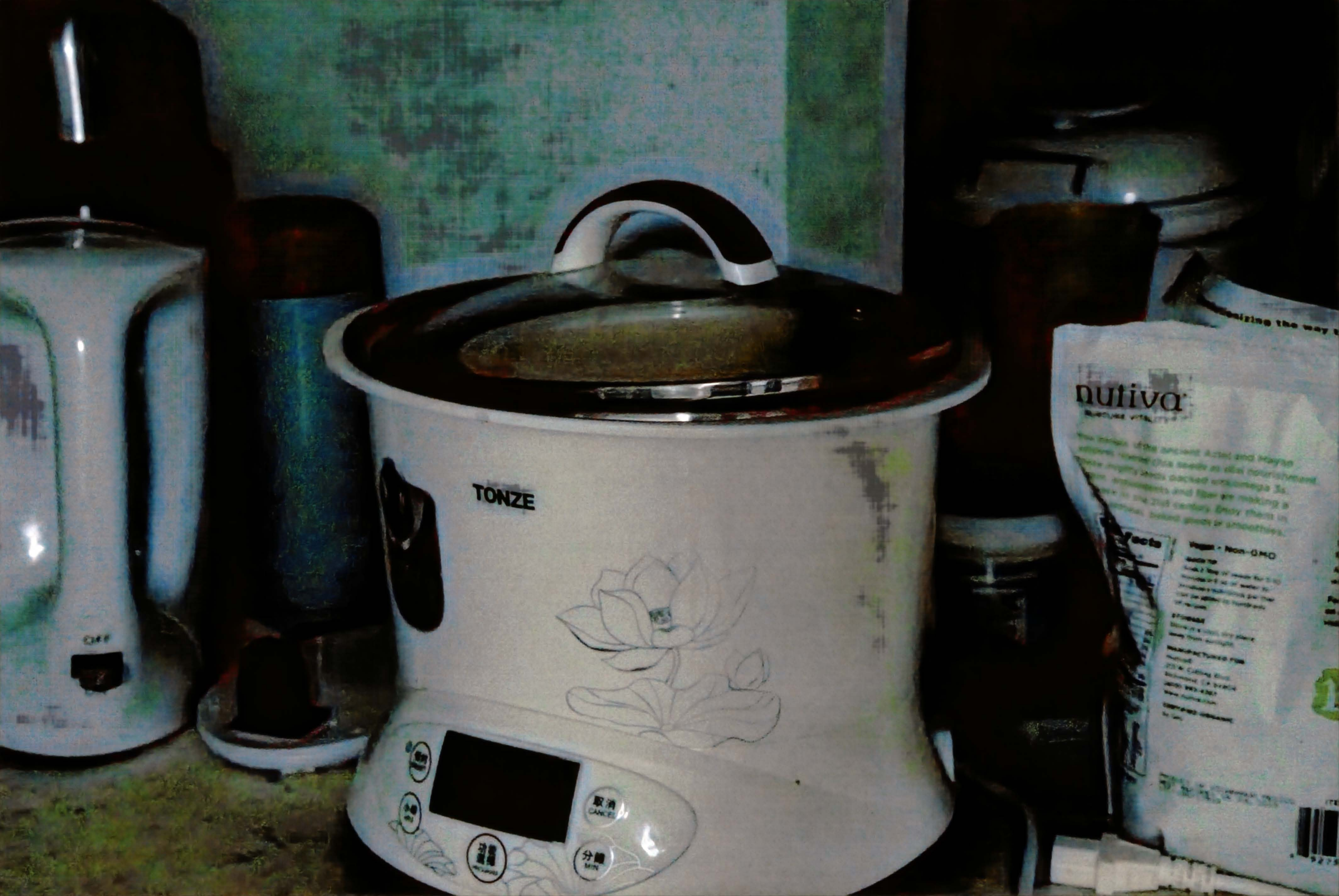}
		}
		\subfigure[EEMEFN]{
			\includegraphics[width=18mm]{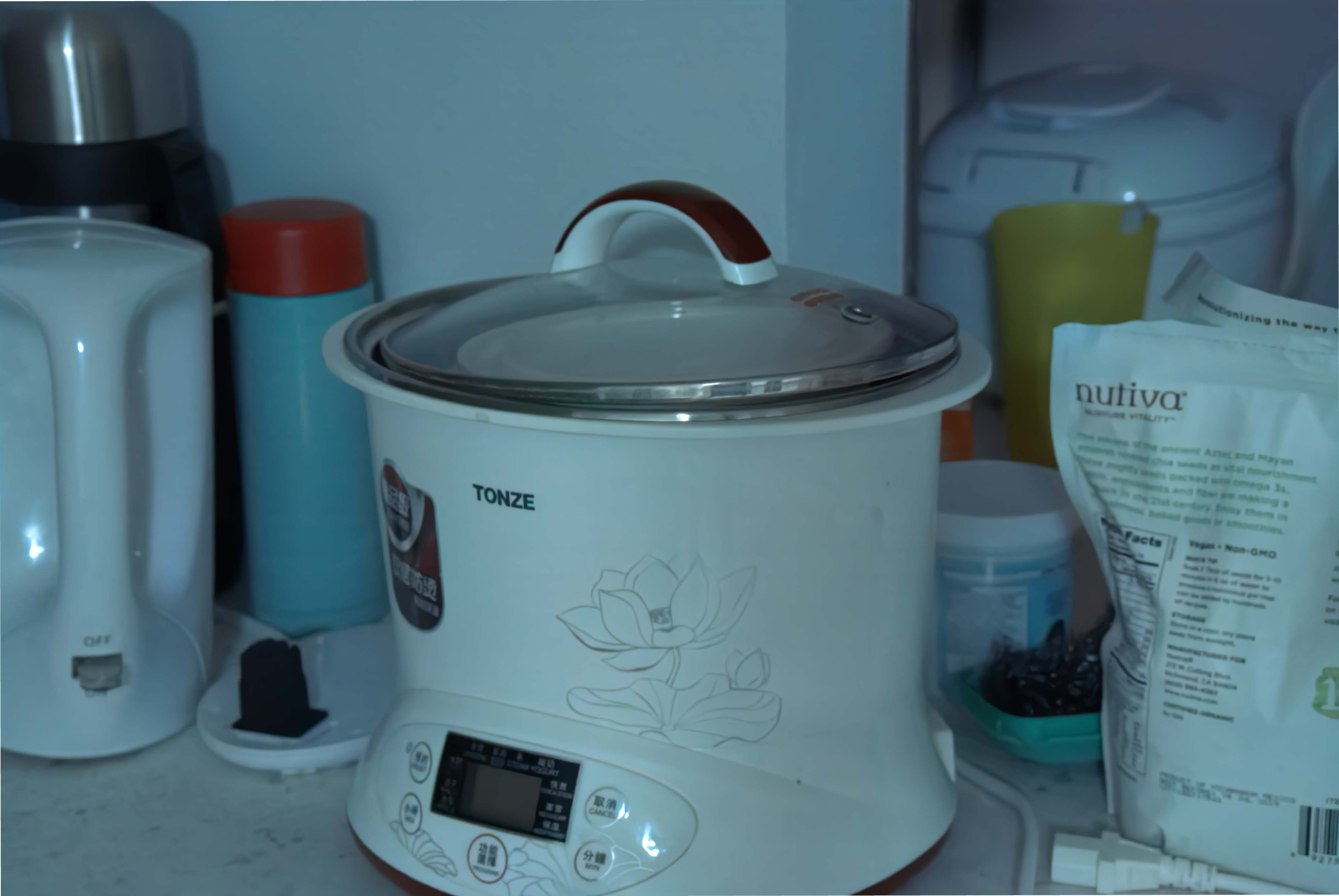}
		}
		\subfigure[ELD]{
			\includegraphics[width=18mm]{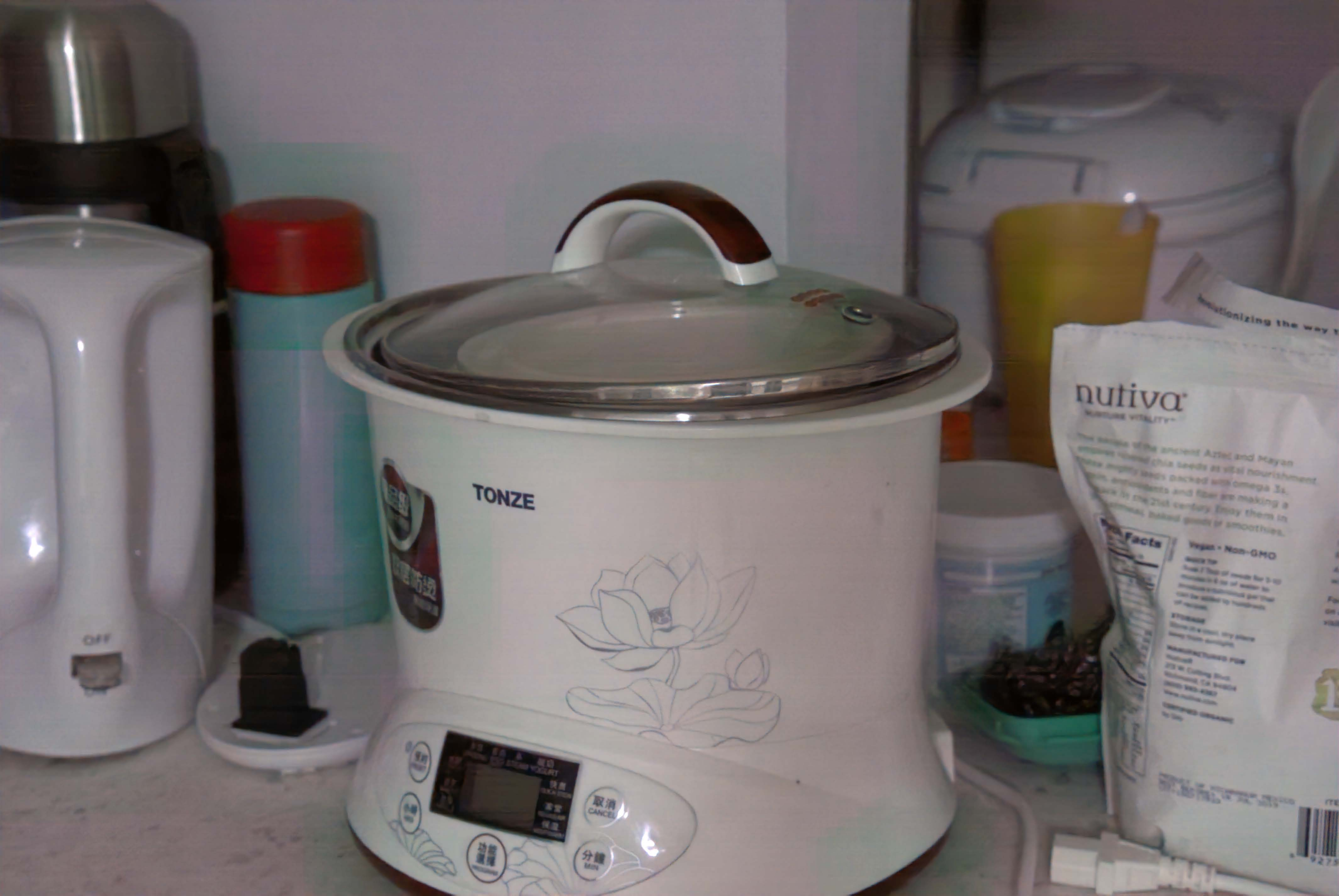}
		}
		\subfigure[REENet]{
			\includegraphics[width=18mm]{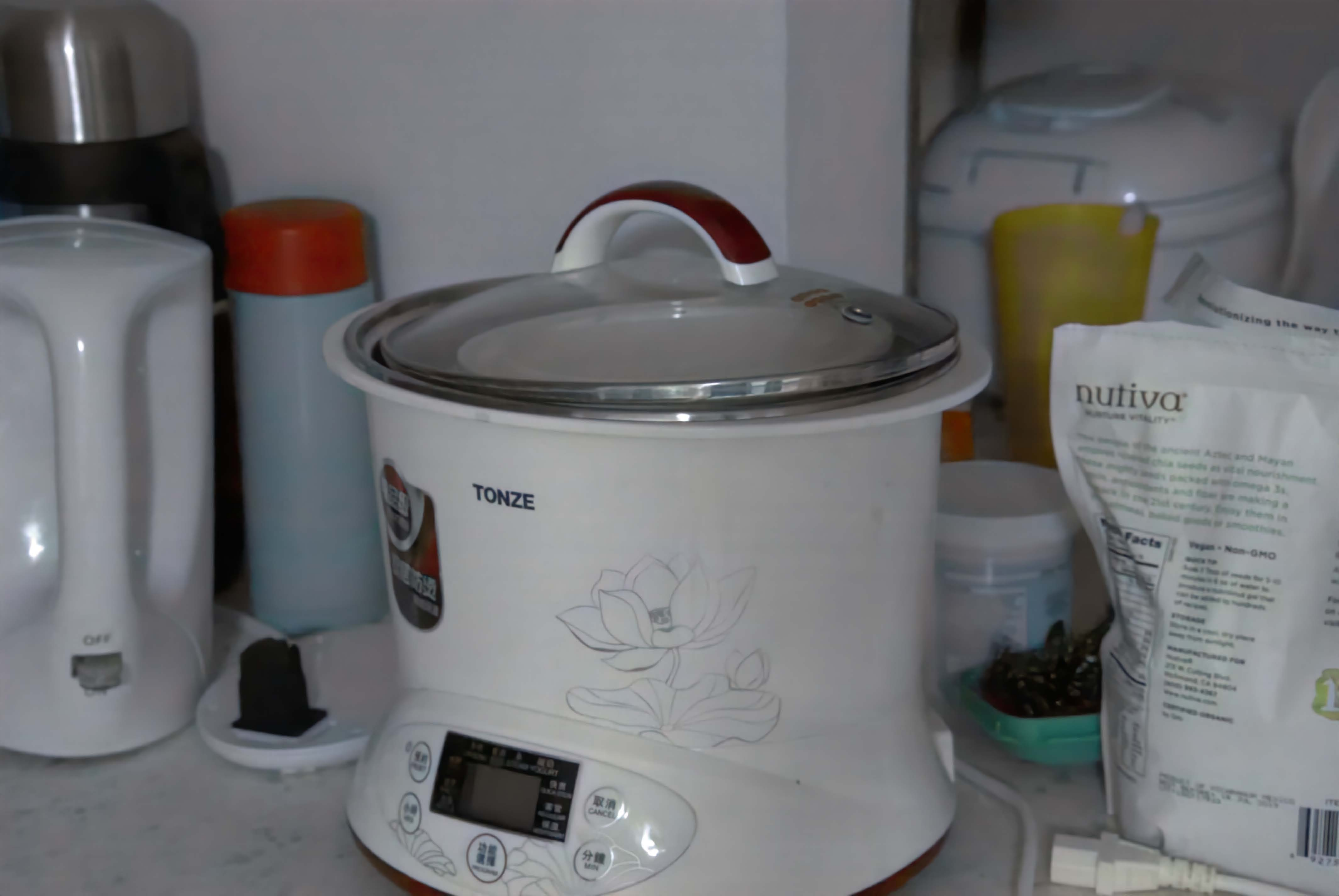}
		}
		\caption{
			Qualitative results of benchmark. Apparently utilizing more characteristics of RAW improves the final quality of results with less noise and artifacts. {The right part of the input is brightened.}
		}
		\label{fig:benchmark}
	\end{figure*}
	
	\noindent \textbf{Quantization}.
	We compare the methods taking the images with different quantization levels as their input as shown in {Fig.~\ref{fig:L}-(c) and Table~\ref{table:Q} (corresponding to Table III in the supplementary material)}.
	{It is observed that, more fine-grained quantization levels only lead to relatively small gains if the compression is implemented after brightening, called \emph{Brighten then Quantize} strategy, as shown in the top three comparisons of Table~\ref{table:Q} (corresponding to Table III in the supplementary material) with a performance gain under 0.48 dB in PSNR and competitive performances in other metrics.}
	However, if quantizing the data into 8-bit format before brightening, called {\emph{Quantize then Brighten}} strategy, tremendous performance drops are observed in Table~\ref{table:quan}.
	The performance gap originates from the dynamic range stretching that makes the brightened dark region have more fine-grained quantization levels and preserve more detailed signals.
	These results demonstrate the importance of compressing low-light images following {\emph{Brighten then Quantize}} strategy.
	\vspace{1mm}
	
	\noindent \textbf{White Balance}.
	The white balance parameters recorded in the meta-data of RAW files also can contribute to low-light image enhancement.
	{The experimental results in Table~\ref{table:W} (corresponding to Table V in the supplementary material) and Fig.~\ref{fig:L}~(d)} show the potential to improve the performance of RAW-based and the proposed RAW-guiding methods REENet by pre-processing RAW images with white balance parameters. A gain over 0.08 dB in PSNR is observed, meanwhile SSIM and NIQE improve slightly.
	In our comparisons, to utilize these parameters during training, we amplify the unpacked 4-channel linear data with the parameters.
	
	{
	We also study the related utilization in the RAW-to-RAW approaches and figure out how much the pre-processing can help bridge the gap between RAW and sRGB in Table~\ref{table:meta}. 
	R2R$_l$ and R2R$_s$ are two RAW-to-RAW based methods. 
	They are both end-to-end trained to target the ground-truth RAW data and then process them into sRGB images with Libraw, which uses short and long-exposure white balance parameters, respectively. 
	It is observed that, short-exposure white balance parameters lead to a performance drop while the long-exposure ones improve the low-light enhancement performance. 
	The performance gap comes from two reasons: 
	1) R2R$_l$ takes the same white balance parameters as the ground truth, which leads to similar reconstructed results to the ground truth; 
	2) the exposure time will affect the accuracy of the light metering in a camera, and the light metering with short exposure might be inaccurate \textit{w.r.t.} ground truth~\cite{Chen_2019_ICCV}.
	}
	
	\begin{figure*}[tb]
		\centering
		\subfigure[The overall architecture of REENet]{
			\includegraphics[height=50mm]{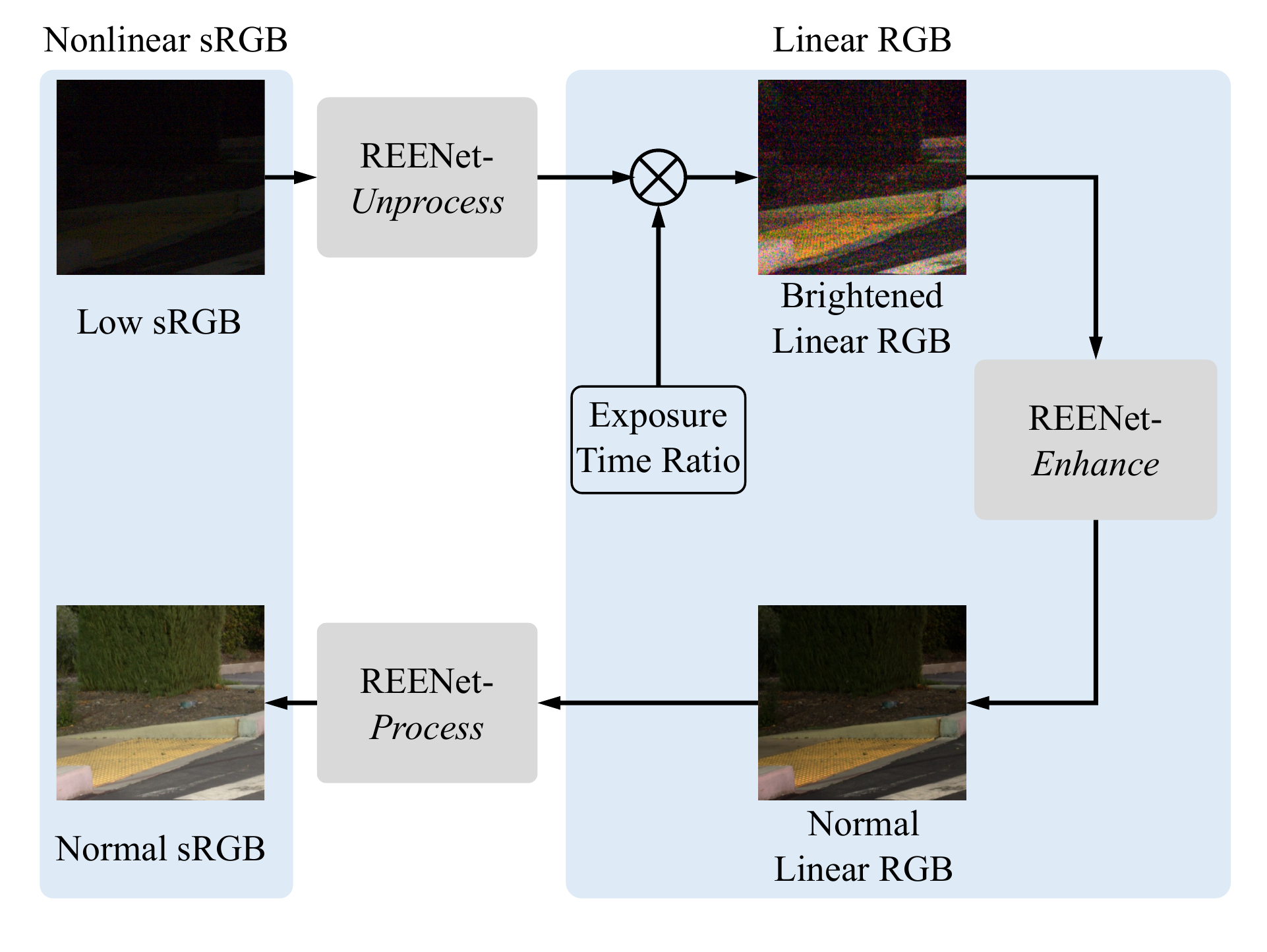}
		}
		\subfigure[The training strategy of REENet]{
			\includegraphics[height=50mm]{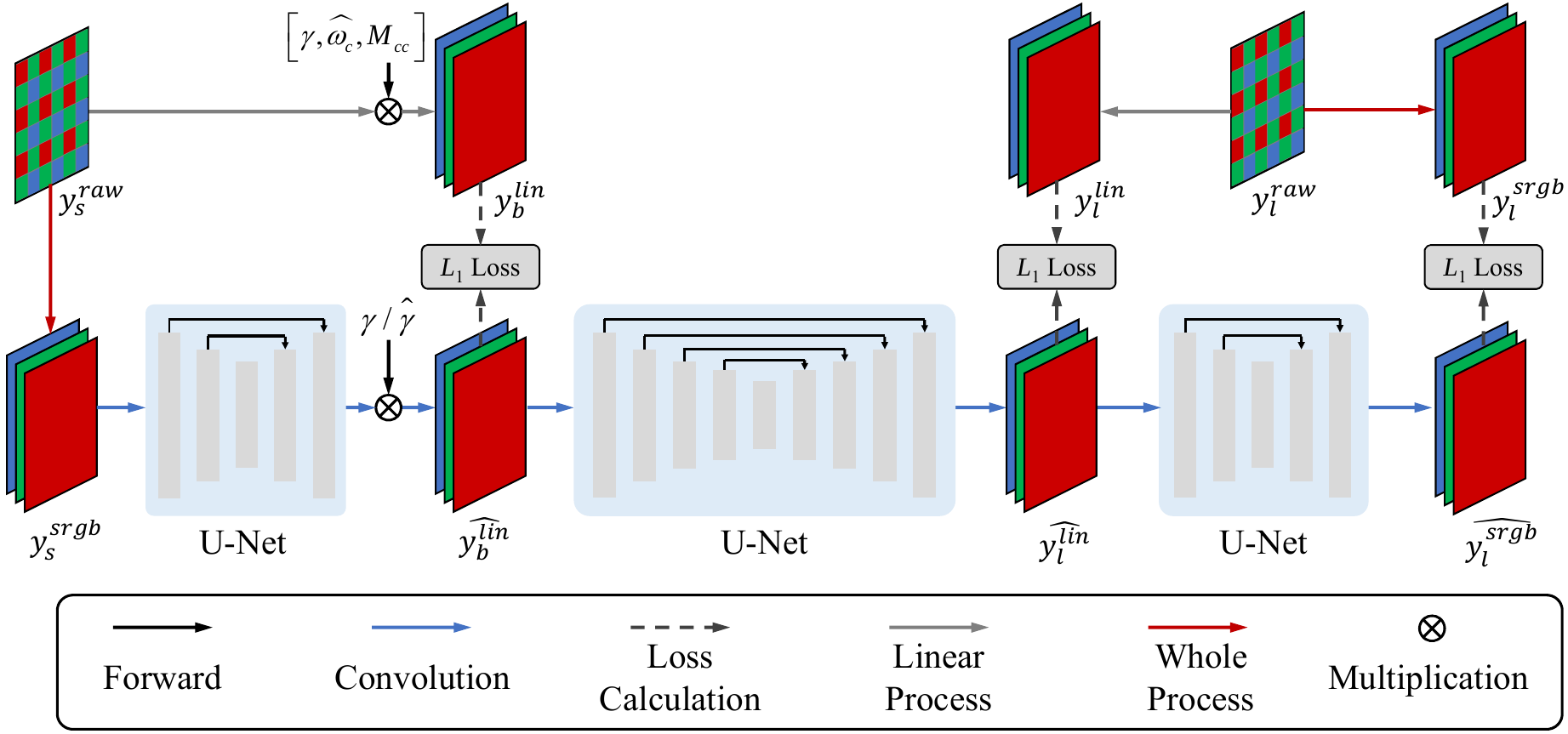}
		}
		
		\caption{
			Illustration for the architecture of our RAW-guiding Exposure Enhancement Network (REENet).
			REENet makes full use of low/normal light RAW images in the training.
			Functionally, the three parts of the REENet, \textit{i.e.} \textit{Unprocess}, \textit{Enhance} and \textit{Process},
			performs unprocessing, enhancement and processing, respectively.
		}
		\label{fig:arc}
	\end{figure*}
	
	\vspace{1mm}
	\noindent \textbf{Comparisons to State-of-the-art Methods}.
	 the above-mentioned baselines, we also evaluate several state-of-the-art sRGB-based methods including HE~\cite{HE1}, Dehazing~\cite{dong2011fast}, MF~\cite{FU201682}, MSR~\cite{Multi_scale_retinex}, LIME~\cite{Guo_2017_Lime}, BIMEF~\cite{BIMEF}, BPDHE~\cite{BPDHE}, LLNet~\cite{LLNet}, SICE~\cite{cai2018learning}, KinD~\cite{Kind}, DeepUPE~\cite{deepUPE} and Zero-DCE~\cite{Zero-DCE},  and RAW-based methods including EEMEFN~\cite{EEMEFN} and ELD~\cite{Wei_2020_CVPR} on SID-Sony dataset and provide systematic benchmark results using various metrics including PSNR, SSIM~\cite{SSIM}, VIF~\cite{VIF}, NIQE~\cite{Mittal2013MakingA} and LPIPS~\cite{lpips}, shown in Table~\ref{table:benchmark}.

	Apparently, there is still a huge performance gap between RAW-based and sRGB based approaches, mainly caused by the absence of \emph{linearity}. 
	Among RAW-based methods, the one equipped with the ground truth meta-data shows better performance, and when the ground truth \emph{exposure time} label is absent, 
	the performance drops a lot because it is quite difficult for the enhancement model to predict the illumination level accurately.
	The effect of \textit{white balance} and \textit{quantization} with \emph{Brighten then Qunatify} strategy is relatively small but still benefits the enhancement.
	Among the sRGB-based methods, proposed REENet with RAW guiding strategy shows superior performance, 
	and there are also performance drops in PSNR when some characteristics are absent 
	but their measures are still higher than other methods that also takes sRGB images as input.
	{Note that $\text{REENet}_{8bit}$ follows the traditional route -- \textit{Quantize then Brighten}, 
	and E adopts \emph{Brighten then Quantize} strategy for better quality.}
	If \textit{Quantize then Brighten} is adopted for E, \textit{i.e.} the same settings as $\text{REENet}_{8bit}$, the PSNR will drop to 16.89 dB as shown in Table~\ref{table:quan}.
	\vspace{1mm}
	
	\noindent \textbf{Qualitative Evaluation}.
	The corresponding qualitative results are shown in Fig.~\ref{fig:benchmark}, where we only provide relatively reasonable results. Apparently, the methods utilizing white balance parameters show accurate colors, \textit{e.g.} L+E+Q+W and L+E+Q. Note that all sRGB-based methods have applied white balance as a part of processing, which corrects the weight of RGB channels. Linearity and explicit exposure time induce the correct illumination, \textit{e.g.} L+E+Q and L+Q, and suppress artifacts in Q and baseline. More fine-grained quantization levels make the results change little, \textit{e.g.} L+E+Q and L+E. 
	As shown in the bottom panel, EEMEFN's results have an obvious color bias.
	ELD achieves better visual quality by using additional synthetic data.
	Comparatively, our REENet restores visually pleasant colors and details.
	Note that, REENet does not need RAW files during the testing phase.
	
	\section{REENet: Raw-guiding Exposure Enhancement Network}
	\label{sec:Net}
	
	From the benchmark results, the critical roles that the properties of RAW files play in the low-light enhancement are obviously observed, especially for linearity and exposure time.
	Therefore, we are inspired to construct a RAW-guiding Exposure Enhancement Network~(REENet) to fully utilize characteristics of RAW files, fully considering the advantages of RAW files as well as its inaccessibility of RAW images if we do not hope to rebuild the ISP process during the testing process.
	To achieve this, 
	REENet only gets access to RAW images in the training process.
	With the guidance of RAW images in the training, REENet learns to project the nonlinear sRGB images into the linear domain, which is proven to be a better paradigm than directly learning to enhance images in the nonlinear domain.
	Furthermore, with the difficulty in reversing the total process in mind, 
	REENet performs the enhancement in the linear RGB domain.
	We adopt the linear process to produce the linear RGB images, which are defined in Sec.~\ref{sec:pipe}.
	Then, with the wealth of the meta-data of RAW files and the linearity, the gap between sRGB images and RAW images can be largely bridged.
	
	{It is noted that, because of the ill-posed nature of the low-light enhancement task, the perfect ground truth is quite hard to define. In \textit{FEM}, the low-light image enhancement mainly focuses on suppressing noise and revealing detailed signals with a target (or given) exposure level. Although the ground truth images may not be perfect in providing a golden exposure level, the abundant information in the RAW image captured with a long exposure time also provides useful guidance for deriving a more effective enhancement model.}

	\subsection{Model Architecture}
	
	As shown in Fig.~\ref{fig:arc}~(a).
	REENet consists of three sub-modules:
	\begin{itemize}
		\item \textit{Unprocess} to project sRGB images into the linear RGB domain;
		\item \textit{Enhance} to suppress the amplified noise and color bias in the brightened images, which are adjusted by being multiplied with the exposure time ratio (ground truth or estimated);
		\item \textit{Process} to project the the enhanced results back into the nonlinear sRGB domain.
	\end{itemize}
	Note that, although the pipeline introduced in Section~\ref{sec:pipe} is simplified,
	our developed \textit{Unprocess} and \textit{Process} are flexible and general frameworks to transform the signals between  linear/nonlinear domains, which helps bridge the gap between the linear domain in image processing systems and the sRGB domain.
	\vspace{1mm}
	
	\noindent 1) \textbf{\textit{Unprocess}: Transfer Nonlinear Data into Linear Domain}.
	Since sRGB images do not include meta-data,
	the conventional enhancement method~\cite{brooks2019unprocessing} projects the processed nonlinear sRGB images back into the linear domain via hand-crafted approaches.
	Hence, the designed inversion process inevitably has a gap with the real processing approaches in various real applications, leading to inaccurate estimation.
	The gap might be further magnified especially, as shown in Fig.~\ref{fig:gamma}, when the exposure time ratio is multiplied.
	Therefore, an end-to-end convolutional neural network, \textit{i.e.} a U-Net~\cite{U-Net} ${\sigma}^{'}(\cdot)$ is adopted for that.
	{More exactly, given the processed input $f_{s}^{srgb}$ and linear target $y_{s}^{lin}$, \textit{Unprocess} aims to predict a brightened linear RGB image:
	\begin{align}
		\widehat{y}_{b}^{lin} = \gamma^{'} \widehat{y}_{s}^{lin} = \gamma^{'} \widehat{\sigma}^{'}(f_{s}^{srgb}),
		\label{equ:unpro}
	\end{align}
	where $\gamma^{'} = \gamma$ if the meta-data of the short-exposure RAW image is available during the testing phase, or $\gamma^{'} = \hat{\gamma}$ if not, and $f_{s}^{srgb}$ is quantized from $y_{s}^{srgb}$ with 8 bits or 16 bits per pixel. The gap of $f_{s}^{srgb}$ and $y_{s}^{srgb}$ depends on the number of quantization levels, whose impact has been explored in our experiments.}
	\vspace{1mm}
	
	\noindent 2) \textbf{\textit{Enhance}: Normal-Light Image Reconstruction}.
    The brightened linear RGB images, amplified by the exposure time ratio, are:
	\begin{equation}
		y_{b}^{lin} = y_{s}^{lin} \times \gamma =
		\left( \begin{array}{c}
			\widehat{w}_{r} \gamma y_{s,r}^{raw}\\ 
			\quad \\
			\widehat{w}_{g} \gamma y_{s,g}^{raw}\\  
			\quad \\
			\widehat{w}_{b} \gamma y_{s,b}^{raw}
		\end{array} \right)
		M_{cc}.
	\end{equation}
	Compared to the long-exposed linear RGB images:
	\begin{equation}
		y_{l}^{lin} =
		\left( \begin{array}{c}
			w_{r} y_{l,r}^{raw}\\ 
			\quad \\
			w_{g} y_{l,g}^{raw}\\  
			\quad \\
			w_{b} y_{l,b}^{raw}
		\end{array} \right)
		M_{cc} =
		\left( \begin{array}{c}
			w_{r} \gamma x_{s,r}\\ 
			\quad \\
			w_{g} \gamma x_{s,g}\\  
			\quad \\
			w_{b} \gamma x_{s,b}
		\end{array} \right)
		M_{cc},
	\end{equation}
	and according to Eqn.~\eqref{equ:brightened}, 
	{\textit{Enhance} targets to suppress the noise, whose noise levels are ${\gamma^{'}}^{2} \lambda_{read}$ and $\gamma^{'} \lambda_{shot}$, respectively,
	and aims to compensate for the color casting caused by the inaccurate white balance $\widehat{W}$.}
	Keeping the excellent modeling capacities of convolutional networks for image/video denoising~\cite{H2012Image,Liu_2018_CVPR_Workshops} and color correction~\cite{SID,Chen_2019_ICCV} in mind, 
	a U-Net is adopted to build the architecture of \textit{Enhance} .
	For simplicity, we use $g(\cdot)$ to denote the fitted denoising and color restoration processes. 
	Given the brightened inputs $y_{s}^{lin} \times \gamma^{'}$ and long-exposed linear targets $y_{l}^{lin}$, \textit{Enhance} aims to estimate $\widehat{y}_{l}^{lin} = \widehat{g}(y_{s}^{lin} \times \gamma^{'})$.
	\vspace{1mm}
	
	\noindent 3) \textbf{\textit{Process}: Transfer Linear Data into Nonlinear Domain}. 
	Similar to \textit{Unprocess}, another U-Net is utilized for modeling the nonlinear process.
	To be exact, given the long-exposure linear input $y_{l}^{lin}$ and the corresponding nonlinear sRGB target $y_{l}^{srgb}$, 
	\textit{Process} outputs $\widehat{y}_{l}^{srgb} = \widehat{\sigma}(y_{l}^{lin})$.
	
	To summarize, our REENet predict the exposure time adjusted result of the input in real applications via: 
	\begin{equation}
		\widehat{y}_{l}^{srgb}=\widehat{\sigma}(\widehat{g}(\widehat{\sigma}^{'}(y_{s}^{srgb})\times \gamma^{'})).
	\end{equation} 
	Note that, the testing phase can work without RAW files as input.
	\subsection{Experimental results}
	\label{sec:Exp}

	\begin{table*}[tb]
		\begin{center}
			\caption{
				Quantitative evaluation comparing  traditional methods and the proposed method.
				The best result is denoted in bold. 
			}
			\begin{tabular}{c|cccccccccc}
				\hline
				Method 	& PSNR$\uparrow$ & {PSNR$\ast\uparrow$}  
				& SSIM$\uparrow$ & {SSIM$\ast\uparrow$} 
				&   VIF$\uparrow$  & {VIF$\ast\uparrow$}
				& NIQE$\downarrow$ & {NIQE$\ast\downarrow$}
				& {LPIPS}$\downarrow$ & {LPIPS$\ast\downarrow$}\\
				\hline
				HE\cite{HE1} 		& 5.90 & {5.90}
				& 0.028 & {0.028}
				& 0.095  & {0.095}
				&  7.71 & {7.71}
				& {0.968} & {0.968}
				\\
				BPDHE~\cite{BPDHE}
				&10.67  & {10.79}
				& 0.072 & {0.188}
				& 0.051  & {0.052}
				& 16.65 & {16.66}
				& {0.969} & {0.970}\\
				Dehazing~\cite{dong2011fast}   
				&12.81 & {15.01}
				& 0.103 & {0.404}
				& 0.077  & {0.103}
				& 8.09 & {6.37}
				& {0.784} & {0.889}\\
				MSR~\cite{Multi_scale_retinex}
				&  10.04 & {10.04}
				& 0.070 & {0.327}
				& 0.116 & {0.116}
				& 6.33 & {6.33}
				& {1.031} & {1.031}\\
				MF~\cite{FU201682}	
				& 13.87 & {14.17}
				& 0.111  & {0.387}
				&0.108  & {0.108}
				& 6.34 & {6.39}
				& {0.950} & {0.960}\\
				LIME~\cite{Guo_2017_Lime} 
				& 12.59  & {12.79}
				& 0.102  & {0.372}
				&0.118 & {0.119}
				& 6.06 & {6.10}
				& {0.980} & {0.983}\\
				BIMEF~\cite{BIMEF}		 
				& 13.06  & {14.95}
				&  0.110 & {0.410}
				&0.086  & {0.104}
				&  7.67 & {9.30}
				& {0.798} & {0.890}\\
				{REENet$_{8bit}$}       
				& { 25.75} & {25.75}
				& {0.808} & {0.808}
				& {0.135} &{0.135}
				& {6.23} & {6.23}
				& {0.424} & {0.424}\\
				REENet			
				& \textbf{28.42} & {\textbf{28.42}}
				& \textbf{0.880} & \textbf{0.880}
				&\textbf{0.139} &\textbf{0.139}
				& \textbf{5.60} & \textbf{5.60}
				& {\textbf{0.322}} & {\textbf{0.322}}\\
				\hline			
			\end{tabular}
			\label{table:comp1}
		\end{center}
	\end{table*}
	\begin{figure*}[tbp]
		\centering
		\subfigure[HE]{
			\includegraphics[width=30mm]{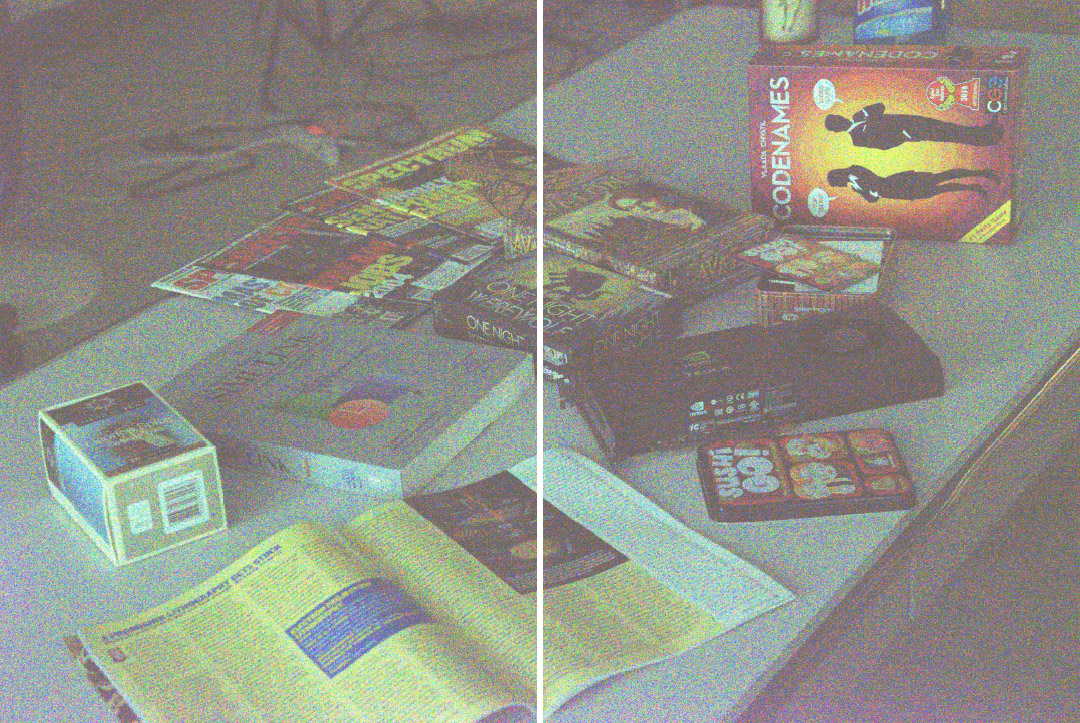}
		}
		\subfigure[BPDHE]{
			\includegraphics[width=30mm]{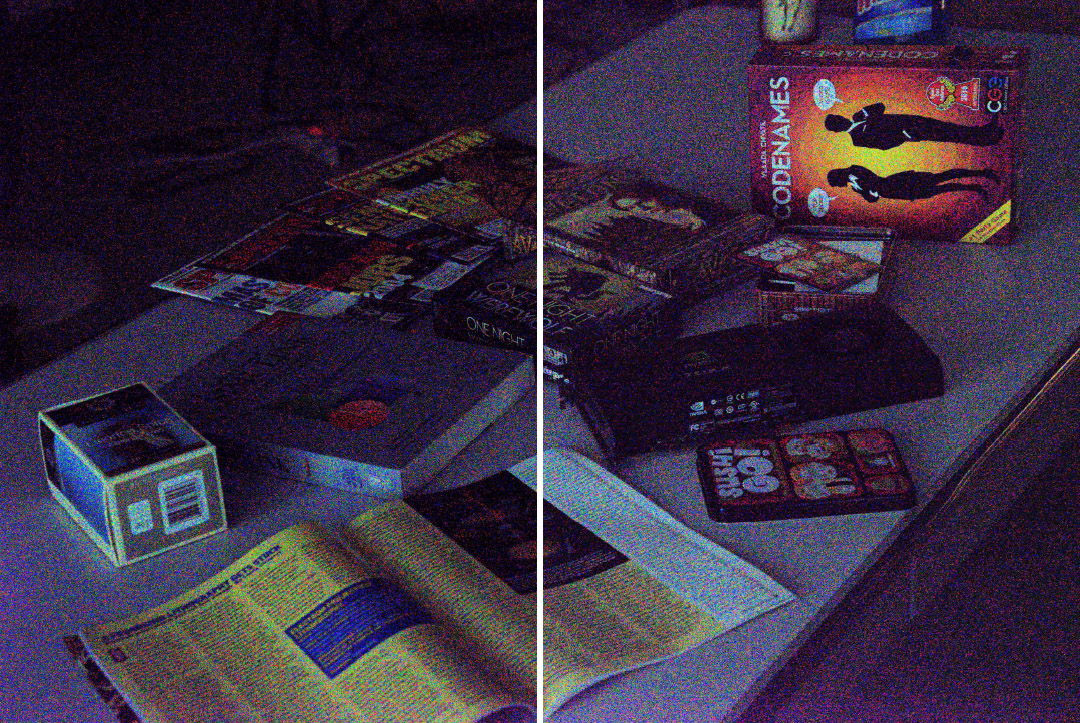}
		}
		\subfigure[Dehazing]{
			\includegraphics[width=30mm]{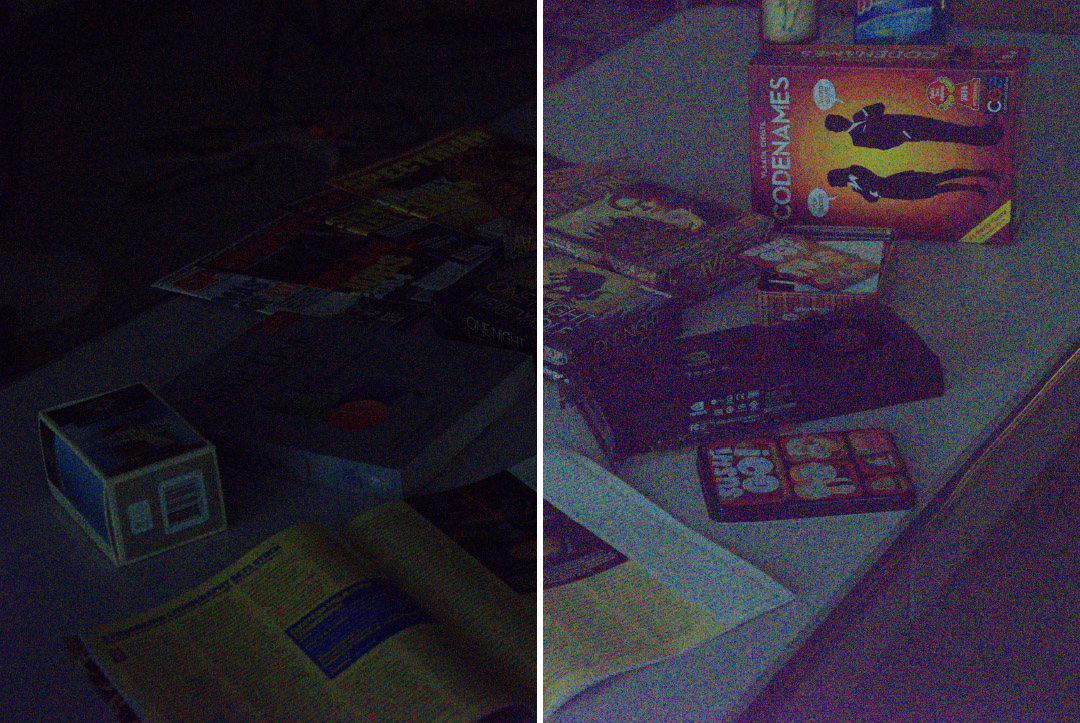}
		}
		\subfigure[MSR]{
			\includegraphics[width=30mm]{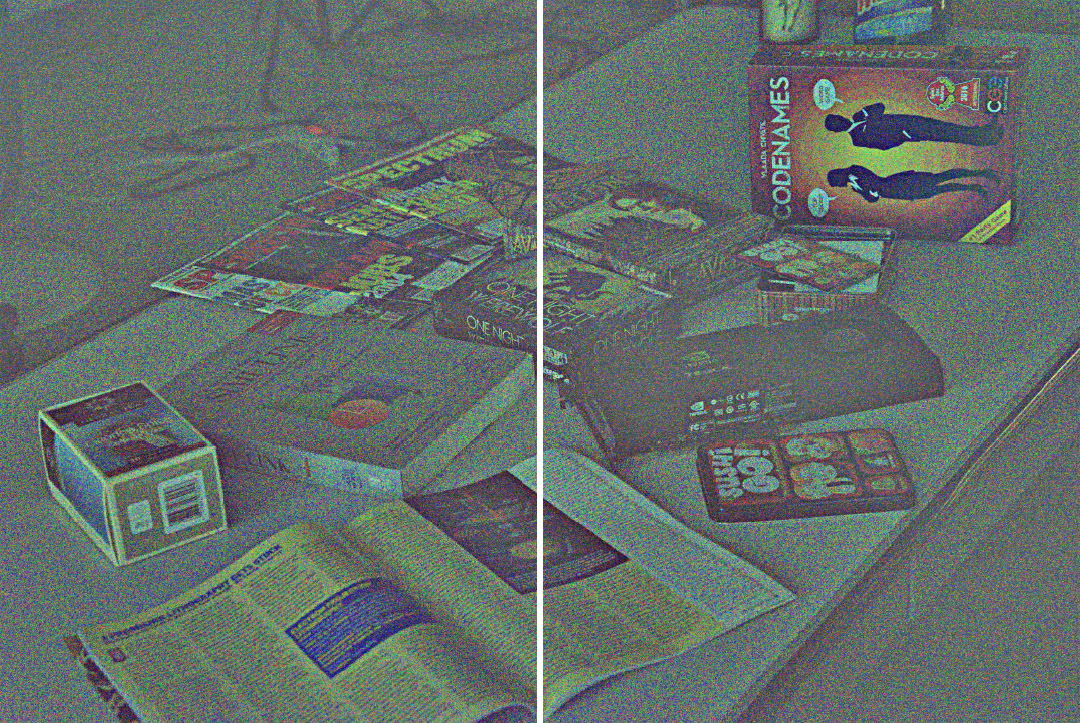}
		}
		\subfigure[MF]{
			\includegraphics[width=30mm]{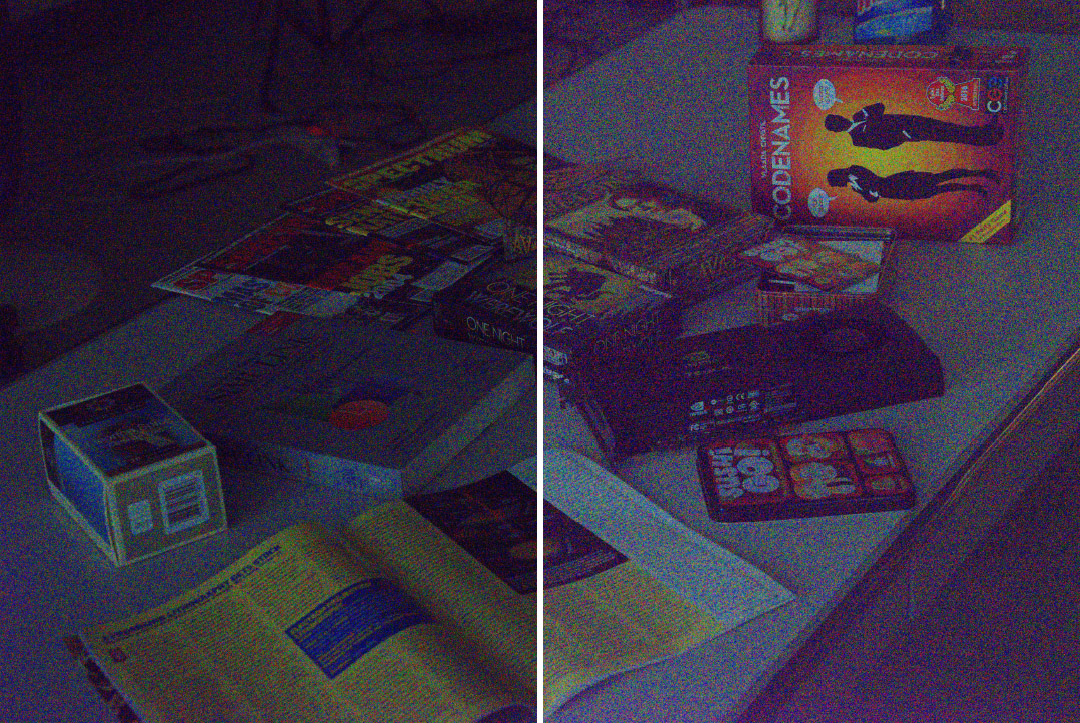}
		}
		\subfigure[LIME]{
			\includegraphics[width=30mm]{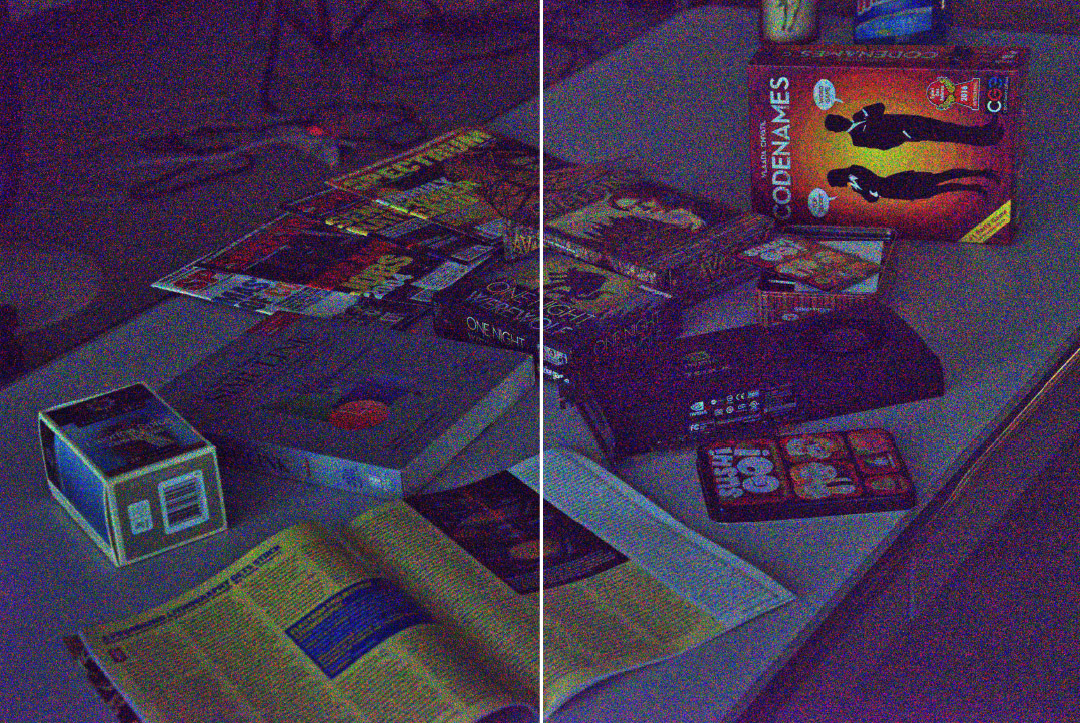}
		}
		\subfigure[BIMEF]{
			\includegraphics[width=30mm]{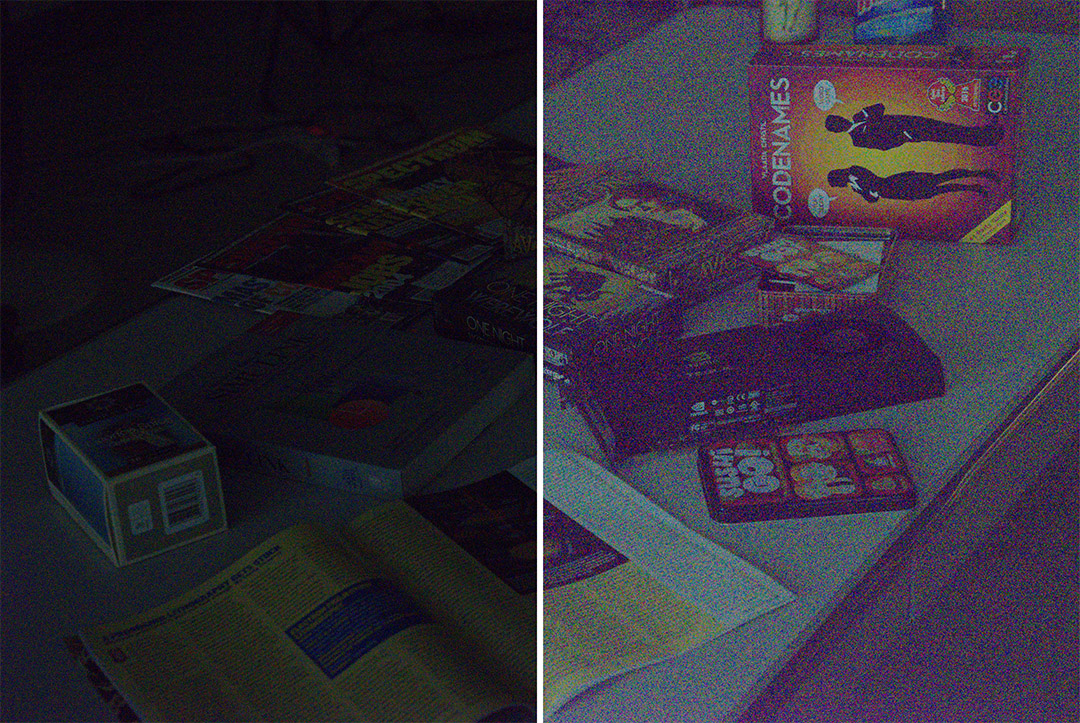}
		}
		\subfigure[REENet$_{8bit}$]{
			\includegraphics[width=30mm]{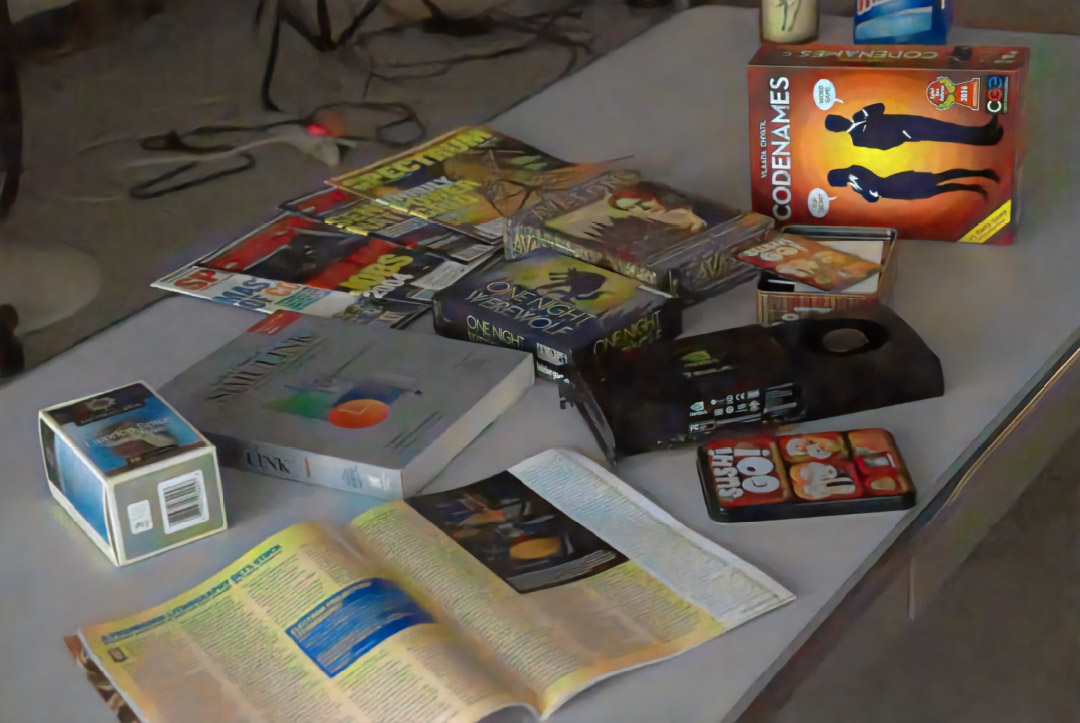}
		}
		\subfigure[REENet]{
			\includegraphics[width=30mm]{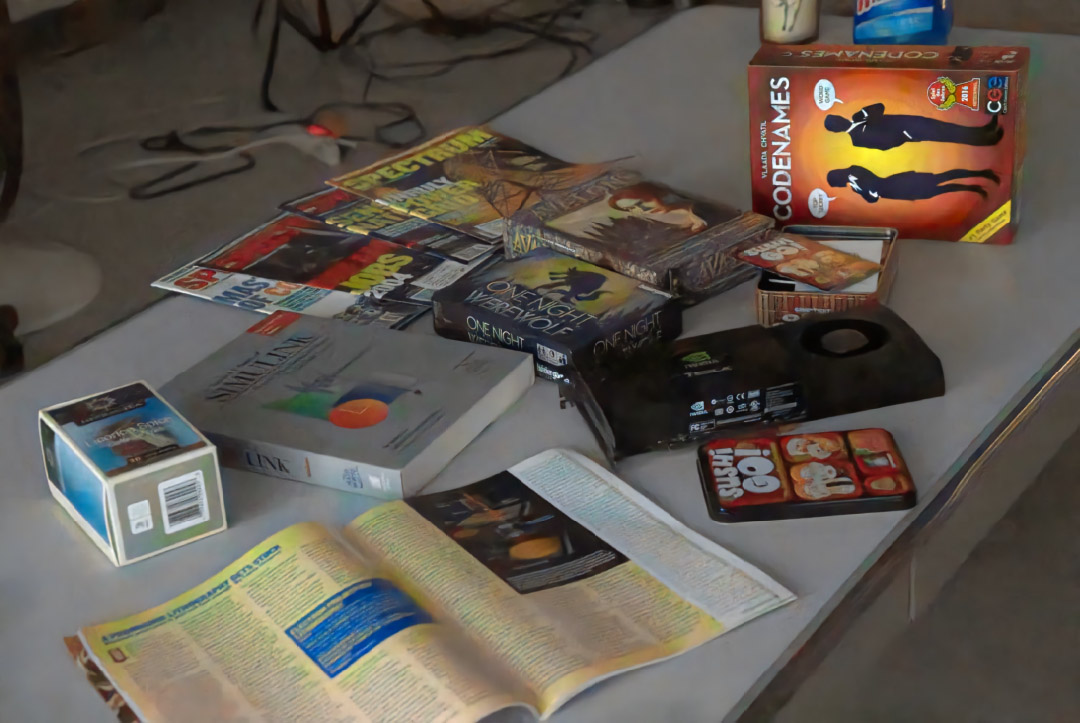}
		}
		\subfigure[Input / GT]{
			\includegraphics[width=30mm]{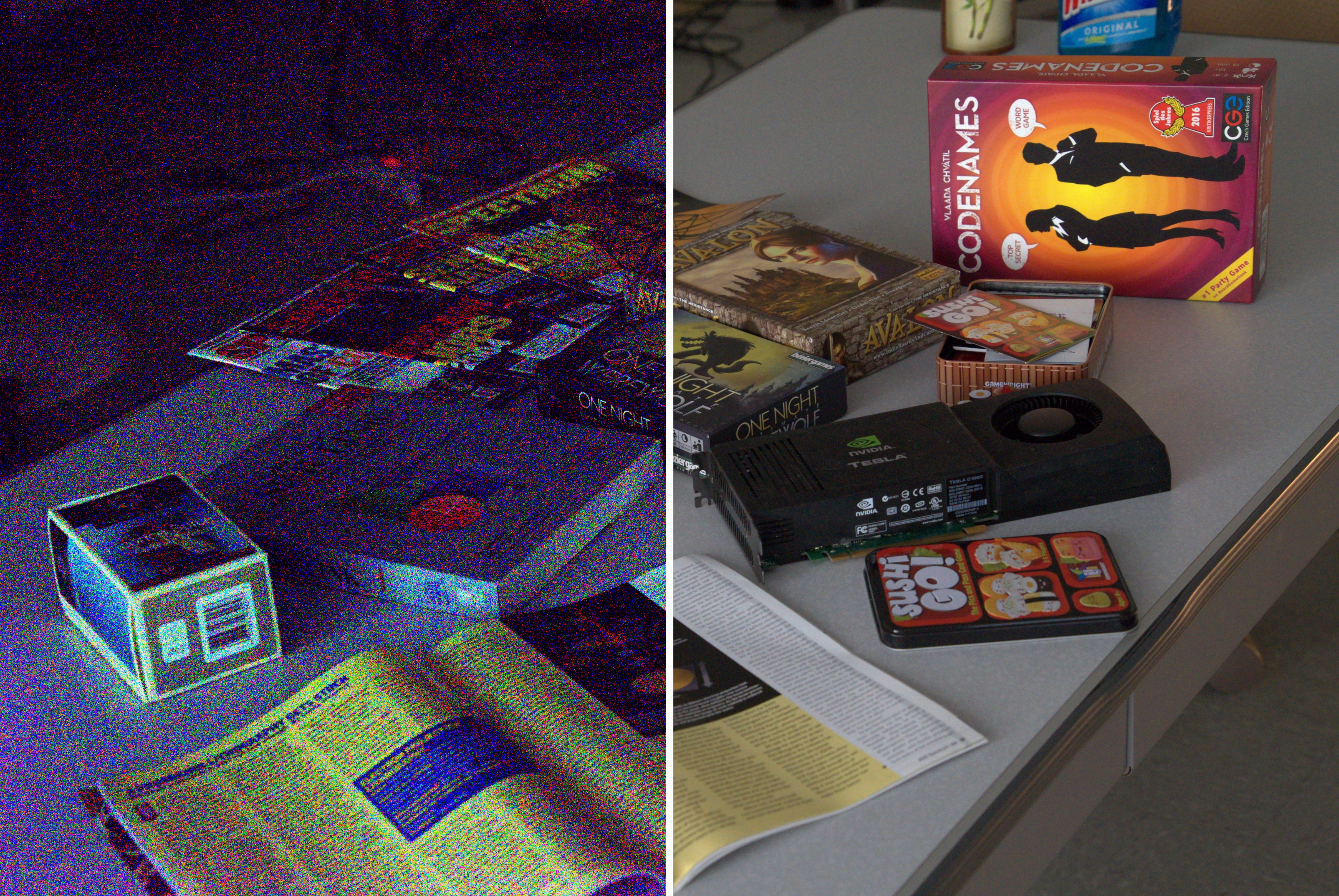}
		}
		\caption{
			{Qualitative evaluation comparing traditional methods and the proposed method. \textbf{Left Panel}: Original results of competing methods. \textbf{Right Panel}: Gamma corrected results with aligned brightness. {The last image is composed of the brightened input at the left and the \textit{Ground Truth} at the right. Note that the input is almost totally invisible without brightening.}}
		}
		\label{fig:comp1}
	\end{figure*}

	\begin{figure*}[!t]
		\centering
		\subfigure[LLNet]{
			\begin{minipage}[t]{20mm}
				\centering
				\includegraphics[width=20mm]{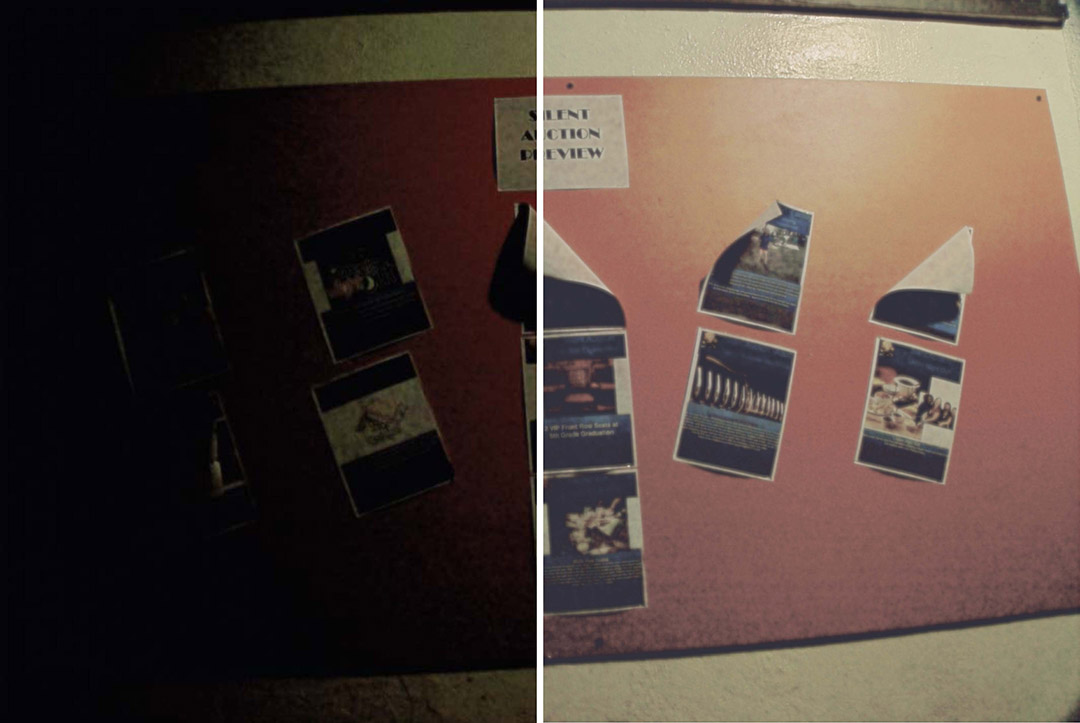}\\
				\vspace{0.2cm}
				\includegraphics[width=20mm]{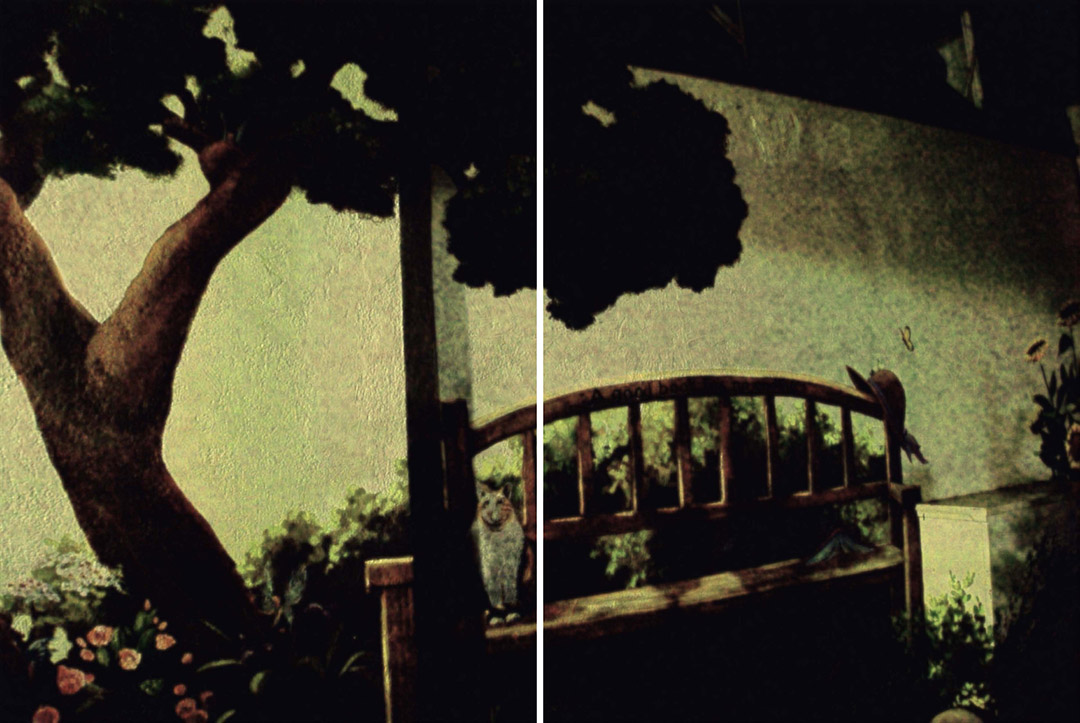}\\
				\vspace{0.2cm}
				\includegraphics[width=20mm]{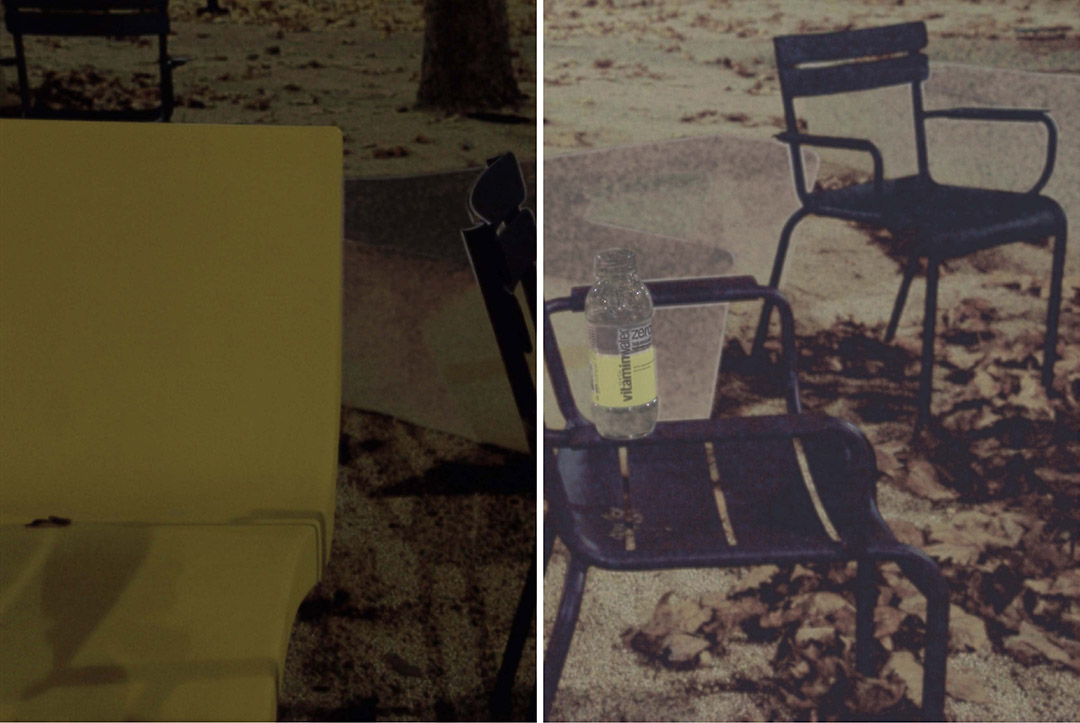}\\
				\vspace{0.2cm}
			\end{minipage}
		}
		\subfigure[SICE]{
			\begin{minipage}[t]{20mm}
				\centering
				\includegraphics[width=20mm]{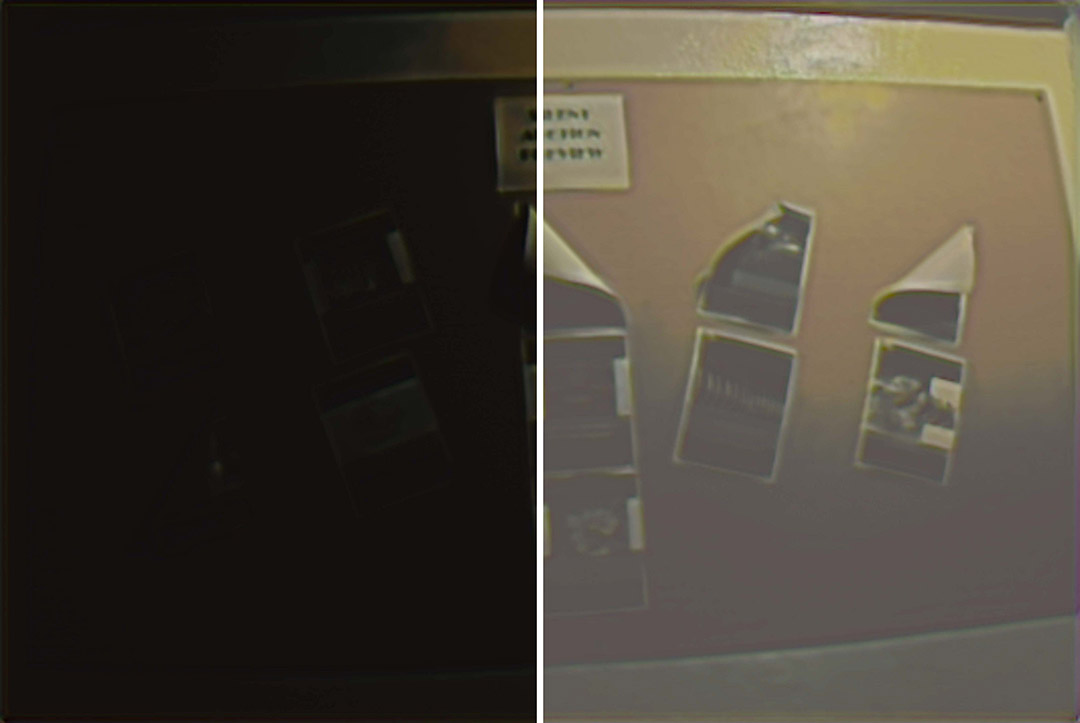}\\
				\vspace{0.2cm}
				\includegraphics[width=20mm]{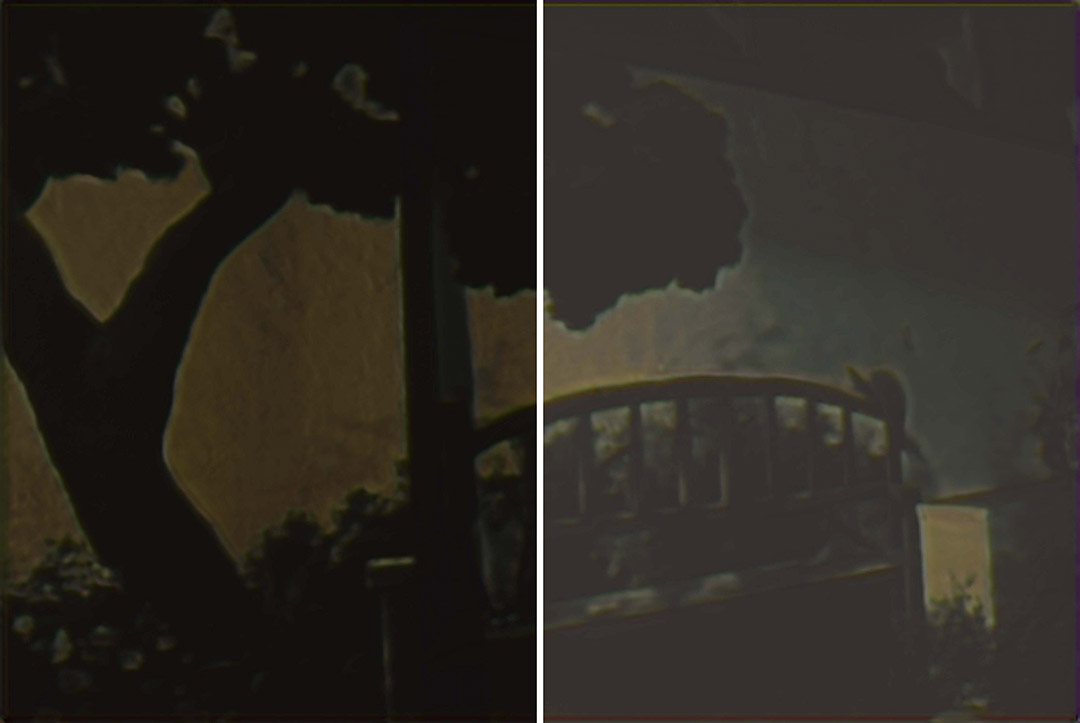}\\
				\vspace{0.2cm}
				\includegraphics[width=20mm]{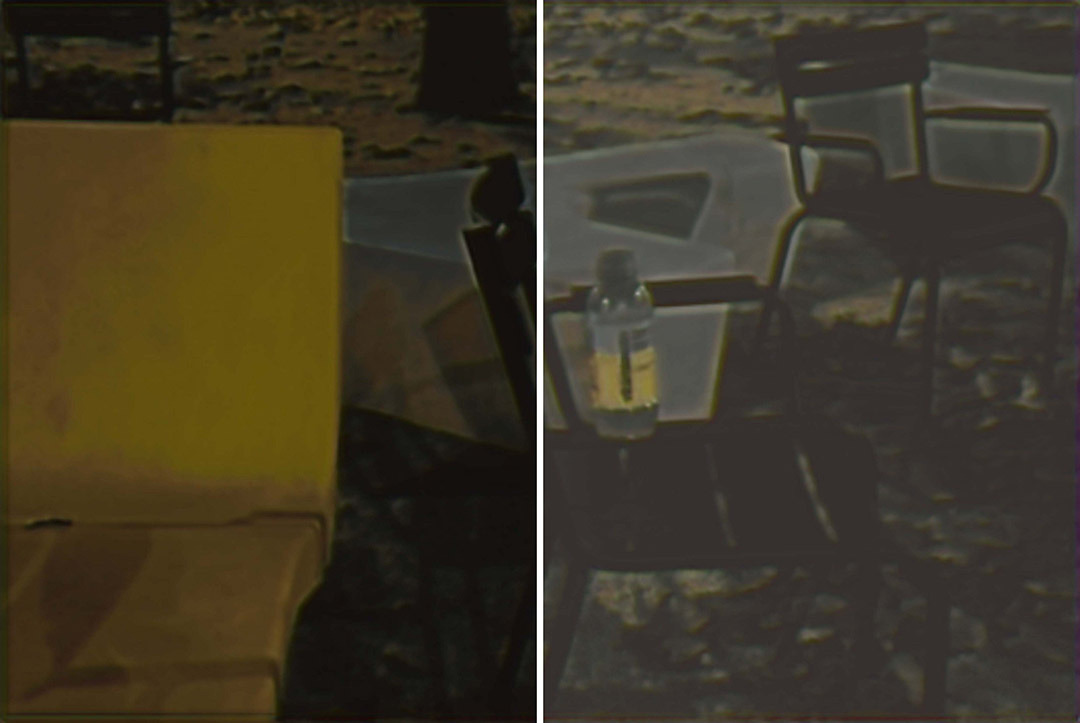}\\
				\vspace{0.2cm}
			\end{minipage}
		}
		\subfigure[KinD]{
			\begin{minipage}[t]{20mm}
				\centering
				\includegraphics[width=20mm]{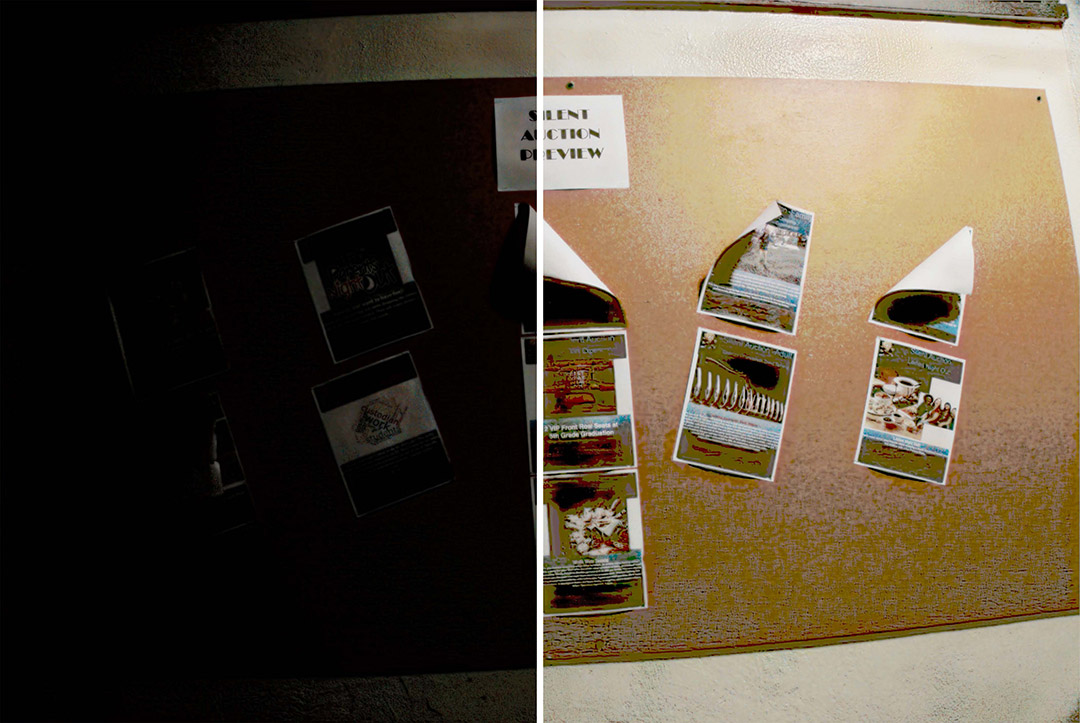}\\
				\vspace{0.2cm}
				\includegraphics[width=20mm]{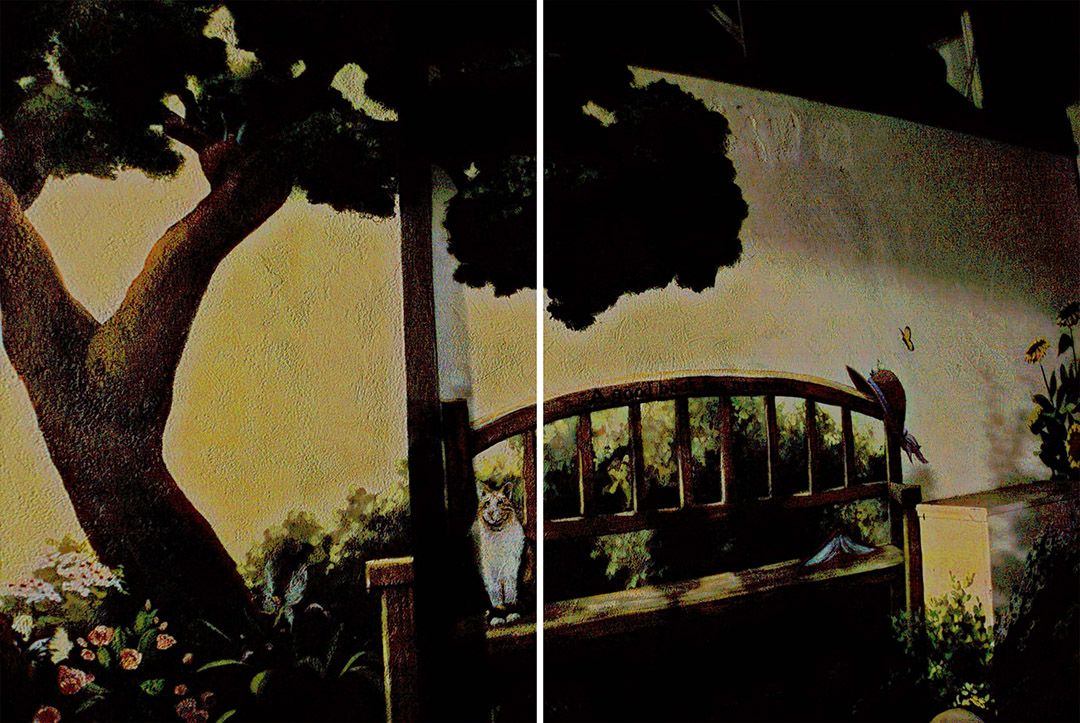}\\
				\vspace{0.2cm}
				\includegraphics[width=20mm]{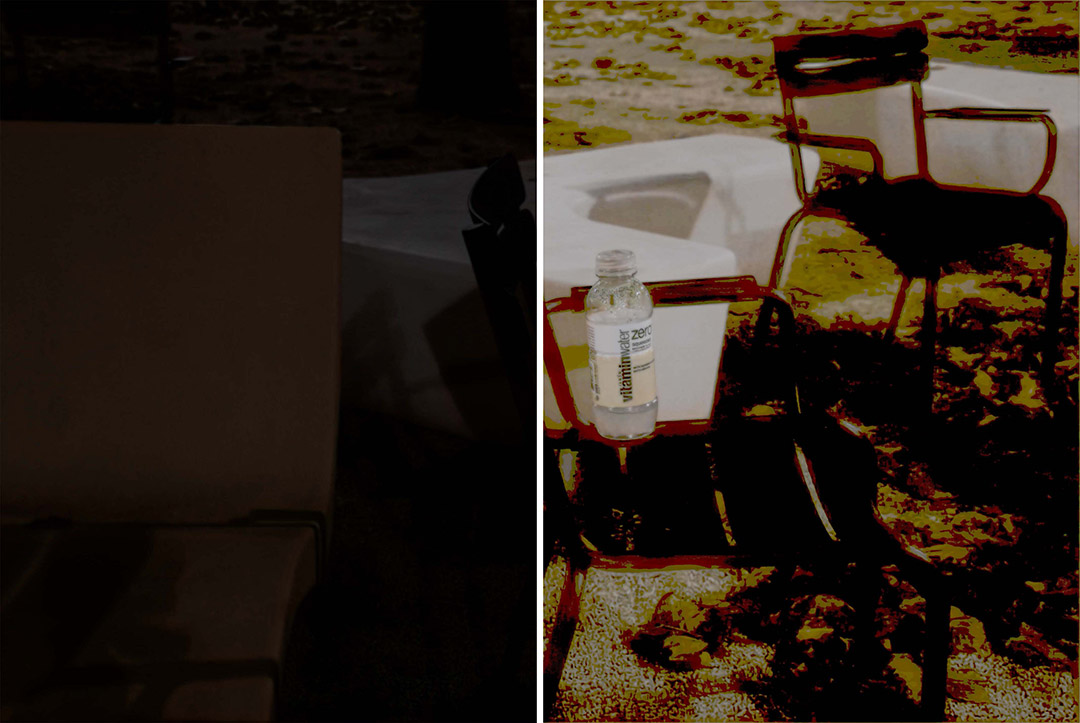}\\
				\vspace{0.2cm}
			\end{minipage}
		}
		\subfigure[DeepUPE]{
			\begin{minipage}[t]{20mm}
				\centering
				\includegraphics[width=20mm]{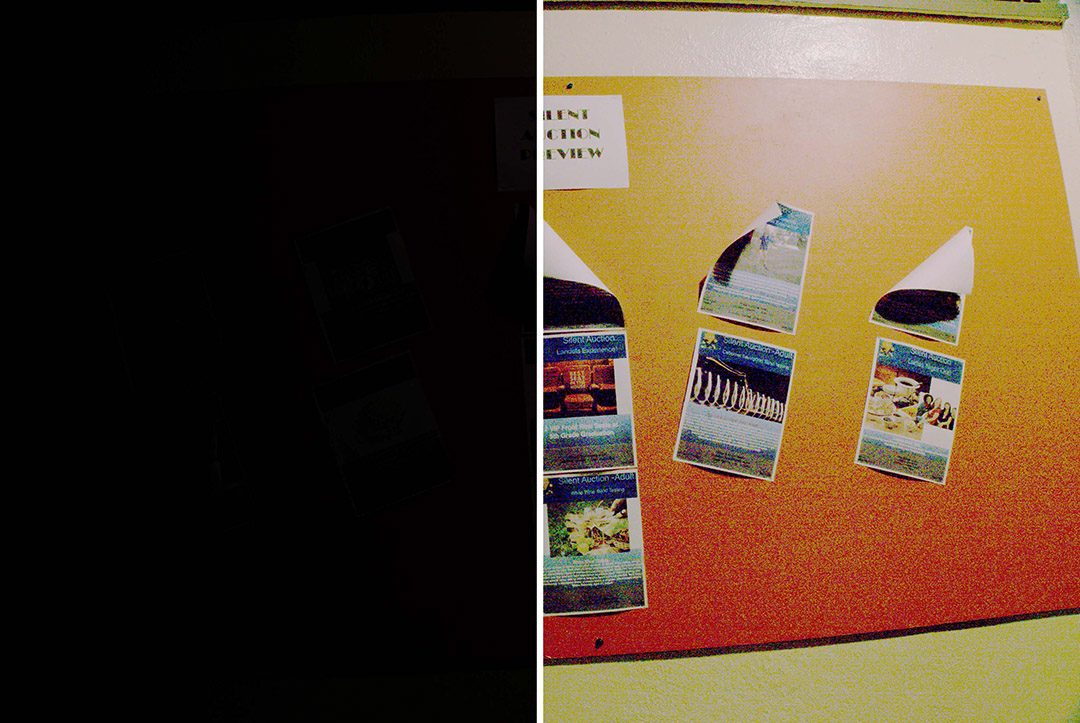}\\
				\vspace{0.2cm}
				\includegraphics[width=20mm]{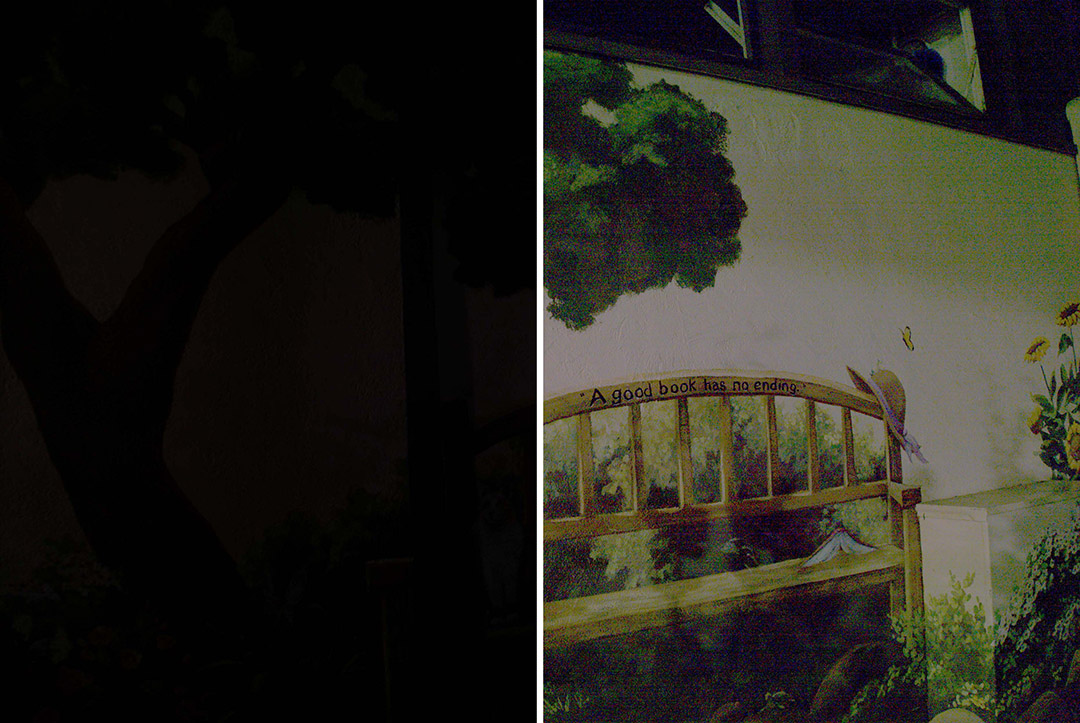}\\
				\vspace{0.2cm}
				\includegraphics[width=20mm]{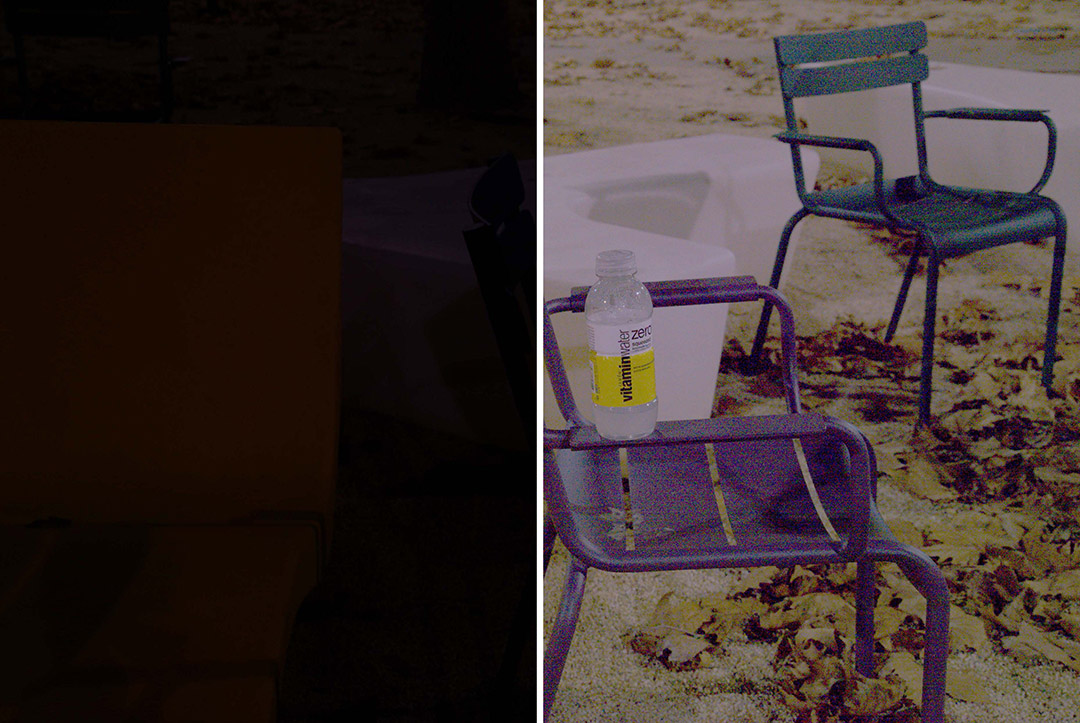}\\
				\vspace{0.2cm}
			\end{minipage}
		}
		\subfigure[REEnet$_{8bit}$]{
			\begin{minipage}[t]{20mm}
				\centering
				\includegraphics[width=20mm]{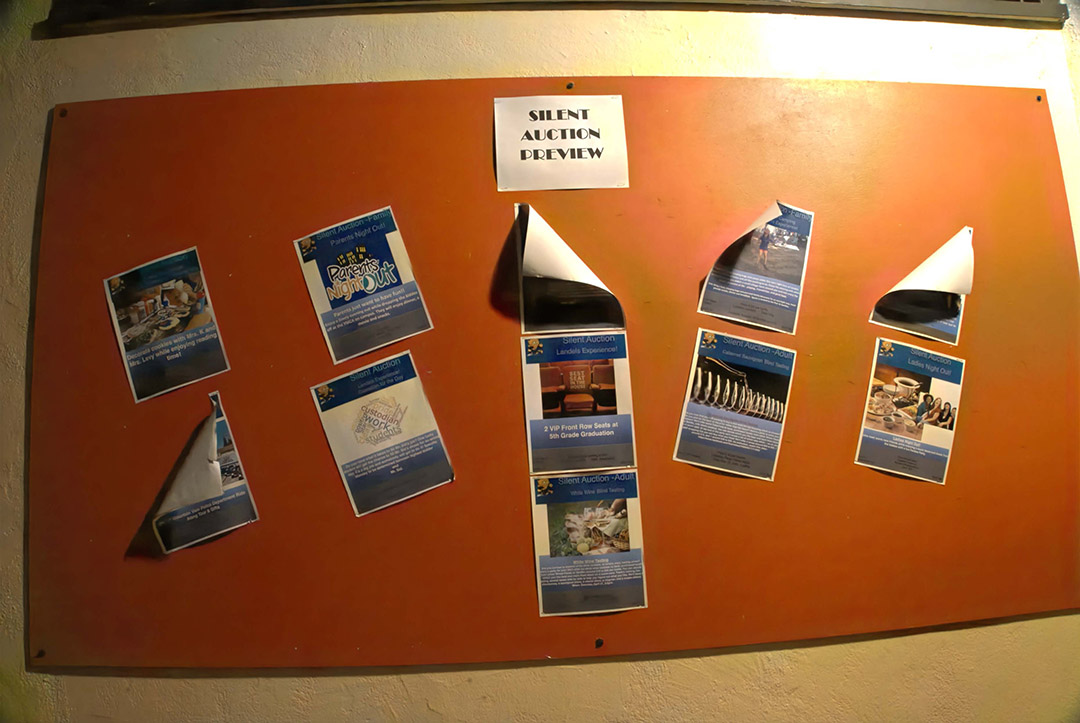}\\
				\vspace{0.2cm}
				\includegraphics[width=20mm]{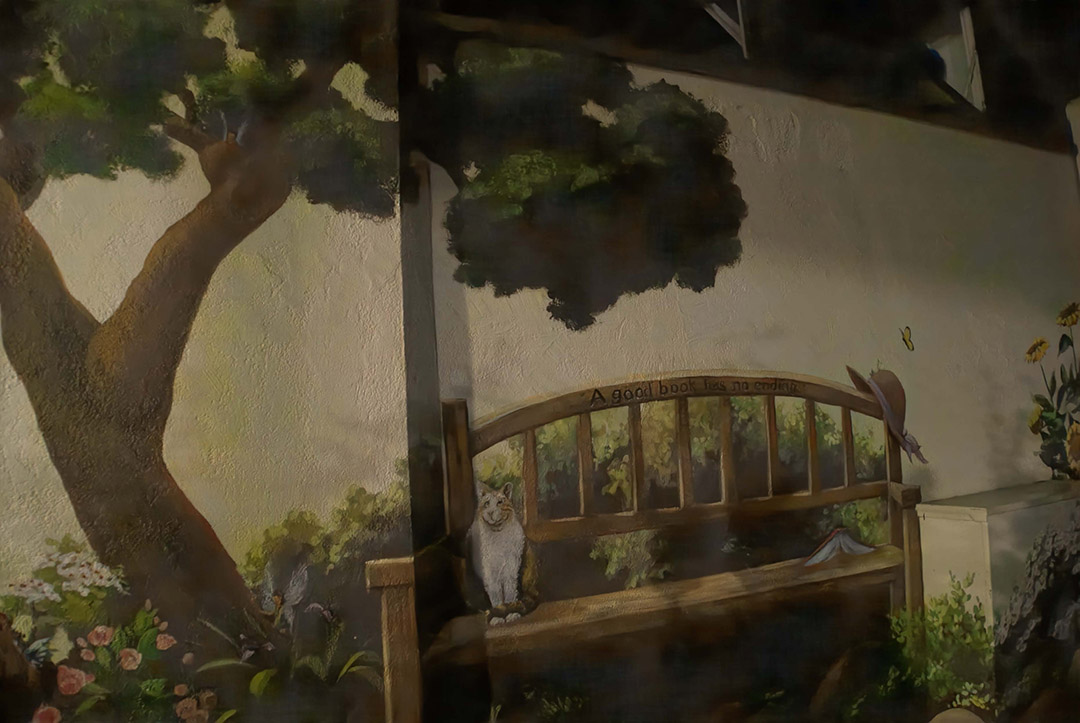}\\
				\vspace{0.2cm}
				\includegraphics[width=20mm]{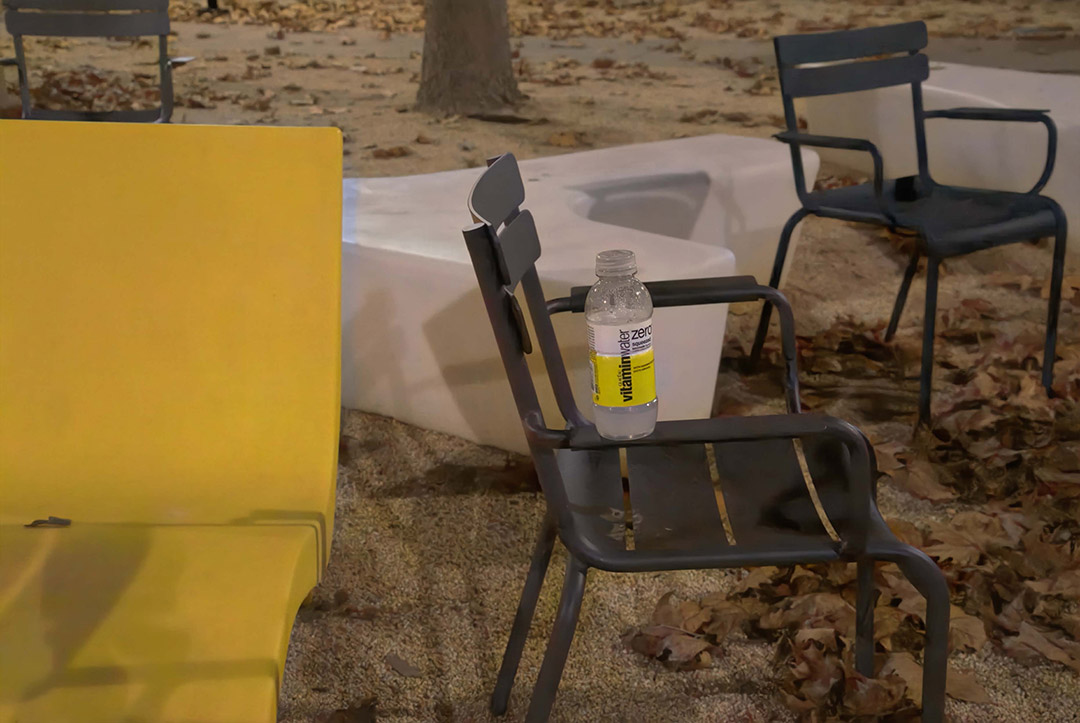}\\
				\vspace{0.2cm}
			\end{minipage}
		}
		\subfigure[REEnet]{
			\begin{minipage}[t]{20mm}
				\centering
				\includegraphics[width=20mm]{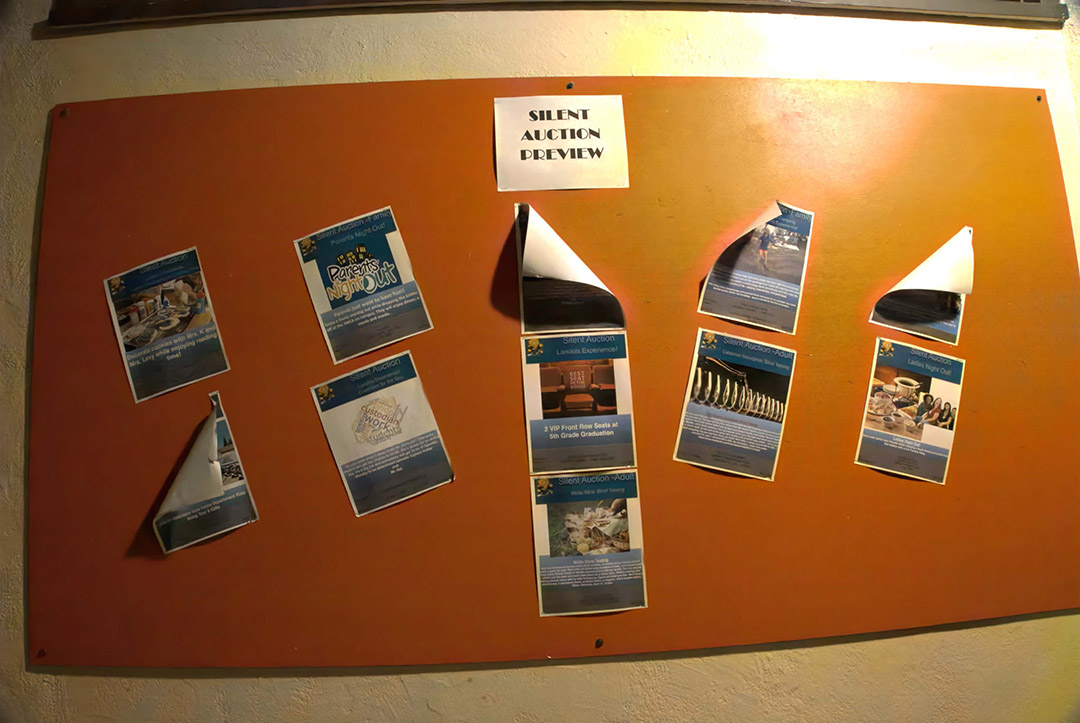}\\
				\vspace{0.2cm}
				\includegraphics[width=20mm]{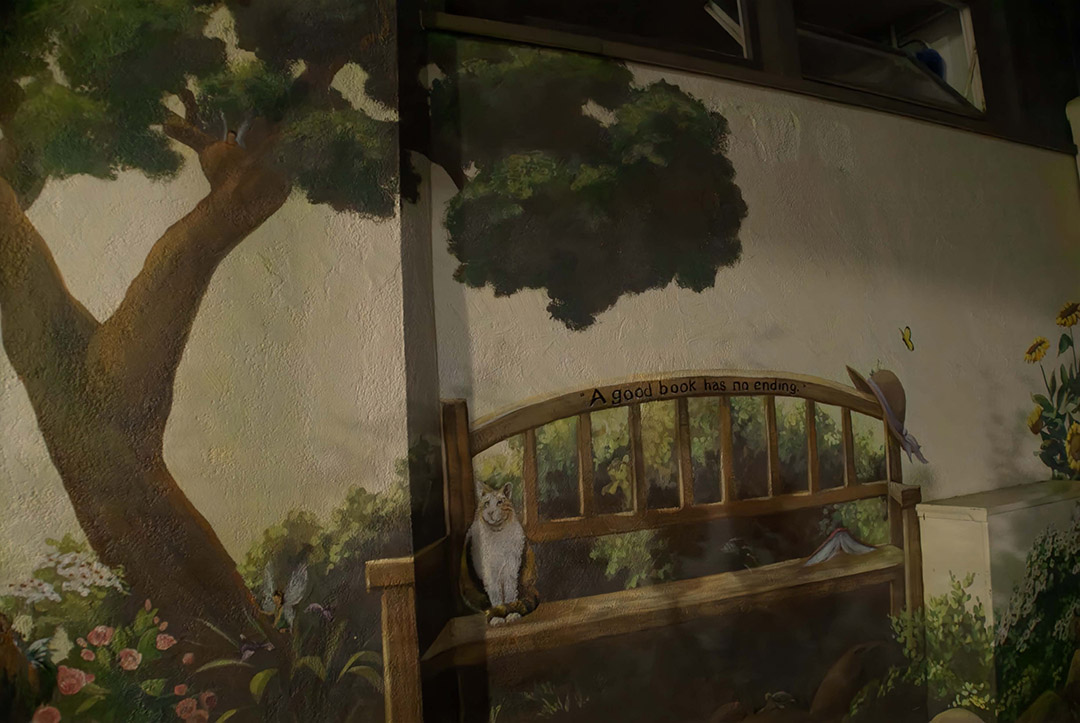}\\
				\vspace{0.2cm}
				\includegraphics[width=20mm]{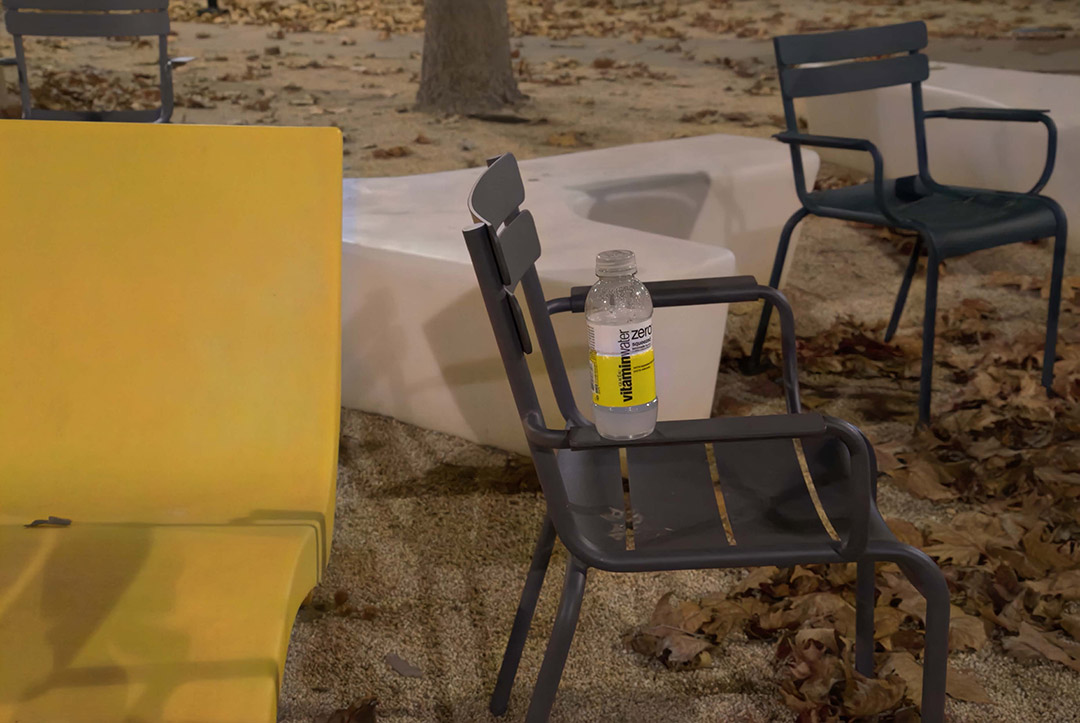}\\
				\vspace{0.2cm}
			\end{minipage}
		}
		\subfigure[Input / GT]{
			\begin{minipage}[t]{20mm}
				\centering
				\includegraphics[width=20mm]{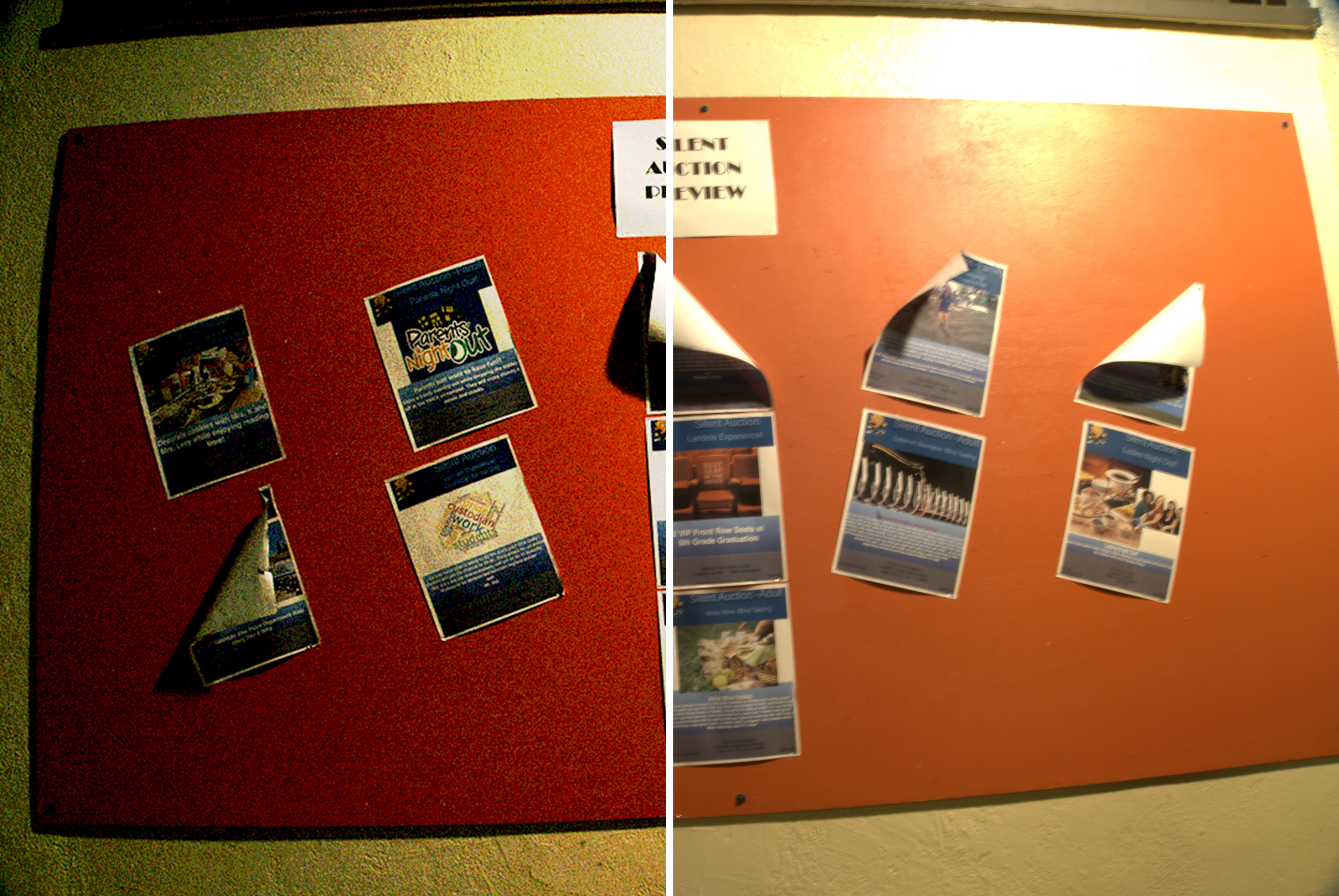}\\
				\vspace{0.2cm}
				\includegraphics[width=20mm]{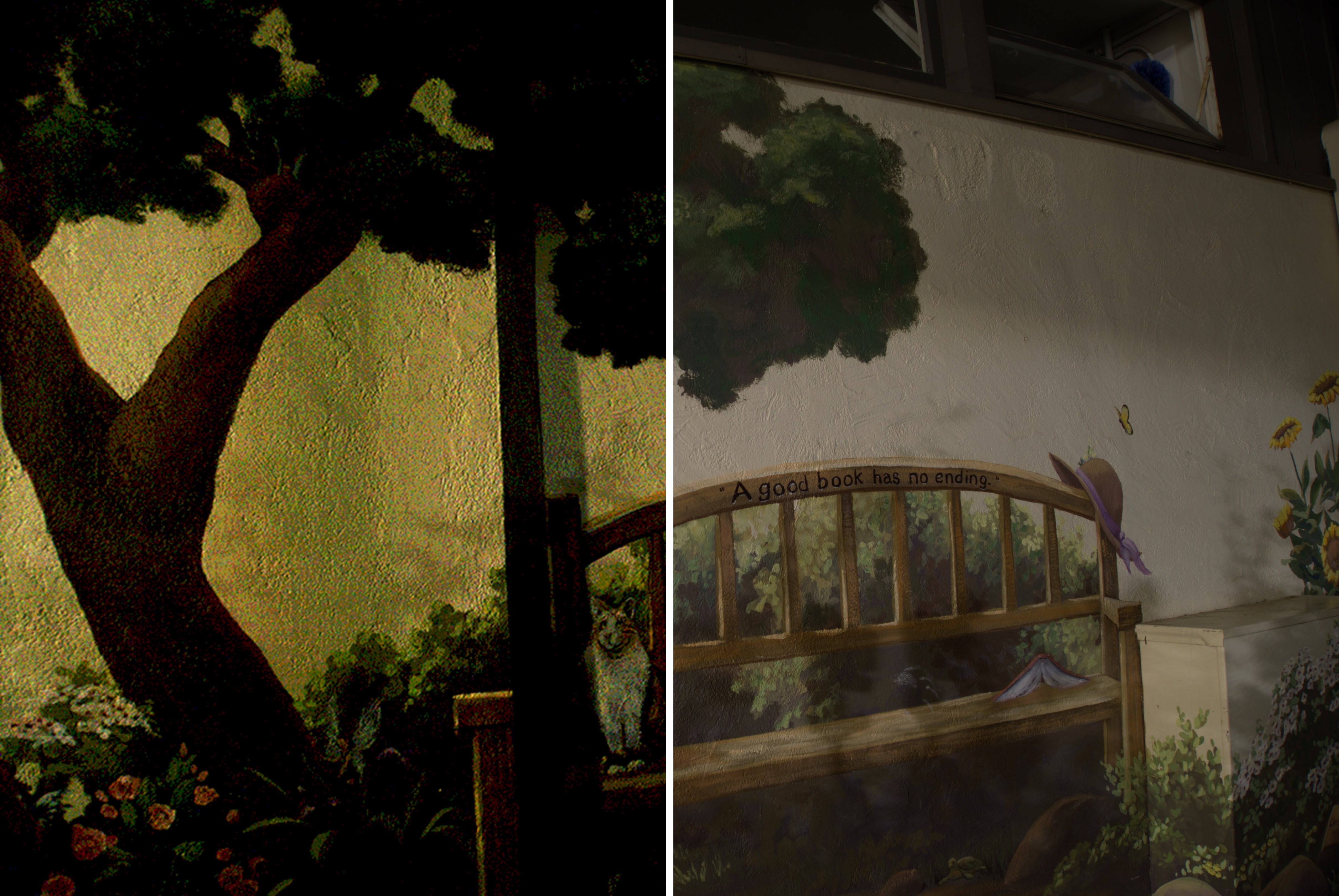}\\
				\vspace{0.2cm}
				\includegraphics[width=20mm]{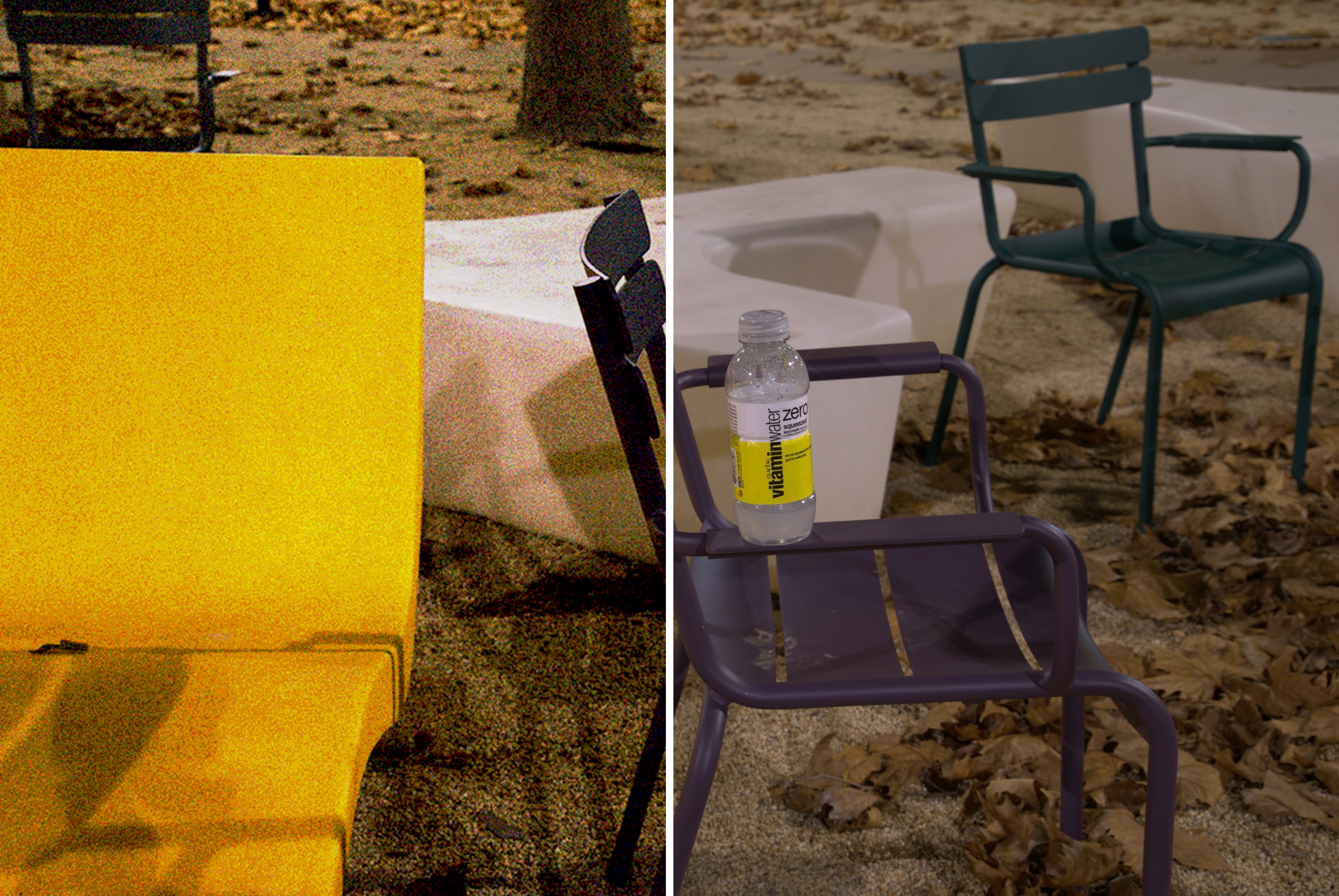}\\
				\vspace{0.2cm}
			\end{minipage}
		}
		\caption{
			{Qualitative evaluation comparing learning-based methods and the proposed method. \textbf{Left Panel}: Original results of competing methods. \textbf{Right Panel}: Gamma corrected results with aligned brightness. {The last image is composed of the brightened input at the left and the \textit{Ground Truth} at the right. Note that the input is almost totally invisible without brightening.}}
		}
		\label{fig:comp2}
	\end{figure*}
	
	\noindent \textbf{Training Details}. 
	During the training process, the nonlinear low/normal-light sRGB images processed by Libraw are taken as inputs and ground truths, respectively.
	The RAW images are also employed as training guidance.
	Adam optimizer~\cite{Adam} and $L_1$ loss are adopted for training. 
    The patch size and  batch size are set to 512$\times$512 and 1, respectively.
	The output results of all three sub-networks are clipped to $[0,1]$.
	All sub-networks of REENet are pre-trained for 3000 epochs independently.
	We set the learning rate to $10^{-4}$ at the beginning and $10^{-5}$ after 2,000 epochs.
	After that, all sub-networks are trained jointly with the learning rate $10^{-5}$ for 1,000 epochs. 
	REENet is trained on Intel(R) Xeon(R) E5-2650 2.20GHz CPU and an Nvidia RTX 2080Ti GPU in Python and Tensorflow.
	Because the input images have very large resolutions, 
	we crop $4256\times2848$ images into $2128\times1424$ patches during the testing if needed.
    To avoid the blocking artifacts, we pad 200 pixels in the patch cropping for each patch. {Because of the extremely dark settings of SID, we adopt 16-bit sRGB images as input to produce higher quality results, and also provide results of the proposed method trained and tested on 8-bit sRGB, named REENet$_{8bit}$.}
    
			\begin{table*}[t]
		\begin{center}
			\caption{
				Quantitative evaluation comparing  learning-based methods and the proposed method.
				The best result is denoted in bold. 
			}
			\begin{tabular}{c|cccccccccc}
				\hline
				Method  
				& PSNR$\uparrow$  & {PSNR$\ast\uparrow$}
				&SSIM$\uparrow$ & {SSIM$\ast\uparrow$}
				&  VIF$\uparrow$  & {VIF$\ast\uparrow$}
				&  NIQE$\downarrow$ & {NIQE$\ast\downarrow$}
				& {LPIPS$\downarrow$} &{LPIPS$\ast\downarrow$}\\
				\hline
				LLNet~\cite{LLNet}
				&14.21 & {17.91}
				&0.221 & {0.547}
				&0.047 & {0.064}
				& 7.65 & {6.64}
				& {0.693} & {0.667}
				\\
				SICE~\cite{cai2018learning}	
				&14.26 &{17.31}
				& 0.366 & {0.621}
				& 0.011 & {0.049}
				& 13.25 & {15.27}
				& {0.766} & {0.715}
				\\
				KinD~\cite{Kind}		
				&13.50  & {16.71}
				& 0.109 & {0.361}
				& 0.048  & {0.076}
				&  9.68 & {7.11}
				& {0.718} & {0.703}
				\\
				DeepUPE~\cite{deepUPE} 		
				&12.10 & {15.16}
				& 0.070 & {0.455}
				&  0.028 & {0.116}
				&  11.28 & {7.99}
				& {0.772} & {0.887}
				\\
				{REENet$_{8bit}$}    
				& { 25.75} & {25.75}
				& {0.808} &{0.808}
				& {0.135} &{0.135}
				& {6.23} &{6.23}
				& {0.424} & {0.424}
				\\
				REENet	
				&\textbf{28.42} & {\textbf{28.42}}
				&  \textbf{0.880} &\textbf{0.880}
				&  \textbf{0.139}  &\textbf{0.139}
				& \textbf{5.60} & \textbf{5.60}
				& {\textbf{0.322}} &{\textbf{0.322}}
				\\
				\hline
			\end{tabular}
			\label{table:comp2}
		\end{center}
	\end{table*}

	\noindent \textbf{Comparison to Conventional Methods}.
	Our methods are compared with conventional methods:
    Dehazing~\cite{dong2011fast}, HE~\cite{HE1},  MSR~\cite{Multi_scale_retinex}, MF~\cite{FU201682},  BIMEF~\cite{BIMEF}, LIME~\cite{Guo_2017_Lime}, BPDHE~\cite{BPDHE}. 	
	In the quantitative evaluation, we adopt PSNR, VIF~\cite{VIF}, SSIM~\cite{SSIM} and LPIPS~\cite{lpips} as the full-reference metrics, 
    and NIQE~\cite{Mittal2013MakingA} as the no-reference metric.
    {
    Considering that some methods do not aim to produce the targeted illumination, 
    we adjust the brightness of these results with Gamma correction, where each image chooses the Gamma curve with the best PSNR to produce the final result. 
    Scores with brightness-aligned results are signified with $\ast$.
    }
	The comparison results are presented in Table~\ref{table:comp1}.
	
	\begin{figure*}[t]
        \centering
        \subfigure[w/o RAW]{
			\includegraphics[width=35mm]{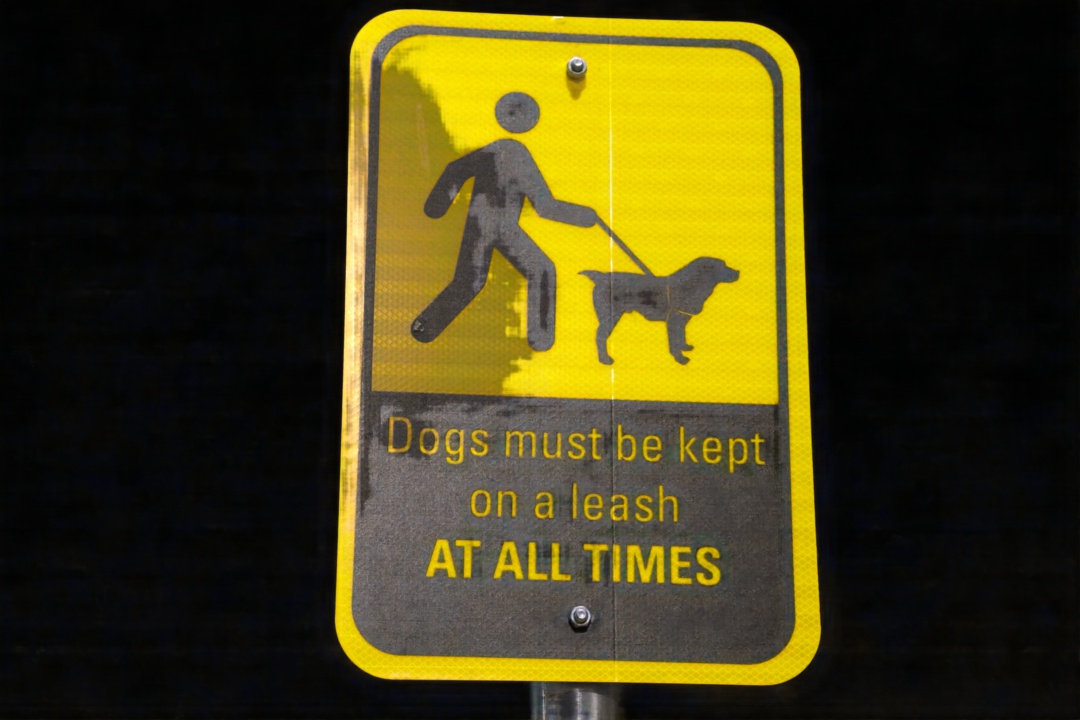}
		}
		\subfigure[w/o RAW, w/ MS-SSIM loss]{
		    \includegraphics[width=35mm]{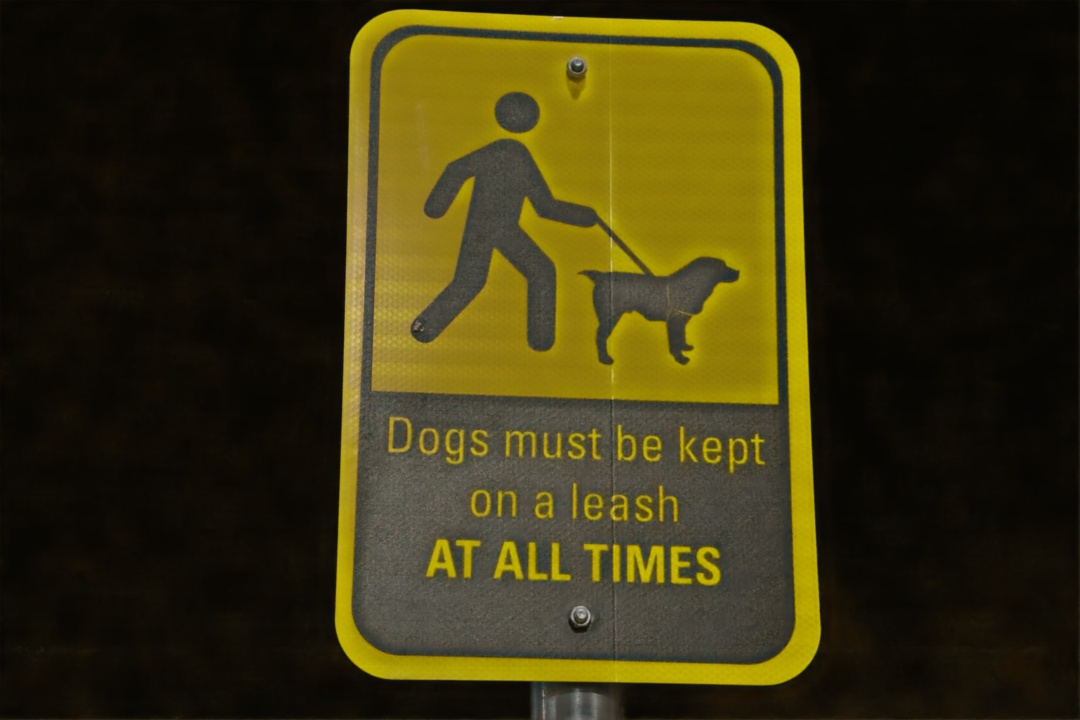}
		}
		\subfigure[w/o RAW, w/ estimated $\hat{\gamma}$]{
		    \includegraphics[width=35mm]{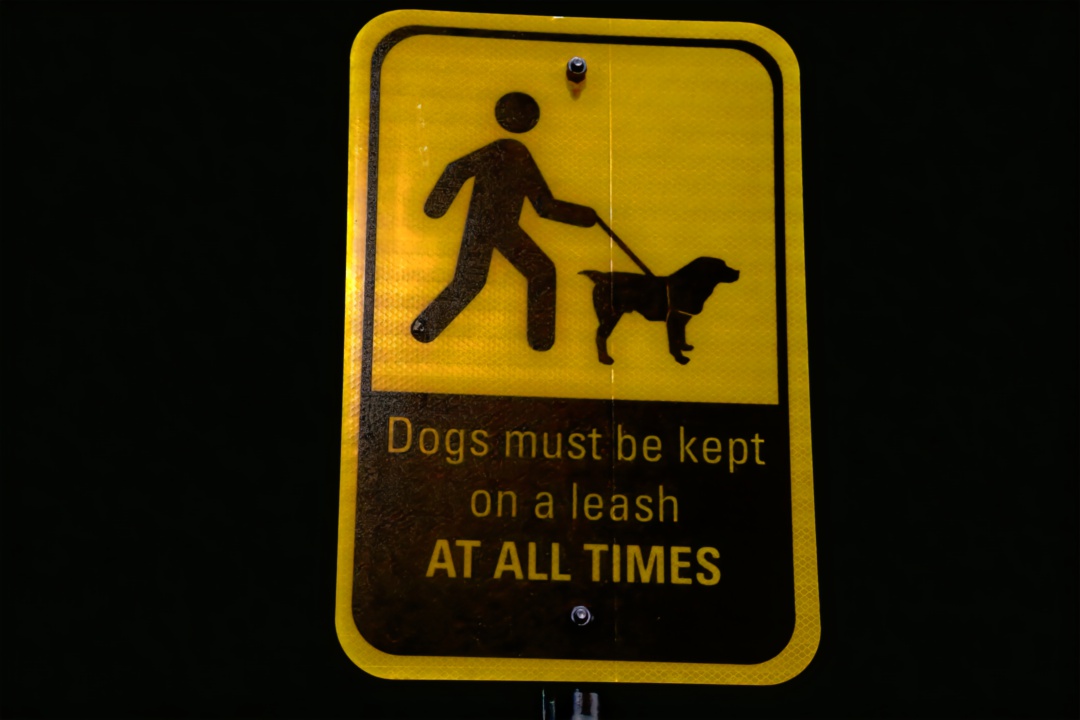}
		}
		\subfigure[w/ estimated $\hat{\gamma}$]{
		    \includegraphics[width=35mm]{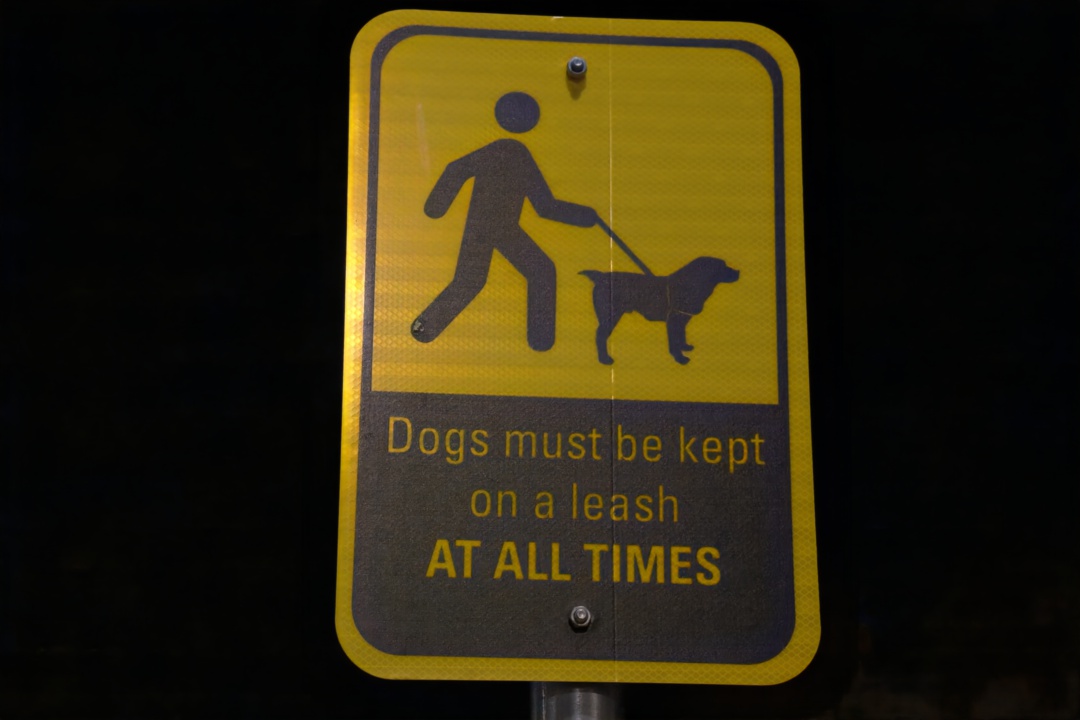}
		}
		\\
		\subfigure[w/o \textit{linear process}]{
		    \includegraphics[width=35mm]{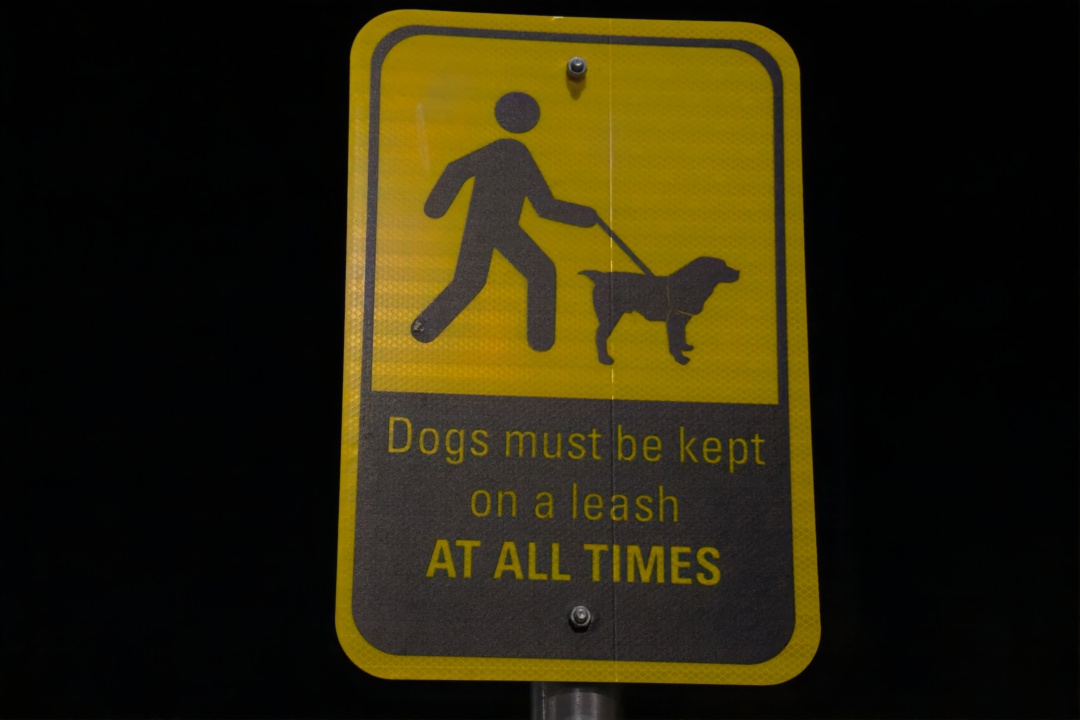}
		}
		\subfigure[w/ handcraft inverse Gamma]{
		    \includegraphics[width=35mm]{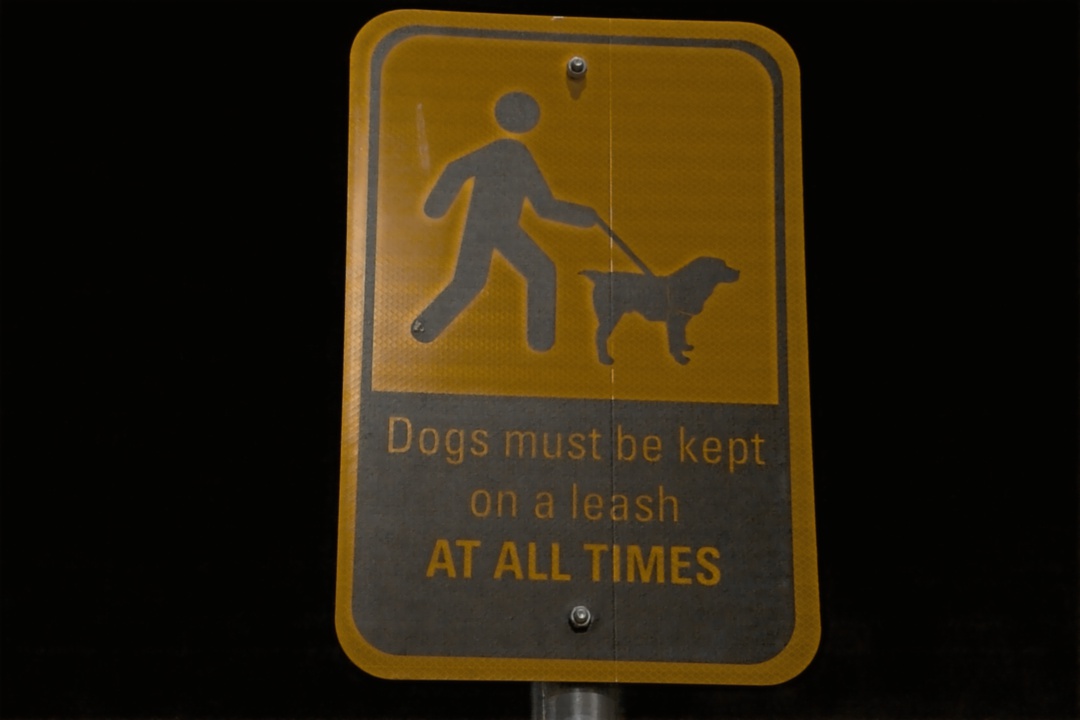}
		}
		\subfigure[REENet]{
		    \includegraphics[width=35mm]{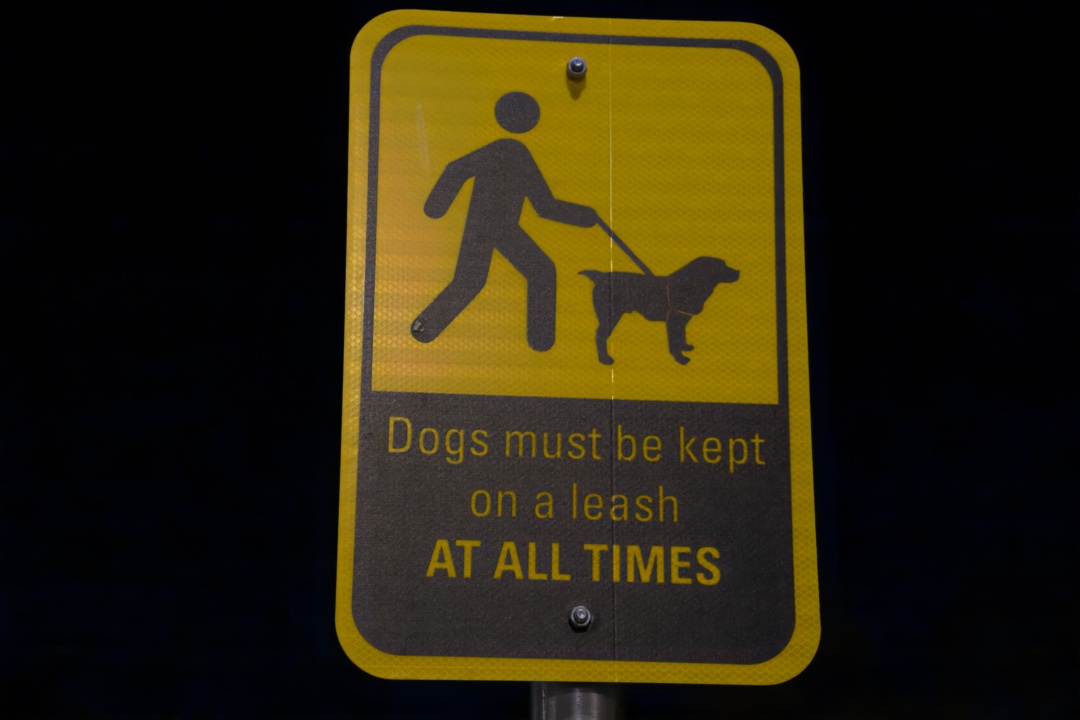}
		}
		\subfigure[Ground Truth]{
		    \includegraphics[width=35mm]{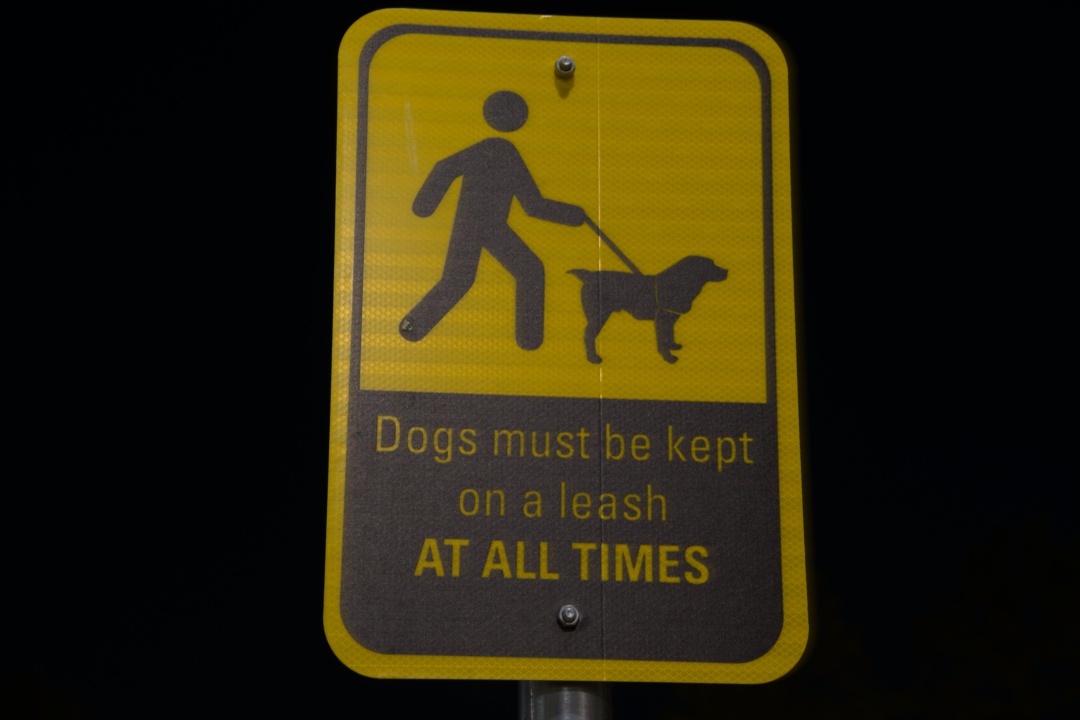}
		}
        \caption{{Visual comparison of different settings described in Table~\ref{table:abla}. The obvious subjective superiority of the proposed REENet proves the efficiency of our design.}}
        \label{fig:abla}
    \end{figure*}
	
	It is demonstrated that, our REENet achieves better performances than conventional methods on SID dataset in all metrics.
	We also show qualitative results in Fig.~\ref{fig:comp1}. 
	It is showed that, conventional methods might brighten images uniformly with observed under/over-exposure regions.
	Besides, SID dataset's images are extremely under-exposed.
	Hence, the enhanced results might include intensive noise and severe color casting or insufficient illumination. 
	{After using a Gamma correction to adjust the brightness, there is still obvious noise and color casting in the results of other methods.}
	
	\begin{figure}[b]
	    \centering
		\subfigure[Fidelity]{
			\includegraphics[width=40mm]{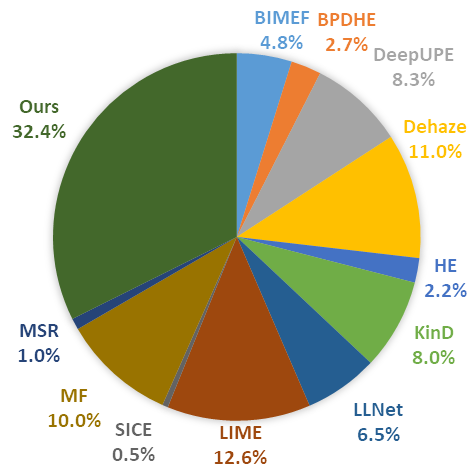}
		}
		\subfigure[Aesthetics]{
		    \includegraphics[width=40mm]{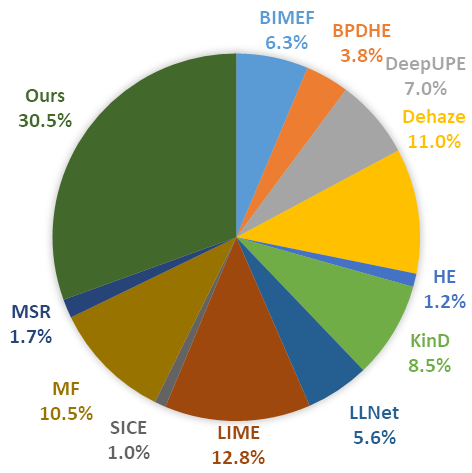}
		}
		\label{fig:user}
		\caption{The preference ratio on fidelity and aesthetics of different methods.}
	\end{figure}
    \vspace{1mm}
	
	\noindent \textbf{Comparison to Learning-Based Methods}. 
	The performances of different learning-based methods taking sRGB low-light images are compared on SID dataset, including SICE~\cite{cai2018learning}, LLNet~\cite{LLNet},  DeepUPE~\cite{deepUPE}, and KinD~\cite{Kind}.
	{When testing KinD, the brightening parameter is set to maximum allowed 5.0.}
	We rescale the resolution of input images when testing SICE because of the GPU memory limit.
	The quantitative results are illustrated in Table~\ref{table:comp2}, with the same setting of metrics.
	It is demonstrated that, if all methods take the processed sRGB images as input during the testing phase, 
	our REENet achieves better results on SID dataset.
	The qualitative results are presented in Fig.~\ref{fig:comp2}.
	It can be seen from Fig.~\ref{fig:comp2} that, without the help of the ratio, the results of these methods are severely under-exposed on SID.
	Comparatively, we obtain visually pleasant results.
	The results further demonstrate the superiority of our method and show the role that the exposure time adjustment play in helping improve the generalized enhancement performance.
	
	\noindent {\textbf{User Study}}.
	{
	Besides full-reference image quality metrics PSNR, SSIM, VIF, LPIPS, and non-reference image quality metric NIQE, we further perform the user study to evaluate the image quality of enhanced results by different methods.
	Besides the four cases shown in Fig.~\ref{fig:comp1} and Fig.~\ref{fig:comp2}, extra six cases are selected from the testing set to add up to 10 cases shown to the participants.
	Each subject is asked to select 3 from the 12 results that best match the target image (Fidelity) and have the best visual quality (Aesthetics).
	A total of 20 volunteers participate in this study and 400 selections are tallied. 
	As shown in Fig.~\ref{fig:user}, the proposed REENet obtains the best average preference ratio of 32.4\% and 30.5\% for both the fidelity and aesthetics, respectively, outperforming other methods.
	Note that, for a fair comparison, we use REENet$_{8bit}$ here and align the brightness of results of other methods.
	The user study quantitatively verifies the superiority of our method.
	}
	\begin{table}[b]
		\begin{center}
			\caption{
				Ablation study of our network architecture design.
			}
			\begin{tabular}{c|c}
				\hline
				$\qquad$ Architecture $\qquad$ & $\qquad$ PSNR$\uparrow$ $\qquad$\\
				\hline
				w/o RAW 									&$\qquad 22.11 \qquad$\\
				$\qquad$w/o RAW, with MS-SSIM loss$\qquad$&$\qquad 21.88 \qquad$\\
				w/o RAW, with estimated ratio $\hat{\gamma}$
				&$\qquad 24.27 \qquad$\\
				\hline
				w/ estimated ratio $\hat{\gamma}$			&$\qquad 25.23 \qquad$\\
				w/o \textit{linear process}						&$\qquad 28.17 \qquad$\\
				{w/ handcrafted inverse Gamma} & {$\qquad 23.90 \qquad$}\\
				REENet										&$\qquad \textbf{28.42} \qquad$\\
				\hline
			\end{tabular}
			\label{table:abla}
		\end{center}
	\end{table}
    
	\noindent \textbf{Ablation Studies}.
	We first perform the ablation study to evaluate the effectiveness of our architecture design in Table \ref{table:abla} {and Fig.~\ref{fig:abla}}.
	Firstly, we consider several versions \textit{i.e.} (the top three methods) only making use of processed sRGB low/normal-light pairs in both training and testing.
	It is observed that, very low PSNRs are obtained, even with the advanced loss, \textit{i.e.} MS-SSIM, and the estimated exposure time ratio $\hat{\gamma}$ via the mean pixel values.
	
	The experiment with the estimated exposure time ratio $\hat{\gamma}$ instead of the ground truth demonstrates the effect of utilizing the meta-data.
	The results using original RAW images reflect the importance of adopting linearly preprocessing RAW images.
	We can see the drop in the measures, as the adopted \textit{linear process} is effective in filling in the gap between RAW image and processed sRGB ones, which helps reduce the difficulty in simulating the whole processing system.
	{Using handcrafted inverse Gamma algorithm~\cite{lin2004radiometric} instead of U-Net as \textit{Unprocess} module will also cause a performance drop because unprocessing becomes less flexible and effective to deal with extremely dark conditions.}
	
	The quantitative results of each subnet are also provided in Table \ref{table:subnets}.
	The gaps among these subnets show how different stages in our design affect the enhancement performance.
	It is observed from the results, \textit{Unprocess} and \textit{Process} make efforts in an accurate nonlinear mapping, which leads to a small performance drop in measures. 
	The large gap between the Row. 1-2 and Row. 3 demonstrates that, it is quite challenging to predict the normal-light linear RGB images from the brightened images, which are severely degraded by intensive noise and color casting.
	\begin{table}[t]
		\begin{center}
			\caption{
				Ablation study for sub-networks taking different inputs.
			}
			\begin{tabular}{c|ccc|c}
				\hline
				\multirow{2}{*}{Inputs} & \multicolumn{3}{c|}{REENet}
				& \multirow{2}{*}{PSNR$\uparrow$} \\ \cline{2-4} 
				& \textit{Unprocess}&\textit{Enhance}&\textit{Process}
				& \\
				\hline
				Low sRGB
				&$\checkmark\quad$&$\checkmark\quad$&$\checkmark$
				& $ 28.42 $ \\
				
				Brightened Linear RGB
				&$\times\quad$&$\checkmark\quad$&$\checkmark$
				& $28.50 $ \\
				
				Normal Linear RGB
				&$\times\quad$&$\times\quad$&$\checkmark$
				& $ 42.68 $ \\
				\hline
			\end{tabular}
			\label{table:subnets}
		\end{center}
	\end{table}
	
\begin{figure}[b]
        \centering
        \subfigure[Brightened Input]{
			\includegraphics[width=25mm]{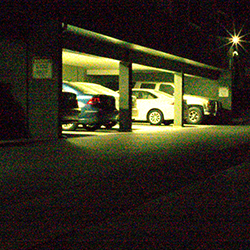}
		}
		\subfigure[Our Result]{
		    \includegraphics[width=25mm]{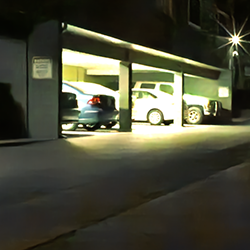}
		}
		\subfigure[Ground Truth]{
		    \includegraphics[width=25mm]{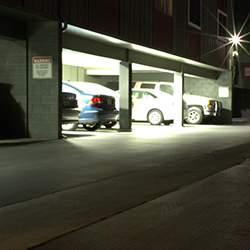}
		}
        \caption{A failure case of the proposed method. There is obvious color casting in the input which looks more yellow, and in our result, this degradation is not handled perfectly.}
        \label{fig:fail}
    \end{figure}

	\vspace{1mm}
	\noindent {\textbf{Failure Case}}.
	{
	A failure case of our method is shown in Fig.~\ref{fig:fail}. Once the input image is heavily degraded with color casting and proposed REENet can effectively enhance the illumination and suppress noise but still with obvious color casting.
	}
    
    \vspace{1mm}
    \noindent {\textbf{Limitations}}.
    {We have compared the running time of different methods as shown in Table~\ref{table:time}. Note that we adopt GPU to accelerate the running method if the code supports it. Due to the high resolution of the SID dataset, inference time is longer and the proposed REENet has a middle-level running time consumption.}
    
    {Besides, due to the highly camera-specific intensive noise in images captured in extremely dark environments and the diversity of ISP, our REENet cannot guarantee a promising performance when directly being applied to low-light images captured from another camera whose noise model is far away from our training set. We will address the issue in our future work.}
	
	\begin{table}[t]
		\begin{center}
			\caption{
				{Running Time (RT) evaluation comparing the proposed method and others. Due to the high resolution of the SID dataset, inference time is longer than usual and the proposed REENet has a middle-level velocity.}
			}
			\begin{tabular}{c|cc}
				\hline
				Method& RT (Sec.)$\downarrow$
				& Platform\\
				\hline
				HE~\cite{HE1}	&5.04 & MATLAB (CPU)\\
				BPDHE~\cite{BPDHE}&7.92 & MATLAB (CPU)\\
				Dehazing~\cite{dong2011fast}&19.74 & MATLAB (CPU)\\
				MSR~\cite{Multi_scale_retinex}&39.76 & MATLAB (CPU)\\
				MF~\cite{FU201682}&32.27 & MATLAB (CPU)\\
				LIME~\cite{Guo_2017_Lime}&8.49 & MATLAB (CPU)\\
				BIMEF~\cite{BIMEF}&23.5 & Matlab (CPU)\\
				LLNet~\cite{LLNet}&326.98 & Tensorflow (CPU)\\
				SICE~\cite{cai2018learning}&53.21 & Tensorflow (CPU) \\
				DeepUPE~\cite{deepUPE}& 52.67 & Tensorflow (GPU)\\
				KinD~\cite{Kind}& 6.31 & Tensorflow (GPU)\\
				REENet& 8.31 & Tensorflow (GPU)\\
				\hline
			\end{tabular}
			\label{table:time}
		\end{center}
	\end{table}

	\section{Conclusion}
	\label{sec:conc}
	In this paper, we make the first benchmarking effort to investigate the superiority of RAW for low light enhancement in detail.
	The characteristics of RAW files \textit{i.e.} linearity, the access to meta-data,
	fine-grained information (more abundant intensities and colors),
	inconvenience for display and lack of efficiency are detailed and
	their effects on the low-light enhancement are illustrated with quantitative results.
	For a fair evaluation, we take a novel view to regard low-light enhancement in a Factorized Enhancement Model (FEM) and obtain a precise and explicit description to decompose the ambiguities of this task into several measurable factors.
	Based on useful insights obtained from the benchmarking results, the proposed REENet adopts RAW-guiding strategy,  overcomes the issues brought by the nonlinearity of the sRGB images and the unavailability of RAW images in many applications, 
	and outperforms many state-of-the-art sRGB-based approaches.
	Our framework only needs to use RAW images during the training phase
	and offers better results with only sRGB inputs in testing,
	hence our results absorb the RAW's information as much as possible in the training
	but do not rely on the RAW input and adjust the ISP process in real applications.
	Experimental results show the superior performance of our method 
	and the rationality of our model design.

	\bibliographystyle{IEEEtran}
	\bibliography{mybib}

\begin{thebibliography}{10}
\providecommand{\url}[1]{#1}
\csname url@samestyle\endcsname
\providecommand{\newblock}{\relax}
\providecommand{\bibinfo}[2]{#2}
\providecommand{\BIBentrySTDinterwordspacing}{\spaceskip=0pt\relax}
\providecommand{\BIBentryALTinterwordstretchfactor}{4}
\providecommand{\BIBentryALTinterwordspacing}{\spaceskip=\fontdimen2\font plus
\BIBentryALTinterwordstretchfactor\fontdimen3\font minus
  \fontdimen4\font\relax}
\providecommand{\BIBforeignlanguage}[2]{{%
\expandafter\ifx\csname l@#1\endcsname\relax
\typeout{** WARNING: IEEEtran.bst: No hyphenation pattern has been}%
\typeout{** loaded for the language `#1'. Using the pattern for}%
\typeout{** the default language instead.}%
\else
\language=\csname l@#1\endcsname
\fi
#2}}
\providecommand{\BIBdecl}{\relax}
\BIBdecl

\bibitem{SID}
C.~Chen, Q.~Chen, J.~Xu, and V.~Koltun, ``Learning to see in the dark,'' in
  \emph{Proc.~IEEE/CVF Int'l Conf.~Comput. Vision and Pattern Recognit.}, Jun.
  2018, pp. 3291--3300.

\bibitem{Jiang_2019_ICCV}
H.~Jiang and Y.~Zheng, ``Learning to see moving objects in the dark,'' in
  \emph{Proc.~IEEE/CVF Int'l Conf.~Comput. Vision}, Oct. 2019, pp. 7323--7332.

\bibitem{CIEXYZNet}
M.~Afifi, A.~Abdelhamed, A.~Abuolaim, A.~Punnappurath, and M.~S. Brown, ``{CIE}
  {XYZ} net: Unprocessing images for low-level computer vision tasks,''
  \emph{{IEEE} Trans. Pattern Anal. Mach. Intell.}, 2021.

\bibitem{Xu_2020_CVPR}
K.~Xu, X.~Yang, B.~Yin, and R.~W. Lau, ``Learning to restore low-light images
  via decomposition-and-enhancement,'' in \emph{Proc.~IEEE/CVF Int'l
  Conf.~Comput. Vision and Pattern Recognit.}, Jun. 2020, pp. 2278--2287.

\bibitem{BPDHE}
H.~Ibrahim and N.~S.~P. Kong, ``Brightness preserving dynamic histogram
  equalization for image contrast enhancement,'' \emph{{IEEE} Trans. Consum.
  Electron.}, vol.~53, no.~4, pp. 1752--1758, Nov. 2007.

\bibitem{arici2009wahe}
T.~Arici, S.~Dikbas, and Y.~Altunbasak, ``A histogram modification framework
  and its application for image contrast enhancement,'' \emph{{IEEE} Trans.
  Image Process.}, vol.~18, no.~9, pp. 1921--1935, Sep. 2009.

\bibitem{dong2011fast}
X.~Dong, G.~Wang, Y.~Pang, W.~Li, J.~Wen, W.~Meng, and Y.~Lu, ``Fast efficient
  algorithm for enhancement of low lighting video,'' in \emph{Proc.~IEEE Int'l
  Conf.~Multimedia and Expo}, Jul. 2011, pp. 1--6.

\bibitem{Multi_scale_retinex}
D.~J. Jobson, Z.-U. Rahman, and G.~A. Woodell, ``A multiscale retinex for
  bridging the gap between color images and the human observation of scenes,''
  \emph{{IEEE} Trans. Image Process.}, vol.~6, no.~7, pp. 965--976, Jul. 1997.

\bibitem{LLNet}
K.~G. Lore, A.~Akintayo, and S.~Sarkar, ``{LLNet}: A deep autoencoder approach
  to natural low-light image enhancement,'' \emph{Pattern Recognition},
  vol.~61, pp. 650--662, Jan. 2017.

\bibitem{Kind}
Y.~{Zhang}, J.~{Zhang}, and X.~{Guo}, ``{Kindling the darkness: a practical
  low-light image enhancer},'' in \emph{Proc.~ACM Int'l Conf. Multimedia}, Oct.
  2019, pp. 1632--1640.

\bibitem{Chen_2019_ICCV}
C.~Chen, Q.~Chen, M.~Do, and V.~Koltun, ``Seeing motion in the dark,'' in
  \emph{Proc.~IEEE/CVF Int'l Conf.~Comput. Vision}, Oct. 2019, pp. 3184--3193.

\bibitem{Brown_ICCV_2019}
M.~S. Brown, ``Understanding color and the in-camera image processing pipeline
  for computer vision,'' in \emph{Proc.~IEEE/CVF Int'l Conf.~Comput. Vision
  Tutorial}, Oct. 2019.

\bibitem{DeepISP}
E.~Schwartz, R.~Giryes, and A.~M. Bronstein, ``{DeepISP}: Toward learning an
  end-to-end image processing pipeline,'' \emph{{IEEE} Trans. Image Process.},
  vol.~28, no.~2, pp. 912--923, Feb. 2019.

\bibitem{Wei_2020_CVPR}
K.~Wei, Y.~Fu, J.~Yang, and H.~Huang, ``A physics-based noise formation model
  for extreme low-light raw denoising,'' in \emph{Proc.~IEEE/CVF Int'l
  Conf.~Comput. Vision and Pattern Recognit.}, Jun. 2020, pp. 2755--2764.

\bibitem{EEMEFN}
M.~Zhu, P.~Pan, W.~Chen, and Y.~Yang, ``{EEMEFN}: Low-light image enhancement
  via edge-enhanced multi-exposure fusion network,'' in \emph{Proc.~AAAI Conf.
  on Artif. Intell.}, Feb. 2020, pp. 13\,106--13\,113.

\bibitem{HE1}
S.~M. Pizer, R.~E. Johnston, J.~P. Ericksen, B.~C. Yankaskas, and K.~E. Muller,
  ``Contrast-limited adaptive histogram equalization: speed and
  effectiveness,'' in \emph{Proc. Conf. Vis. in Biomed. Comput.}, May 1990, pp.
  337--345.

\bibitem{HE2}
M.~Abdullah-Al-Wadud, M.~Kabir, M.~Dewan, and O.~Chae, ``A dynamic histogram
  equalization for image contrast enhancement,'' \emph{{IEEE} Trans. Consum.
  Electron.}, vol.~53, no.~2, pp. 593--600, May 2007.

\bibitem{HE3}
C.~Lee, J.-H. Kim, C.~Lee, and C.-S. Kim, ``Optimized brightness compensation
  and contrast enhancement for transmissive liquid crystal displays,''
  \emph{{IEEE} Trans. Circuits Syst. Video Technol.}, vol.~24, no.~4, pp.
  576--590, Apr. 2014.

\bibitem{DICM}
C.~Lee, C.~Lee, and C.-S. Kim, ``Contrast enhancement based on layered
  difference representation of {2D} histograms,'' \emph{{IEEE} Trans. Image
  Process.}, vol.~22, no.~12, pp. 5372--5384, Dec. 2013.

\bibitem{nakai2013dheci}
K.~Nakai, Y.~Hoshi, and A.~Taguchi, ``Color image contrast enhacement method
  based on differential intensity/saturation gray-levels histograms,'' in
  \emph{Proc. Int'l Symp. Intell. Signal Process. and Communication Systems},
  Nov. 2013, pp. 445--449.

\bibitem{dehaze2}
X.~Zhang, P.~Shen, L.~Luo, L.~Zhang, and J.~Song, ``Enhancement and noise
  reduction of very low light level images,'' in \emph{Proc.~IEEE Int'l
  Conf.~Pattern Recognit.}, Nov. 2012, pp. 2034--2037.

\bibitem{dehaze1}
L.~Li, R.~Wang, W.~Wang, and W.~Gao, ``A low-light image enhancement method for
  both denoising and contrast enlarging,'' in \emph{Proc.~IEEE Int'l
  Conf.~Image Process.}, 2015, pp. 3730--3734.

\bibitem{pdpf}
Q.~Zhang, Y.~Nie, L.~Zhang, and C.~Xiao, ``Underexposed video enhancement via
  perception-driven progressive fusion,'' \emph{{IEEE} Trans. Vis. Comput.
  Graphics}, vol.~22, no.~6, pp. 1773--1785, Jun. 2016.

\bibitem{celik2011cvc}
T.~Celik and T.~Tjahjadi, ``Contextual and variational contrast enhancement,''
  \emph{{IEEE} Trans. Image Process.}, vol.~20, no.~12, pp. 3431--3441, Dec.
  2011.

\bibitem{plm}
S.-Y. Yu and H.~Zhu, ``Low-illumination image enhancement algorithm based on a
  physical lighting model,'' \emph{{IEEE} Trans. Circuits Syst. Video
  Technol.}, vol.~29, no.~1, pp. 28--37, Jan. 2019.

\bibitem{BIMEF}
Z.~Ying, G.~Li, and W.~Gao, ``A bio-inspired multi-exposure fusion framework
  for low-light image enhancement,'' \emph{arXiv e-prints}, Nov. 2017.

\bibitem{Single_scale_retinex}
D.~J. Jobson, Z.-U. Rahman, and G.~A. Woodell, ``Properties and performance of
  a center/surround retinex,'' \emph{{IEEE} Trans. Image Process.}, vol.~6,
  no.~3, pp. 451--462, Mar. 1997.

\bibitem{lee2013amsr}
C.-H. Lee, J.-L. Shih, C.-C. Lien, and C.-C. Han, ``Adaptive multiscale retinex
  for image contrast enhancement,'' in \emph{Proc. Int'l Conf. Signal-Image
  Technol. \& Internet-Based Systems}, Dec. 2013, pp. 43--50.

\bibitem{NPE}
S.~Wang, J.~Zheng, H.-M. Hu, and B.~Li, ``Naturalness preserved enhancement
  algorithm for non-uniform illumination images,'' \emph{{IEEE} Trans. Image
  Process.}, vol.~22, no.~9, pp. 3538--3548, Sep. 2013.

\bibitem{novel_retinex}
X.~Fu, Y.~Sun, M.~LiWang, Y.~Huang, X.-P. Zhang, and X.~Ding, ``A novel retinex
  based approach for image enhancement with illumination adjustment,'' in
  \emph{Proc.~IEEE Int'l Conf.~Acoust., Speech, and Signal Process.}, May 2014,
  pp. 1190--1194.

\bibitem{FU201682}
X.~Fu, D.~Zeng, Y.~Huang, Y.~Liao, X.~Ding, and J.~Paisley, ``A fusion-based
  enhancing method for weakly illuminated images,'' \emph{Signal Process.},
  vol. 129, pp. 82--96, Dec. 2016.

\bibitem{Li_2017_SRRM}
M.~Li, J.~Liu, W.~Yang, X.~Sun, and Z.~Guo, ``Structure-revealing low-light
  image enhancement via robust retinex model,'' \emph{{IEEE} Trans. Image
  Process.}, vol.~27, no.~6, pp. 2828--2841, Jun. 2018.

\bibitem{Ren_2018_SD}
X.~Ren, M.~Li, W.-H. Cheng, and J.~Liu, ``Joint enhancement and denoising
  method via sequential decomposition,'' in \emph{IEEE Int'l Symp. Circuits and
  Systems}, Apr. 2018, pp. 1--5.

\bibitem{MSRNet_2017}
L.~Shen, Z.~Yue, F.~Feng, Q.~Chen, S.~Liu, and J.~Ma, ``{MSR}-net:low-light
  image enhancement using deep convolutional network,'' \emph{arXiv e-prints},
  Nov. 2017.

\bibitem{LLCNN}
L.~Tao, C.~Zhu, G.~Xiang, Y.~Li, H.~Jia, and X.~Xie, ``{LLCNN}: A convolutional
  neural network for low-light image enhancement,'' in \emph{Proc.~IEEE~Vis.
  Commun. and Image Process.}, Dec. 2017, pp. 1--4.

\bibitem{Lv2018MBLLEN}
F.~Lv, F.~Lu, J.~Wu, and C.~Lim, ``{MBLLEN}: Low-light image/video enhancement
  using cnns,'' in \emph{Brit. Mach. Vision Conf.}, Sep. 2018, p. 220.

\bibitem{cai2018learning}
J.~Cai, S.~Gu, and L.~Zhang, ``Learning a deep single image contrast enhancer
  from multi-exposure images,'' \emph{{IEEE} Trans. Image Process.}, vol.~27,
  no.~4, pp. 2049--2062, Apr. 2018.

\bibitem{deepUPE}
R.~Wang, Q.~Zhang, C.-W. Fu, X.~Shen, W.-S. Zheng, and J.~Jia, ``Underexposed
  photo enhancement using deep illumination estimation,'' in
  \emph{Proc.~IEEE/CVF Int'l Conf.~Comput. Vision and Pattern Recognit.}, Jun.
  2019, pp. 6842--6850.

\bibitem{DRD}
C.~Wei*, W.~Wang*, W.~Yang, and J.~Liu, ``Deep retinex decomposition for
  low-light enhancement,'' in \emph{Brit. Mach. Vision Conf.}, Sep. 2018, p.
  155.

\bibitem{jiang2019enlightengan}
Y.~Jiang, X.~Gong, D.~Liu, Y.~Cheng, C.~Fang, X.~Shen, J.~Yang, P.~Zhou, and
  Z.~Wang, ``{EnlightenGAN}: Deep light enhancement without paired
  supervision,'' \emph{{IEEE} Trans. Image Process.}, vol.~30, pp. 2340--2349,
  Jan. 2021.

\bibitem{Zero-DCE}
C.~Guo, C.~Li, J.~Guo, C.~C. Loy, J.~Hou, S.~Kwong, and R.~Cong,
  ``Zero-reference deep curve estimation for low-light image enhancement,'' in
  \emph{Proc.~IEEE/CVF Int'l Conf.~Comput. Vision and Pattern Recognit.}, Jun.
  2020, pp. 1777--1786.

\bibitem{Yang_2020_CVPR}
W.~Yang, S.~Wang, Y.~Fang, Y.~Wang, and J.~Liu, ``From fidelity to perceptual
  quality: A semi-supervised approach for low-light image enhancement,'' in
  \emph{Proc.~IEEE/CVF Int'l Conf.~Comput. Vision and Pattern Recognit.}, Jun.
  2020, pp. 3060--3069.

\bibitem{Gharbi_2017_SIGGRAPH}
M.~Gharbi, J.~Chen, J.~T. Barron, S.~W. Hasinoff, and F.~Durand, ``Deep
  bilateral learning for real-time image enhancement,'' \emph{{ACM} Trans.
  Graph.}, vol.~36, no.~4, pp. 118:1--118:12, Jul. 2017.

\bibitem{Yang_2018_CVPR}
X.~Yang, K.~Xu, Y.~Song, Q.~Zhang, X.~Wei, and R.~W.~H. Lau, ``Image correction
  via deep reciprocating {HDR} transformation,'' in \emph{Proc.~IEEE/CVF Int'l
  Conf.~Comput. Vision and Pattern Recognit.}, Jun. 2018, pp. 1798--1807.

\bibitem{Afifi_2021_CVPR}
M.~Afifi, K.~G. Derpanis, B.~Ommer, and M.~S. Brown, ``Learning multi-scale
  photo exposure correction,'' in \emph{Proc.~IEEE/CVF Int'l Conf.~Comput.
  Vision and Pattern Recognit.}, Jun. 2021, pp. 9157--9167.

\bibitem{Wolf_2021_CVPR}
V.~Wolf, A.~Lugmayr, M.~Danelljan, L.~V. Gool, and R.~Timofte, ``{DeFlow}:
  Learning complex image degradations from unpaired data with conditional
  flows,'' in \emph{Proc.~IEEE/CVF Int'l Conf.~Comput. Vision and Pattern
  Recognit.}, Jun. 2021, pp. 94--103.

\bibitem{brooks2019unprocessing}
T.~Brooks, B.~Mildenhall, T.~Xue, J.~Chen, D.~Sharlet, and J.~T. Barron,
  ``Unprocessing images for learned raw denoising,'' in \emph{Proc.~IEEE/CVF
  Int'l Conf.~Comput. Vision and Pattern Recognit.}, Jun. 2019, pp.
  11\,036--11\,045.

\bibitem{CycleISP}
S.~W. Zamir, A.~Arora, S.~Khan, M.~Hayat, F.~S. Khan, M.-H. Yang, and L.~Shao,
  ``{CycleISP}: Real image restoration via improved data synthesis,'' in
  \emph{Proc.~IEEE/CVF Int'l Conf.~Comput. Vision and Pattern Recognit.}, Jun.
  2020, pp. 2693--2702.

\bibitem{Hasinoff2014}
S.~W. Hasinoff, \emph{Photon, Poisson Noise}.\hskip 1em plus 0.5em minus
  0.4em\relax Boston, MA: Springer US, 2014, pp. 608--610.

\bibitem{Foi_TIP_2008}
A.~Foi, M.~Trimeche, V.~Katkovnik, and K.~Egiazarian, ``Practical
  poissonian-gaussian noise modeling and fitting for single-image raw-data,''
  \emph{{IEEE} Trans. Image Process.}, vol.~17, no.~10, pp. 1737--1754, Oct.
  2008.

\bibitem{U-Net}
O.~Ronneberger, P.~Fischer, and T.~Brox, ``{U-Net}: Convolutional networks for
  biomedical image segmentation,'' in \emph{Proc. Med. Image Comput. and
  Computer-Assisted Intervention}, vol. 9351, Oct. 2015, pp. 234--241.

\bibitem{Guo_2017_Lime}
X.~Guo, Y.~Li, and H.~Ling, ``{LIME}: Low-light image enhancement via
  illumination map estimation,'' \emph{{IEEE} Trans. Image Process.}, vol.~26,
  no.~2, pp. 982--993, Feb. 2017.

\bibitem{SSIM}
Z.~Wang, A.~C. Bovik, H.~R. Sheikh, and E.~P. Simoncell, ``Image quality
  assessment: From error visibility to structural similarity,'' \emph{{IEEE}
  Trans. Image Process.}, vol.~15, no.~2, pp. 430--444, Feb. 2004.

\bibitem{VIF}
H.~R. Sheikh and A.~C. Bovik, ``Image information and visual quality,''
  \emph{{IEEE} Trans. Image Process.}, vol.~15, no.~2, pp. 430--444, Feb. 2006.

\bibitem{Mittal2013MakingA}
A.~Mittal, R.~Soundararajan, and A.~C. Bovik, ``Making a “completely blind”
  image quality analyzer,'' \emph{{IEEE} Signal Process. Lett.}, vol.~20,
  no.~3, pp. 209--212, Mar. 2013.

\bibitem{lpips}
R.~Zhang, P.~Isola, A.~A. Efros, E.~Shechtman, and O.~Wang, ``The unreasonable
  effectiveness of deep features as a perceptual metric,'' in
  \emph{Proc.~IEEE/CVF Int'l Conf.~Comput. Vision and Pattern Recognit.}, Jun.
  2018, pp. 586--595.

\bibitem{H2012Image}
H.~C. Burger, C.~J. Schuler, and S.~Harmeling, ``Image denoising: Can plain
  neural networks compete with {BM3D}?'' in \emph{Proc.~IEEE/CVF Int'l
  Conf.~Comput. Vision and Pattern Recognit.}, Jun. 2012, pp. 2392--2399.

\bibitem{Liu_2018_CVPR_Workshops}
P.~Liu, H.~Zhang, K.~Zhang, L.~Lin, and W.~Zuo, ``Multi-level wavelet-cnn for
  image restoration,'' in \emph{Proc.~IEEE/CVF Int'l Conf.~Comput. Vision and
  Pattern Recognit. Workshop}, Jun. 2018, pp. 773--782.

\bibitem{Adam}
D.~P. Kingma and J.~Ba, ``Adam: {A} method for stochastic optimization,'' in
  \emph{Proc.~Int'l Conf.~Learn. Representations}, May 2015.

\bibitem{lin2004radiometric}
S.~Lin, J.~Gu, S.~Yamazaki, and H.~Shum, ``Radiometric calibration from a
  single image,'' in \emph{Proc.~IEEE Int'l Conf.~Comput. Vision and Pattern
  Recognit.}, Jun. 2004, pp. 938--945.

\end{thebibliography}
	\begin{IEEEbiography}[{\includegraphics[width=1in,height=1.25in,keepaspectratio]{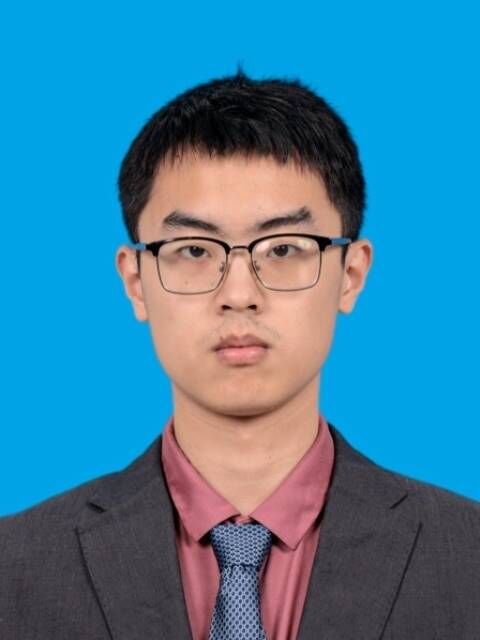}}]
    {Haofeng Huang} (Student Member, IEEE)
    received the B.S. degree in computer science from Peking University, Beijing, China in 2021, where he is currently
working toward the Ph.D. degree with the Wangxuan
Institute of Computer Technology, Peking
University. His current research interests include deep-learning based image/video comprssion, image/video coding for machines, and intelligent visual enhancement.
\end{IEEEbiography}

\begin{IEEEbiography}[{\includegraphics[width=1in,height=1.25in,clip,keepaspectratio]{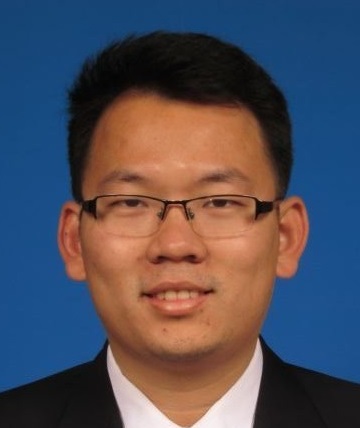}}]
	{Wenhan Yang} (Member, IEEE)
	received the B.S degree and Ph.D. degree (Hons.) in computer science from Peking University, Beijing, China, in 2012 and 2018. He is currently a postdoctoral research fellow with the Department of Computer Science, City University of Hong Kong. Dr. His current research interests include image/video processing/restoration, bad weather restoration, human-machine collaborative coding. He has authored over 100 technical articles in refereed journals and proceedings, and holds 9 granted patents.
	He received the IEEE ICME-2020 Best Paper Award, the IFTC 2017 Best Paper Award, and the IEEE CVPR-2018 UG2 Challenge First Runner-up Award.
	He was the Candidate of CSIG Best Doctoral Dissertation Award in 2019. He served as the Area Chair of IEEE ICME-2021, and the Organizer of IEEE CVPR-2019/2020/2021 UG2+ Challenge and Workshop.
\end{IEEEbiography}

\begin{IEEEbiography}[{\includegraphics[width=1in,height=1.25in,keepaspectratio]{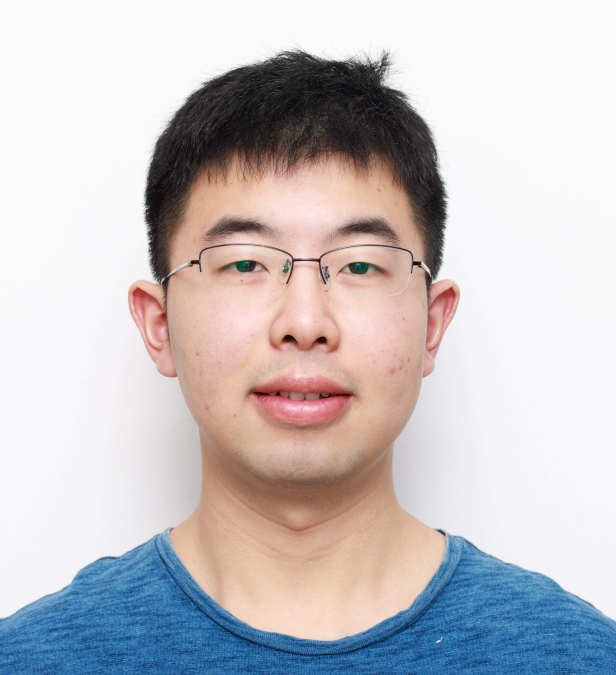}}]
     {Yueyu Hu} (Graduate Student Member, IEEE)
     received the B.S. degree and M.S. degree in computer science from Peking University, Beijing, China, in 2018 and 2021, respectively. He is currently working toward the Ph.D. degree at New York University, New York, NY. His current research interests include machine learning inspired 2D and 3D image compression and processing. He received the Best Paper Award at IEEE ICME-2020.
\end{IEEEbiography}

\begin{IEEEbiography}[{\includegraphics[width=1in,height=1.25in,clip,keepaspectratio]{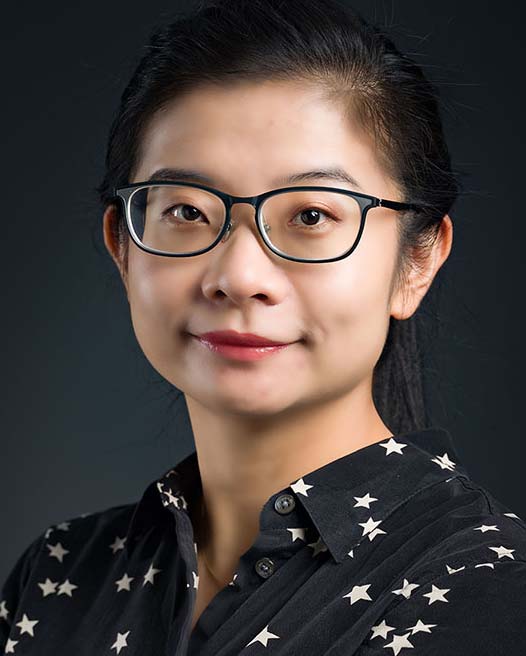}}]
	{Jiaying Liu} (Senior Member, IEEE)
	received the Ph.D. degree (Hons.) in computer science from Peking University, Beijing, China, 2010. She is currently an Associate Professor, Boya Young Fellow with the Wangxuan Institute of Computer Technology, Peking University, China. She has authored more than 100 technical articles in refereed journals and proceedings, and holds 60 granted patents. Her current research interests include multimedia signal processing, compression, and computer vision. She is a senior member of IEEE, CSIG and CCF. She was a visiting scholar with the University of Southern California, Los Angeles, California, from 2007 to 2008. She was a visiting researcher with Microsoft Research Asia, in 2015 supported by the Star Track Young Faculties Award. She has served as a member of Multimedia Systems and Applications Technical Committee (MSA TC), and Visual Signal Processing and Communications Technical Committee (VSPC TC) in IEEE Circuits and Systems Society. She received the IEEE ICME 2020 Best Paper Award and IEEE MMSP 2015 Top10\% Paper Award. She has also served as the Associate Editor of the IEEE Trans. on Image Processing, the IEEE Trans. on Circuit System for Video Technology and Journal of Visual Communication and Image Representation, the Technical Program Chair of IEEE ICME-2021/ACM ICMR-2021, the Area Chair of CVPR-2021/ECCV-2020/ICCV-2019, and the CAS Representative at the ICME Steering Committee. She was the APSIPA Distinguished Lecturer (2016-2017).
\end{IEEEbiography}

\begin{IEEEbiography}[{\includegraphics[width=1in,height=1.25in,clip,keepaspectratio]{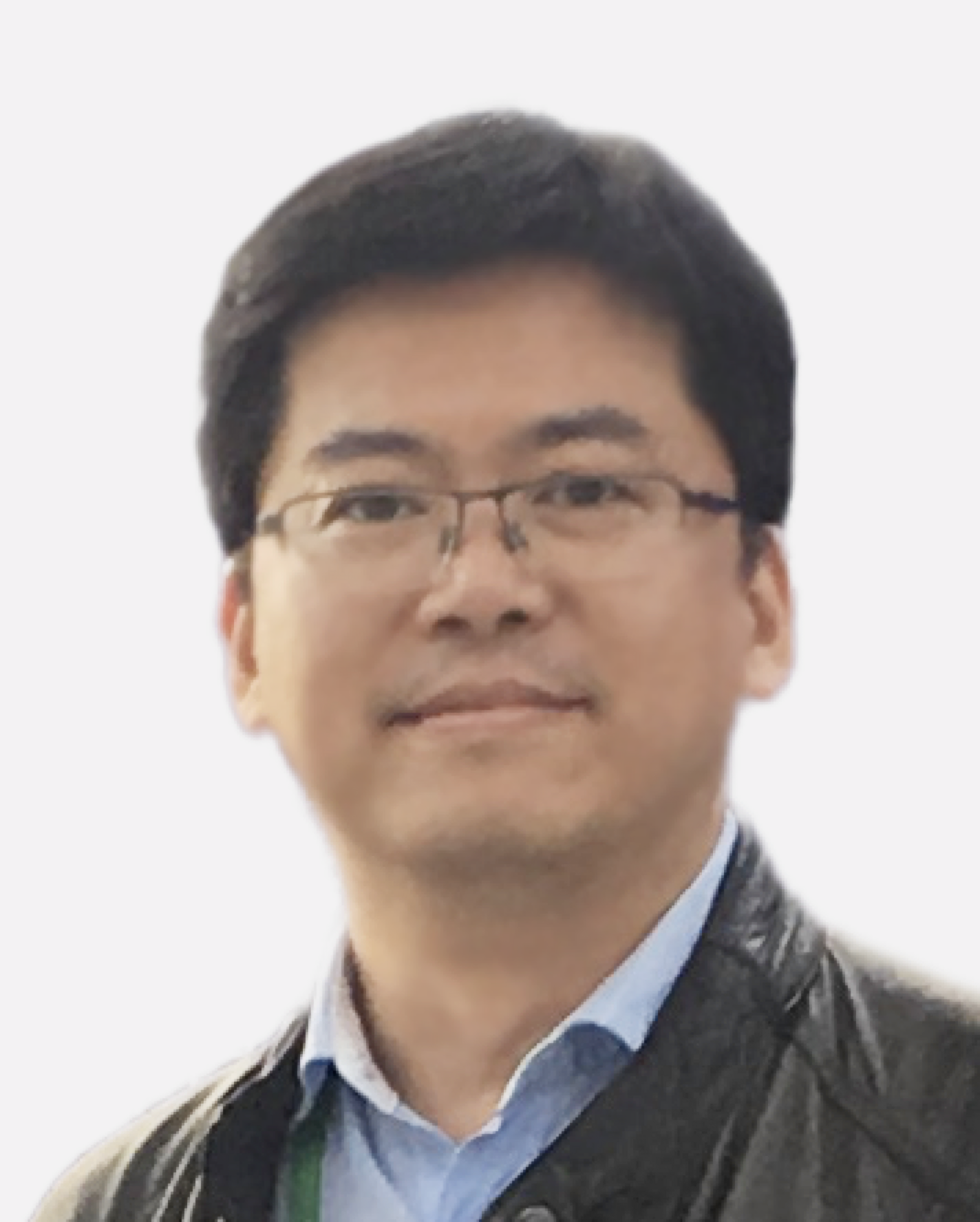}}]
	{Ling-Yu Duan} (Member, IEEE)
	is a Full Professor with the National Engineering Laboratory of Video Technology (NELVT), School of Electronics Engineering and Computer Science, Peking University (PKU), China, and has served as the Associate Director of the Rapid-Rich Object Search Laboratory (ROSE), a joint lab between Nanyang Technological University (NTU), Singapore, and Peking University (PKU), China since 2012. He is also with Peng Cheng Laboratory, Shenzhen, China, since 2019. He received the Ph.D. degree in information technology from The University of Newcastle, Callaghan, Australia, in 2008. His research interests include multimedia indexing, search, and retrieval, mobile visual search, visual feature coding, and video analytics, etc. He has published about 200 research papers. He received the IEEE ICME Best Paper Award in 2019/2020, the IEEE VCIP Best Paper Award in 2019, and EURASIP Journal on Image and Video Processing Best Paper Award in 2015, the Ministry of Education Technology Invention Award (First Prize) in 2016, the National Technology Invention Award (Second Prize) in 2017, China Patent Award for Excellence (2017), the National Information Technology Standardization Technical Committee ``Standardization Work Outstanding Person" Award in 2015. He was a Co-Editor of MPEG Compact Descriptor for Visual Search (CDVS) Standard (ISO/IEC 15938-13) and MPEG Compact Descriptor for Video Analytics (CDVA) standard (ISO/IEC 15938-15). Currently he is an Associate Editor of IEEE Transactions on Multimedia, ACM Transactions on Intelligent Systems and Technology and ACM Transactions on Multimedia Computing, Communications, and Applications, and serves as the area chairs of ACM MM and IEEE ICME.
\end{IEEEbiography}

\end{document}